\documentclass[pdflatex,sn-mathphys-num]{sn-jnl}

\usepackage{graphicx}%
\usepackage{multirow}%
\usepackage{amsmath,amssymb,amsfonts}%
\usepackage{amsthm}%
\usepackage{mathrsfs}%
\usepackage[title]{appendix}%
\usepackage{xcolor}%
\usepackage{textcomp}%
\usepackage{manyfoot}%
\usepackage{booktabs}%
\usepackage{algorithm}%
\usepackage{algorithmicx}%
\usepackage{algpseudocode}%
\usepackage{listings}%
\usepackage{graphicx}
\usepackage{epstopdf,epsfig}
\usepackage{newtxtext}

\usepackage{amsmath}
\usepackage{dsfont}

\usepackage{wrapfig}
\usepackage{xcolor}
\usepackage{tikz}
\usepackage{float}
\usepackage{subcaption}

\usepackage{mathtools}

\usepackage{gensymb}

\usepackage{tikz}
\usepackage{pgfplots}
\usetikzlibrary{arrows.meta}
\usepackage{cancel}
\usepackage{soul}

\theoremstyle{thmstyleone}%
\theoremstyle{thmstyletwo}%

\theoremstyle{thmstylethree}%

\begin{document}

\title{A carrier-wave factored one-way Navier--Stokes method for boundary-layer instability modelling}


\author*[1]{\sur{Elliot James Badcock}}\email{ejb321@ic.ac.uk}

\author[1]{\sur{Shahid Mughal}}\email{s.mughal@imperial.ac.uk}

\affil*[1]{\orgdiv{Department of Mathematics}, \orgname{Imperial College London}, \orgaddress{\street{South Kensington Campus}, \city{London}, \postcode{SW7 2AZ}, \country{United Kingdom}}}


\abstract{We present M-OWNS, a spatial marching method that combines the carrier-wave factoring of the parabolised stability equations (PSE) with a recursive one-way Navier--Stokes (OWNS-R) projection framework. A distinct numerical resolution and efficiency advantage is offered by the approach, in modelling disturbance and instability state evolution.
A spectral resolution comparison analysis shows that {{to leading order,}} for any excited eigenfunction whose eigenvalue lies closer to the carrier wavenumber than to the
origin, the wave-factored system resolves the mode at a coarser streamwise numerical step
size {relative to} the unfactored system.
A non-iterating variant, with the carrier wavenumber determined from the base flow, temporal frequency and spanwise wavenumber alone, achieves {{equivalent resolution accuracy}} at identical per-step cost to unfactored OWNS. For the fixed-carrier variant, M-OWNS reduces the total solve count by factors of two to eight relative to unfactored OWNS across the test cases considered, with larger reductions {{possible}} when the iterated closure {{condition of PSE is suitable.}} The method is validated across incompressible and subsonic
flat-plate boundary-layers, three-dimensional crossflow {{disturbances}},
and a Mach~4.5 hypersonic boundary-layer with four forcing configurations:
eigenfunction inlet forcing, wall suction/blowing, multi-mode freestream forcing and randomised inlet forcing. The wall suction/blowing case is validated against a fully elliptic {{linear harmonic Navier--Stokes}} solver.
For deterministic forcing {{scenarios}}, M-OWNS captures disturbance amplitudes, acoustic radiation fields, and modal synchronisation sequences at coarser streamwise resolution than unfactored OWNS. Under broadband randomised forcing, {{M-OWNS resolves mixed-mode
disturbance development at half the numerical cost relative to standard OWNS.}}}


\maketitle

\section{Introduction}\label{sec:intro}


Predicting laminar--turbulent transition in boundary layers requires models that
can track disturbance evolution from receptivity through amplification across a
developing base flow. The simplest method, local eigenvalue analysis (LST),
assumes parallel flow and examines one streamwise station at a time
\citep{mackBoundaryLayerLinearStability1984a}. At the other extreme, global
methods solve the full elliptic linearised harmonic Navier--Stokes equations (LHNS), capturing
both upstream and downstream disturbance dynamics without invoking a parallel-flow assumption
\citep{theofilisGlobalLinearInstability2011,appelBoundaryLayerInstabilities2021}.
The generality comes at substantial cost, principally through memory
constraints and the need to invert large matrices. For convective
disturbances, where upstream and downstream dynamics evolve independently,
spatial marching methods offer an attractive alternative: the solution is
advanced station by station in the downstream direction at a fraction of the
global cost. The difficulty is that the LHNS {{operator}}
supports upstream-propagating acoustic modes alongside the
downstream-propagating disturbances of interest.
These upstream modes can destabilise a na\"ive spatial march; practical methods must either remove, suppress or neglect them at coarser resolutions.


The parabolised stability equations (PSE) contend with upstream modes through
a WKB-like wave-carrier factoring: the disturbance is decomposed into a rapidly varying carrier
wave and a slowly varying envelope, with the carrier wavenumber~$\alpha_0$
updated iteratively at each streamwise station
\citep{herbertParabolizedStabilityEquations1997,bertolottiAnalysisLinearStability1991}.
The factoring shifts the downstream excited spectrum toward the origin, so the
physically relevant modes can be resolved at coarse streamwise resolution. For single mode disturbances this is highly effective; the iteration drives~$\alpha_0$ toward the wavenumber of the dominant mode, stretching its wavelength. However, the upstream
modes remain in the system. At coarse resolution they fall outside the
integrator's accuracy region and cause no harm, but when fine resolution is
needed (multi-modal/non-modal disturbances, receptivity near the leading edge
\citep{turnerAsymptoticReceptivityAnalysis2006a}, nonlinear calculations) the
accuracy region widens and captures them, destabilising the {{numerical}} marching {{procedure}} 
\citep{liNaturePSEApproximation,liSpectralAnalysisParabolized1997}.
Regularisation strategies (implicit damping \emph{e.g.} backward Euler marching
\citep{anderssonStabilizationProcedureParabolic1998} or neglecting the
streamwise pressure gradient \citep{herbertParabolizedStabilityEquations1997}) permit
finer steps but introduce characteristic errors in the wave ansatz and shape
function. \citet{towneCriticalParabolized2019} trace these errors to the
step-size restriction and the regularisation required to enforce it. To our knowledge, the wave-carrier
factoring itself has never been tested independently of the
restriction within a spatial marching method.


The One-Way Navier--Stokes (OWNS) method takes a different approach: rather
than shifting eigenvalues away from the upstream modes, it removes the upstream
modes entirely via one-way projection
\citep{towneOnewaySpatialIntegration2015,rigasOneWayNavierStokes2017}, thus 
eliminating the
PSE step-size restriction, and {{importantly}} retaining the streamwise pressure gradient.
The
projection is controlled by recursion parameters placed in the complex plane,
which determine which spectral regions the projection removes or retains.
These parameters are specific to the flow configuration and forcing frequency,
and must be selected with care. The OWNS-R variant
\citep{zhuRecursiveOnewayNavierStokes2023} made the projection computationally
tractable by evaluating it through a sequence of recursions.
Despite its success in parabolisation,
OWNS does not employ wave-carrier factoring. The discretisation must resolve
eigenvalues at their full wavenumbers from the origin, and accurate prediction
of disturbance amplitudes requires fine streamwise resolution. Fine resolution
in turn widens the integrator's spectral window, demanding that the projection
suppresses a broader portion of the upstream-propagating spectrum. The OWNS-R
recursion has a practical ceiling on the number of recursion
parameters~$N_\beta$, beyond which numerical error accumulates
\citep{sleeman2025greedy}, so a wider spectral window requires more carefully
placed parameters, a non-trivial task for general flows.


The present paper combines the two approaches. The carrier-wave factoring of PSE is applied within the OWNS-R framework, with the carrier wavenumber $\alpha_0$ chosen from the eigenvalue spectrum of the spatial operator or from a representative value of the excited modes; this spectral anchoring is what makes the factoring computationally beneficial~(\S\ref{sec:circle}). We retain the name M-OWNS from \citet{IUTAM2024_ejbmug}: the M refers to this choice of {{wave-}}carrier being tied to the excited spectrum of the operator, not to any modal restriction on the disturbance. Unfactored OWNS corresponds to $\alpha_0 = 0$. The one-way projection removes the upstream-propagating modes that would otherwise destabilise a spatial march in PSE, so the factoring benefit is realised without the minimum step-size restriction of PSE.


{{{A leading-order}} spectral resolution comparison analysis~(\S\ref{sec:circle}) shows that, for any excited eigenfunction whose eigenvalue lies closer to~$\alpha_0$ than to the origin, the {{wave-}}factored system resolves that mode at a coarser step size than the unfactored system. The spectral geometry of boundary-layer flows supports this in two complementary regimes. When the dominant response is composed of discrete modes (Tollmien--Schlichting (T--S), crossflow, or slow and fast hypersonic waves) clustered near the vorticity/entropy continuous-spectrum branch endpoint $\alpha_1$ of Eq.~\eqref{eqn:continuous_spectra}, a carrier at or near $\alpha_1$ recentres the numerical scheme on the excited cluster. When several branches are simultaneously excited, such as in receptivity or optimal-disturbance calculations, a carrier placed at a representative centre of the excited spectrum covers the full range at once~(\S\ref{sec:closure}). For single-mode calculations, $\alpha_0$ can be iterated toward the dominant eigenvalue as in PSE. For multi-modal or receptivity calculations, fixing $\alpha_0$ at a suitable spectral centre avoids iteration entirely at identical per-step cost to unfactored OWNS.


The carrier-wave factoring approach is presented within the OWNS-R framework, but the improved spectral resolution achieved is not specific to OWNS.
We use OWNS-R because the one-way projection permits fine spatial marching steps without
the regularisation errors PSE incurs~\citep{towneCriticalParabolized2019}. For
the fixed-carrier variant with the second-order implicit {{(BFD2)}} scheme, M-OWNS reduces
the total solve count by factors of two to eight relative to unfactored OWNS
across the test cases considered; larger reductions follow when a single mode
dominates and the Rayleigh-quotient iteration applies~(\S\ref{sec:closure}).


Section~\ref{sec:eqn} introduces the governing equations and wave-carrier formulation, and \S\ref{sec:projection} describes the OWNS-R projection and M-OWNS algorithm. The method is validated on flows of increasing complexity: subsonic flat-plate boundary layers~(\S\ref{flat_plate_subsonics}), three-dimensional crossflow over a swept cylinder \citep{malikCrossflowDisturbancesThreedimensional1994}~(\S\ref{hiemy_flow}), and a Mach~4.5 hypersonic boundary layer \citep{mackBoundaryLayerLinearStability1984a}~(\S\ref{hypersonic}). The hypersonic case is tested under forcing conditions ranging from single eigenfunction inlet forcing to broadband randomised forcing, including fast acoustic freestream waves and wall suction/blowing. For each deterministic forcing configuration, M-OWNS at coarse streamwise resolution captures disturbance amplitudes, acoustic radiation fields, and modal synchronisation sequences that unfactored OWNS at the same resolution does not. The randomised forcing case demonstrates convergence under broadband excitation, with M-OWNS and OWNS initially resolving complementary parts of the spectrum as resolution is increased. In both iterated and non-iterated forms, {{we demonstrate that}} M-OWNS captures the full field at a coarser spatial resolution {{and reduced computational cost relative to conventional}} OWNS.

\section{Governing Equations and Wave-Carrier Formulation}\label{sec:eqn}

We consider compressible boundary-layer flows governed by the body-fitted Navier--Stokes equations summarised in Appendix~\ref{apendeqnsbf}. The base flow is assumed invariant in the spanwise $(z)$ direction, as appropriate for infinite swept wings. Technical details of the OWNS-R approach are given by \citet{zhuRecursiveOnewayNavierStokes2023}; the present section introduces only the governing equations and the carrier-wave factoring that underpins M-OWNS.

\subsection{Two-way Disturbance Evolution Equation}\label{sec:two}

We begin by decomposing the state vector into $Q$, a known invariant in $z$ and time-independent base flow, and $q$, a small harmonic disturbance with frequency $\omega$ and a spanwise Fourier mode of wavenumber $\beta$ in $z$, 
\begin{equation}\label{eqn:dist_decomp}
    {q}^{\mathrm{tot}}(x,y,z,t) = Q(x,y)  + q(x,y)\exp(i\beta z-i\omega t). 
\end{equation}
Substituting this ansatz into the Navier--Stokes equations (Eqs.~\eqref{eqn:1}-\eqref{eqn:2}) and then linearising with respect to $q$ yields the LHNS, which can be expressed as
\begin{equation}\label{eqn:LHNS}
\mathrm{A}\left(\frac{\partial}{\partial y}\,;\,\beta \right)\frac{\partial q}{\partial x} = \mathrm{B}\left(\frac{\partial}{\partial y}\,;\,\beta,\,\omega\right)q + \mathrm{C}\,\frac{\partial^2q}{\partial x^2}.
\end{equation}
Here, $\mathrm{A}$, $\mathrm{B}$ and $\mathrm{C}$ represent linear operators of size $5\times5$. The steady base flow $Q(x,y)$ and its derivatives are obtained from a standard boundary-layer solver \citep{mughalactive}.

The second-order $\mathrm{C}$ terms arise from the streamwise viscous derivatives in the Navier--Stokes equations. For linear disturbances, these terms may be omitted without loss of accuracy \citep{hajhariri1994}; \citet{zhuRecursiveOnewayNavierStokes2023} confirm this through extensive numerical testing across a variety of flows. In the present work, the $\mathrm{C}$ terms are retained in the implicit marching discretisation but are omitted from the eigenvalue analysis that follows since they do not enter into the definition of the projection operator used in \S\ref{sec:projection}. 
 
After discretising in the wall-normal direction with $n_y$ points, the spatial evolution equation takes the semi-discrete form
\begin{equation}\label{eqn:two-way2}
    \mathbf{A}\frac{\mathrm{d} \phi}{\mathrm{d} x}=\mathbf{B}\phi,
\end{equation}
where $\mathbf{A}$, $\mathbf{B}$ and $\mathbf{C}$ are the banded wall-normal
discretisations of the continuous operators $\mathrm{A}$, $\mathrm{B}$ and $\mathrm{C}$ in
Eq.~\eqref{eqn:LHNS}, each of order $5n_y$. The terms associated with $\mathbf{C}$ are incorporated into the discretisation stencil and therefore enter the marching scheme implicitly, rather than through augmentation of the state vector.

Since the pencil $\left(\mathbf{B},\mathbf{A}\right)$ is regular, the finite and infinite eigenvalues
define disjoint spectral sets. The block diagonalisation of
\citet[Chapter~VI, Theorem~2.12]{stewartMatrixPerturbationTheory}
then yields a decomposition into deflating subspaces associated with each set.
On the finite-eigenvalue subspace, $\mathbf{A}$ is nonsingular, since this subspace excludes the infinite spectrum removed by the deflation. We assume that the eigenvalues of the pencil on this subspace are semisimple, so that the pencil admits a complete set of eigenfunctions. This assumption is used only for the modal-expansion proof in §\ref{sec:proj_operator} that the projection operator parabolises the wave-carrier factored operator; the method itself does not rely on it.

The pencil reduces to the standard eigenvalue problem
\begin{equation}\label{eqn:M_def}
    \mathbf{M}\phi_k = i\lambda_k\,\phi_k, \qquad \mathbf{M} = \mathbf{A}^{-1}\mathbf{B},
\end{equation}
where the inverse is taken on the finite spectral subspace, and $(\phi_k,\lambda_k)$ are an eigenfunction and finite eigenvalue pair.
Let the columns of $\mathbf{V}$ be the corresponding right eigenfunctions $\phi_k$, and let the rows of $\mathbf{U}$ be the associated left eigenfunctions, chosen such that
\begin{equation}\label{eqn:biorthog}
    \mathbf{U}\mathbf{V} = \mathds{I},
\end{equation} on the finite eigenvalue subspace.
The spatial dynamics on this subspace, and eigen-decomposition of $\mathbf{M}$, can then be written as
\begin{equation}\label{eqn:two-way4}
    \frac{\mathrm{d}\phi}{\mathrm{d} x} = \mathbf{M}\phi, \qquad
    \mathbf{M} = \mathbf{V}\,i\boldsymbol{\Delta}\,\mathbf{U},
\end{equation}
where $\boldsymbol{\Delta} = \operatorname{diag}(\lambda_k)$, and we write
$\sigma(\mathbf{M}) = \{\lambda_k\}$ for the finite eigenvalue spectrum.
In practice, $\mathbf{M}$ is not formed explicitly; the numerical methods
of \S\ref{sec:projection} operate directly on the pencil $(\mathbf{B},\mathbf{A})$.

\subsection{Disturbance Energy and Rayleigh Quotient}\label{sec:energy_wavenumber}

On the semi-infinite wall-normal domain, we define the inner product between two disturbance states ${\phi}_1$ and ${\phi}_2$ as
\begin{equation}\label{eqn:inner}
    \left({\phi}_1,{\phi}_2\right)_{\!\mathcal{H}} = \int_0^\infty {\phi}_1^*\,\mathcal{H}\,{\phi}_2\,\mathrm{d}y,
\end{equation}
where $(\cdot)^*$ denotes conjugate transpose and $\mathcal{H}$ is a positive semi-definite Hermitian weight matrix. The formulation that follows is independent of the specific choice of $\mathcal{H}$; any positive semi-definite weight matrix may be used. 
In the present work we adopt the density-weighted kinetic energy norm \citep{changLinearNonlinearPSE1993},
\begin{equation}\label{eqn:energy_norm}
    \mathcal{H} = \operatorname{diag}\!\left(0,\,{\rho},\,{\rho},\,{\rho},\,0\right), \qquad
    E_{\!\mathcal{H}} = \tfrac{1}{2}\int_0^\infty {\rho}\!\left(|u|^2 + |v|^2 + |w|^2\right)\,\mathrm{d}y,
\end{equation}
where ${\rho}=\rho(x,y)$ is the local base flow density. Results have also been verified with the diagonal part of the compressible energy norm {{of}} \citet{chuEnergyTransferSmall1965} and 
\citet{hanifiTransientGrowthCompressible1996}.

Given a solution ${\phi}(x,y)$ of Eq.~\eqref{eqn:two-way2}, we define the Rayleigh quotient
\begin{equation}\label{eqn:wavenumber_extract}
    \alpha(x) = -i\,\dfrac{\left({\phi},\,\dfrac{\mathrm{d} {\phi}^{}}{\mathrm{d} x}\right)_{\!\mathcal{H}}}
    {\left\lVert{\phi}\right\rVert_{\!\mathcal{H}}^2},
\end{equation}
where $\lVert{\phi}\rVert_{\!\mathcal{H}}^2 = ({\phi},{\phi})_{\!\mathcal{H}}$.

\paragraph{Interpretation as a wavenumber}
The Rayleigh quotient $\alpha$ defined by Eq.~\eqref{eqn:wavenumber_extract} does not require
solving the eigenvalue problem~\eqref{eqn:M_def}; it is evaluated directly from the
disturbance field ${\phi}$ and the weight $\mathcal{H}$.  Its interpretation as a
wavenumber depends on the modal composition of the disturbance.

For a disturbance dominated by a single eigenfunction,
${\phi}\approx c_k(x)\,{\phi}_k(x)$, where $c_k(x)$ is a slowly varying
coefficient and $(\phi_k,\lambda_k)$ is an eigenfunction/eigenvalue pair of $\mathbf{M}$ (Eq.~\eqref{eqn:M_def}), the Rayleigh quotient recovers $\alpha = \lambda_k$ exactly when
$c_k$ is constant, and approximately when $c_k$ varies slowly relative to the
wavelength $2\pi/|\lambda_k|$.  The Rayleigh quotient then provides a local
streamwise wavenumber and amplification rate.

When several eigenfunctions carry comparable energy, $\alpha$ is a weighted
average of the excited eigenvalues.  If the eigenfunctions are mutually
orthogonal under $\mathcal{H}$, the cross terms
$({\phi}_j,\,{\phi}_k)_{\!\mathcal{H}}$ vanish and the weights reduce to each
mode's share of $\lVert{\phi}\rVert_{\!\mathcal{H}}^2$. The Rayleigh quotient
is then a weighted average of the excited eigenvalues and lies within their convex hull.

When the eigenfunctions are not orthogonal under $\mathcal{H}$, the cross
terms do not vanish.  They oscillate at the beating wavenumbers
$|\lambda_j - \lambda_k|$ and, when multiple non-orthogonal eigenfunctions
carry significant amplitude, can dominate over the diagonal contributions.
In this regime $\alpha$ may be driven outside the convex hull of the excited
eigenvalues and can no longer be interpreted as a weighted average of
eigenvalues or as a wavenumber of any individual mode.  The Rayleigh quotient
should therefore be regarded as a computational reference quantity
characterising the dominant phase behaviour of the disturbance, rather than a
physically precise wavenumber. 


\subsection{Carrier-Wave Factoring}\label{sec:wkb}

We factor the solution of Eq.~\eqref{eqn:two-way2} as
\begin{equation}\label{eqn:wkb_ansatz}
    {\phi}(x) = \tilde{{\phi}}(x)\exp i\Theta_0(x), \quad \frac{\mathrm{d}\Theta_0}{\mathrm{d}x} = \alpha_0,
\end{equation}
where $\alpha_0$ is a carrier wavenumber and $\tilde{{\phi}}$ is the envelope. The same decomposition is employed by PSE \citep{herbertParabolizedStabilityEquations1997} and adaptive LHNS \citep{guoSolutionAdaptiveApproach1997, francosumarivaAdaptiveHarmonicLinearized2018}: we refer to it as carrier-wave factoring. As a change of variables it is always formally valid, the only requirement being that $\alpha_0$ is sufficiently differentiable for the system at hand. The carrier wavenumber $\alpha_0$ need not equal any eigenvalue in $\sigma\left({\mathbf{M}}\right)$ nor the Rayleigh quotient $\alpha$ defined in Eq.~\eqref{eqn:wavenumber_extract}; specific choices of $\alpha_0$ are discussed in \S\ref{sec:closure}.

Substituting Eq.~\eqref{eqn:wkb_ansatz} into the semi-discrete system Eq.~\eqref{eqn:two-way2} yields the factored system
\begin{equation}\label{eqn:factored_system}
    \mathbf{A}\frac{\mathrm{d}\tilde{{\phi}}}{\mathrm{d} x} = \tilde{\mathbf{B}}\,\tilde{{\phi}},
\end{equation}
where $\tilde{\mathbf{B}} = \mathbf{B} - i\alpha_0\mathbf{A}$. The pressure derivative $\mathrm{d} p/\mathrm{d} x = i\alpha_0\tilde{p} + \mathrm{d}\tilde{p}/\mathrm{d} x$ is retained in its entirety; often in PSE the $\mathrm{d}\tilde{p}/\mathrm{d} x$ term is set to zero in subsonic regions \citep{herbertParabolizedStabilityEquations1997}.

The factoring shifts the spectrum by a scalar and leaves the eigenfunctions unchanged \citep{towneCriticalParabolized2019}. The spatial operator of the factored system is
\begin{equation}\label{eqn:Mtilde}
    \tilde{\mathbf{M}} = \mathbf{V}\,i({\Delta} - \alpha_0\mathbf{\mathds{I}})\,\mathbf{U} = \mathbf{M} - i\alpha_0\,\mathds{I},
\end{equation}
where $\mathbf{VU}=\mathds{I}$ on the finite-eigenvalue subspace, with shifted spectrum 
\begin{equation}\label{eqn:spectral_shift}
    \sigma(\tilde{\mathbf{M}}) = \{\lambda_k - \alpha_0 : \lambda_k \in \sigma(\mathbf{M})\}.
\end{equation}

When the second-order $\mathbf{C}$ terms are retained, the factored system \eqref{eqn:factored_system} becomes
\begin{equation}\label{eqn:factored_C}
\tilde{\mathbf{A}}\frac{\mathrm{d}\tilde{{\phi}}}{\mathrm{d} x} = \tilde{\mathbf{B}}\,\tilde{{\phi}} + \mathbf{C}\frac{\mathrm{d}^2\tilde{{\phi}}}{\mathrm{d} x^2},
\end{equation}
where
\begin{equation}\label{eqn:factored_operators_C}
    \tilde{\mathbf{A}} = \mathbf{A} - 2i\alpha_0\mathbf{C}, \qquad
    \tilde{\mathbf{B}} = \mathbf{B} - i\alpha_0\mathbf{A} + \left(i\frac{\mathrm{d}\alpha_0}{\mathrm{d} x} - \alpha_0^2\right)\mathbf{C}.
\end{equation}

\subsection{Spectral resolution comparison}\label{sec:circle}

To examine the resolution properties of the wave-factored system, consider the
parallel-flow eigenfunction expansion of Eq.~\eqref{eqn:two-way4},
\begin{equation}\label{eqn:parallel_expansion}
    \phi(x) = \sum_k v_k \exp(i\lambda_k x),
\end{equation}
where the eigenfunctions $v_k$ are independent of $x$. A streamwise integration
scheme resolves mode~$k$ only if the spatial step is small relative to its
wavelength. {{To leading order}} this requirement is
\begin{equation}\label{eqn:res_req}
    \Delta x \lesssim \frac{C}{|\lambda_k|},
\end{equation}
where $C$ is an order-one constant set by the accuracy region of the scheme, for
example $C = \pi$ at the Nyquist limit of a second-order centred scheme.
{{We do not assume a particular scheme. The precise accuracy depends on
the position in the complex plane of the \emph{argument} $i\lambda_k\Delta x$, the
value at which the scheme is evaluated, and therefore on its phase as well as its
modulus, and on the discretisation.}}

Introducing the carrier wave gives the envelope
\begin{equation}\label{eqn:envelope_expansion}
    \tilde\phi(x) = \sum_k v_k \exp\!\big(i(\lambda_k - \alpha_0)x\big),
\end{equation}
whose modes oscillate at the residual wavenumbers $|\lambda_k - \alpha_0|$. The
envelope is resolved when
\begin{equation}\label{eqn:res_req_fac}
    \Delta x \lesssim \frac{C}{|\lambda_k - \alpha_0|}.
\end{equation}
The same constant $C$ appears in Eqs.~\eqref{eqn:res_req}
and~\eqref{eqn:res_req_fac} because the two systems are advanced with the same
discretisation and scheme. The argument of the factored system is $i(\lambda_k-\alpha_0)\Delta x$.

{{Equation~\eqref{eqn:spectral_shift} shifts the eigenvalues by $-\alpha_0$;
equivalently, holding the eigenvalues fixed, the scheme's resolved region moves to
$\alpha_0$.}}

\paragraph{{{Leading-order}} spectral resolution condition}
Comparing Eqs.~\eqref{eqn:res_req} and~\eqref{eqn:res_req_fac} at fixed $\Delta x$, factoring relaxes
the resolution requirement of mode~$k$ whenever
\begin{equation}\label{eqn:benefit_condition}
    |\lambda_k - \alpha_0| < |\lambda_k|.
\end{equation}
The locus $|\lambda - \alpha_0| = |\lambda|$ is the perpendicular bisector of the
segment from $0$ to $\alpha_0$. Modes on the $\alpha_0$ side benefit, modes on the
far side require a finer step than the unfactored system, and for
$|\lambda_k| \gg |\alpha_0|$ the factoring is neutral. {{The constant $C$ is common
to both systems, so the leading-order condition carries no explicit
scheme-dependent term.}}

{{We call Eq.~\eqref{eqn:benefit_condition} the spectral resolution condition. It is
a leading-order statement. For modes whose argument lies near the centre of the scheme's accuracy
region the modulus of the wavenumber alone sets the requirement; nearer the edge
the requirement acquires a dependence on the phase of the argument and on the
scheme, as noted at Eq.~\eqref{eqn:res_req}, so the true break-even locus is a
scheme-dependent curve rather than the bisector. This dependence is common to both factored and unfactored systems and is a property of the integration scheme, not of the factoring.}}

\paragraph{Spectral clustering}
In subsonic and supersonic boundary-layer flows, the downstream-propagating
discrete modes and the endpoints of the continuous spectral branches cluster in
the complex $\lambda$-plane. A carrier placed near this cluster sits close to the
excited modes, while the modes far along the branches have large $|\lambda_k|$ and
require a fine step with or without factoring. The benefit thus depends on the
choice of $\alpha_0$. A poorly placed carrier wastes the translation, and since
$\alpha_0 = 0$ recovers the unfactored system, the unfactored march is always
available as a fallback. Any estimate near the cluster serves as a reasonable
$\alpha_0$. Specific choices are discussed in \S\ref{sec:closure}.

{{\paragraph{Reliability of the comparison}
With the carrier near the physically relevant cluster, the modes $(\lambda_k)$ closest to
$\alpha_0$ satisfy $|\lambda_k-\alpha_0| \ll |\lambda_k|$. 
For a fixed step size $\Delta x$, the factored argument
$|(\lambda_k-\alpha_0)\Delta x|$ is then small and lies near the centre of the scheme's accuracy region. In this
limit the modulus governs and the phase- and scheme-dependent corrections are
subdominant, so the comparison is reliable for these modes.
 
Modes within the cluster but at finite distance from $\alpha_0$ need not satisfy
the stronger inequality $|\lambda_k-\alpha_0| \ll |\lambda_k|$. They may still
satisfy the spectral resolution condition $|\lambda_k-\alpha_0| < |\lambda_k|$,
but the argument $|(\lambda_k-\alpha_0)\Delta x|$ is then not necessarily small. The phase- and
scheme-dependent corrections therefore enter both the factored and unfactored
systems, and unless a specific scheme, spectrum, and choice of $\alpha_0$ are
prescribed, neither system can be claimed better resolved at this order.
 
As $\Delta x$ decreases, the arguments of both systems move towards the centre of
the scheme's accuracy region. For modes satisfying $|\lambda_k-\alpha_0| < |\lambda_k|$, the factored
argument $|(\lambda_k-\alpha_0)\Delta x|$ remains smaller than the
unfactored $|\lambda_k\Delta x|$. Consequently, as $\Delta x$ decreases,
the factored system reaches the small-argument regime, in which the
leading-order comparison is asymptotically valid, at coarser streamwise
resolutions than the unfactored march.
This is consistent with
the results reported here, in particular the randomised-forcing study of
\S\ref{sec:random}, where the factored system converges at substantially lower
streamwise resolution than the unfactored march.
 
This argument requires the excited or physically relevant modes to cluster near one $\alpha_0$. A
spectrum spread widely in $\lambda$ cannot be served by a single carrier, and the
unfactored march remains the fallback.}}

\paragraph{Resolution disc and the hypersonic spectrum example}
At a fixed step the condition has a geometric reading. The unfactored system
resolves a disc about the origin (Eq.~\eqref{eqn:res_req}) and the factored system
an equal disc about $\alpha_0$ (Eq.~\eqref{eqn:res_req_fac}), so factoring
translates the resolved disc to where the excited modes lie.
Fig.~\ref{fig:circle_spec} shows this for a Mach~$4.5$ boundary layer with
$\alpha_0 = \omega$. At $\Delta x = C/|\omega|$ the factored disc covers the fast
and slow discrete modes, while the unfactored disc covers only the fast.

\begin{figure}[ht!]
    \centering
    \includegraphics[width=0.6\linewidth]{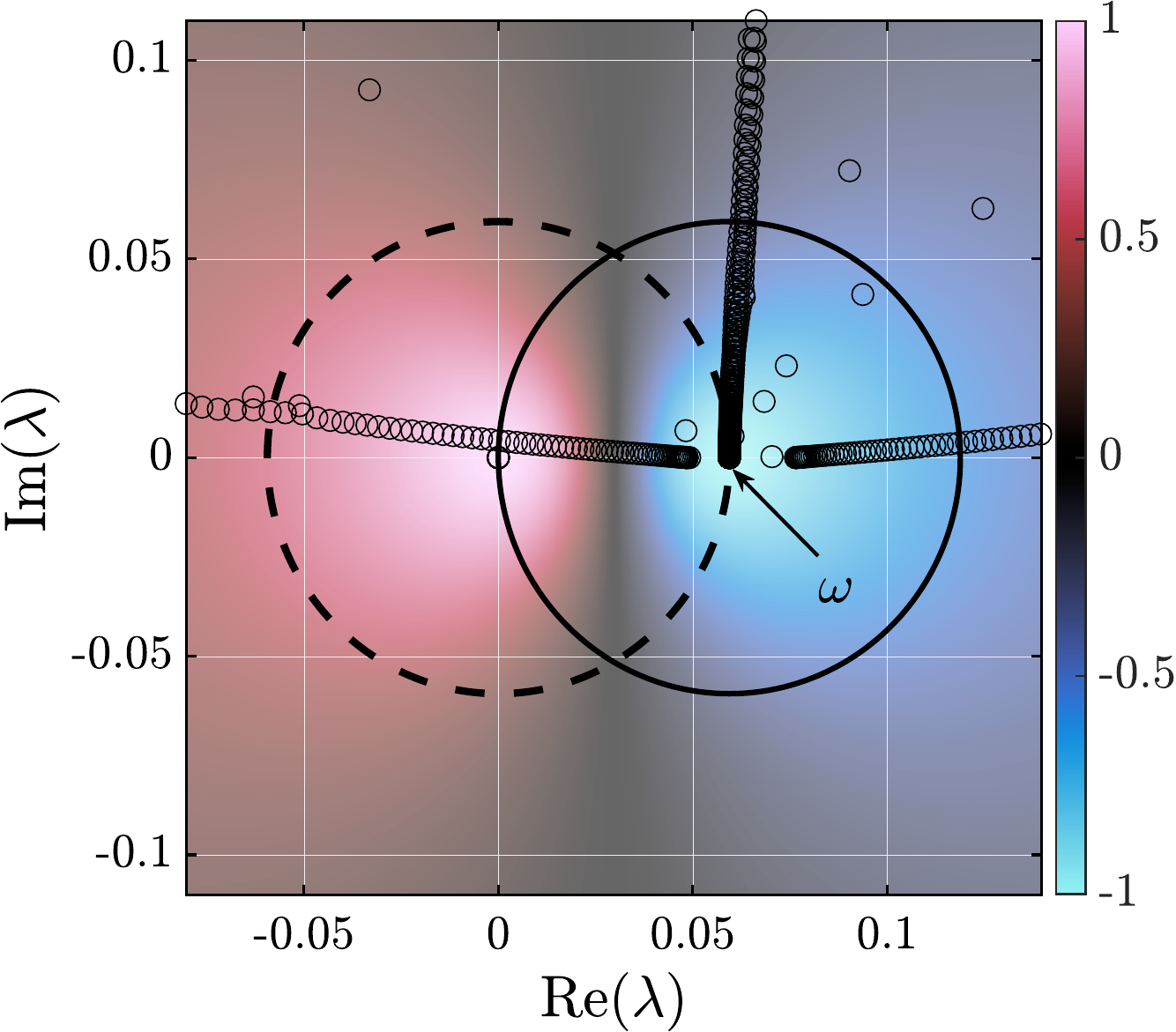}
    \caption{Eigenvalue spectrum for a Mach~$4.5$ boundary layer at frequency
    parameter $F = 220$ (see \S\ref{hypersonic}), with carrier wavenumber
    $\alpha_0 = \omega$. Background:
    $\tanh\!\left(\log\left(|\lambda - \omega|/|\lambda|\right)\right)$, showing
    the leading-order per-mode resolution cost ratio (blue: carrier-wave cheaper;
    red: carrier-wave more expensive; black: break-even). Solid circle: factored
    region $|\lambda - \omega| \lesssim |\omega|$ at step size
    $\Delta x = C/|\omega|$. Dashed circle: unfactored region
    $|\lambda| \lesssim |\omega|$ at the same $\Delta x$.}
    \label{fig:circle_spec}
\end{figure}

\paragraph{Connection to PSE}
The resolved-region picture is related to the PSE minimum step-size restriction.
In PSE the downstream march is formally ill posed because the spectrum contains
upstream-propagating, streamwise-elliptic modes. The march remains stable only
while these modes lie outside the resolved region, and refining $\Delta x$
enlarges that region until they are captured, which fixes the minimum step
\citep{liNaturePSEApproximation, liSpectralAnalysisParabolized1997,
towneCriticalParabolized2019}. Here the upstream-propagating modes are removed by
the one-way projection (\S\ref{sec:projection}), so the resolved region stays a
pure resolution concept and the PSE lower bound on $\Delta x$ does not apply.

{{\paragraph{Scheme stability}
Scheme stability is a separate question from resolution. A retained mode that
lies in the integration scheme's unstable region is spuriously amplified, and
because the carrier shift moves each mode's argument, factoring can move an
otherwise stable mode into that region. This depends on the scheme and the
spectrum and is not removed by the one-way projection. More generally, the modulus criterion Eq.~\eqref{eqn:benefit_condition} makes no distinction between phase and amplitude error, which need not be equally consequential in practice; the spurious amplification just described is the extreme case.}}

\paragraph{Slowly varying eigenfunctions}
Parallel-flow analysis assumes the eigenfunctions ${v}_k$ are
independent of $x$.  In non-parallel flows ${v}_k$ varies with $x$, and the
full computational benefit of the factoring additionally requires that this
variation {{be}} resolved by the spatial discretisation:
\begin{equation}\label{eqn:slow_evec}
    \Delta x\left\|\frac{\mathrm{d}{v}_k}{\mathrm{d}x}\right\|_\mathcal{H} \ll \|{v}_k\|_\mathcal{H}.
\end{equation}
This condition depends on the {{rapidity}} of base flow variation and the step size, not
on $\alpha_0$.  In typical boundary-layer flows the base flow evolves on scales
much longer than the disturbance wavelength, so it is {{usually}} satisfied.
In regions of strong non-parallelism (e.g.\ laminar separation bubbles or near
the leading edge), finer $\Delta x$ may be needed, but this requirement is a
property of the base flow and applies equally to {{both}} factored and unfactored systems.

\subsection{Closure Condition}\label{sec:closure}

The wave-factored system Eq.~\eqref{eqn:factored_system} contains a free parameter: the carrier wavenumber $\alpha_0$.  The resolution comparison in
\S\ref{sec:circle} shows that any computational benefit depends on
$\alpha_0$ lying near the excited spectral cluster: the question is how to
choose $\alpha_0$ in practice.  \citet{bertolottiAnalysisLinearStability1991, herbertBoundarylayerTransitionAnalysis1991} identify that different definitions of $\alpha$
{used historically} in PSE (for example the Rayleigh quotient, or the pointwise maximum of a
disturbance quantity) produce
essentially the same results, because any $\alpha_0$ within the cluster places
the accuracy disc over the relevant eigenvalues. The apparent ambiguity provides
additional freedom in the definition of $\alpha$.

\paragraph{Iterating the Rayleigh quotient}
The standard PSE closure {{condition}} \citep{bertolottiAnalysisLinearStability1991, herbertParabolizedStabilityEquations1997} determines
$\alpha_0$ by iterating the Rayleigh quotient.  Given an estimate
$\alpha_{0}^{(n)}$, the factored system is solved for the envelope
$\tilde{\phi}^{(n)}$ and a corrected carrier wavenumber is extracted as
\begin{equation}\label{eqn:adaptive}
    \alpha_0^{(n+1)} = \alpha_0^{(n)}
    - i\,
    \frac{\left(\tilde{\phi}^{(n)},\;
    \dfrac{\mathrm{d} \tilde{\phi}^{(n)}}{\mathrm{d} x}\right)_{\!\mathcal{H}}}
    {\bigl\lVert\tilde{\phi}^{(n)}\bigr\rVert_{\!\mathcal{H}}^2},
    \qquad n = 0,1,\ldots\;.
\end{equation}
Iteration continues until $|\alpha_0^{(n+1)} - \alpha_0^{(n)}| < \mathcal{E}$
for a prescribed {{non-dimensional}} tolerance~$\mathcal{E}$. 
The same iterative scheme is used in adaptive LHNS
\citep{guoSolutionAdaptiveApproach1997,
francosumarivaAdaptiveHarmonicLinearized2018}.

For a disturbance dominated by a single eigenfunction the iteration converges
to $\alpha_0 \approx \lambda_k$. The carrier {{wave}} tracks the dominant eigenvalue and
the envelope reduces to the eigenfunction modulated by a slowly varying
amplitude.  This is the regime of maximum computational benefit.  Convergence
is typically achieved in 1--3 iterations for $\mathcal{E} = 10^{-9}$.
The envelope can then easily be resolved at a coarse {{spatial grid spacing}}, since the residual wavenumber $|\lambda_k - \alpha_0|$ is small when $\alpha_0\approx\lambda_k$ -- the iteration identifies the $\alpha_0$ that achieves this, rather than imposing slow variation on the envelope. 

\paragraph{Behaviour under multi-modal disturbances}
The iteration relies on the Rayleigh quotient providing a meaningful estimate
of the dominant wavenumber.  When several eigenfunctions carry comparable
energy in a system whose eigenfunctions are approximately orthogonal under
$\mathcal{H}$, the cross terms in the Rayleigh quotient are small and
$\alpha_0$ converges to a value within the convex hull of the excited
eigenvalues (\S\ref{sec:energy_wavenumber}).

When the eigenfunctions are strongly non-orthogonal, cross terms contribute at leading order and the Rayleigh quotient can {{reside}} outside the convex hull of the excited eigenvalues. In this regime it can no longer be interpreted as a weighted-average wavenumber of the excited spectrum.  Used as a closure condition, this has
several consequences.  The converged $\alpha_0$ may lie outside the eigenvalue
cluster, so that the accuracy region is no longer centred over the excited
spectrum and {{any modes that require resolving may fall outside the translated region.}} Further, the iteration may converge to widely different values at
adjacent streamwise stations, breaking smoothness of $\alpha_0(x)$.  In severe
cases it may fail to converge altogether.

\paragraph{Fixed carrier}
An alternative that avoids these difficulties is to fix $\alpha_0$ at a
value known from the base flow.  Several such choices cover the usual
cases.  All share advantages that the carrier is trivially smooth,
requires no iterations, does not involve the Rayleigh quotient, and
leaves the per-step cost identical to unfactored OWNS.

The carrier-wave should lie near the spectral content that dominates the
physics of interest, so that eigenvalues of the excited modes
fall well inside the accuracy disc.
The trade-off is precision in the single-mode regime: the dominant
eigenvalue lies close to $\alpha_{0}$ but not exactly at it, so the
envelope retains a slowly varying residual phase
$\exp(i(\lambda_k - \alpha_{0})x)$ that the iterated variant would
remove.  Since $|\lambda_k - \alpha_{0}| \ll |\alpha_{0}|$ for modes
in the targeted spectral region, this residual is well within the
resolution of any reasonable discretisation.  This is demonstrated in
Fig.~\ref{fig:cyl_ke}, where the fixed-carrier calculation converges
more slowly than the iterated system but faster than the unfactored
system.

For problems dominated by discrete modes and the vorticity/entropy
continuous spectrum, {{we}} set
$\alpha_0 = \alpha_{1} = (\omega - \beta W_\infty)/U_\infty$ from
Eq.~\eqref{eqn:continuous_spectra}, the vorticity/entropy branch
endpoint, where $U_\infty$ and $W_\infty$ are the non-dimensional
freestream streamwise and spanwise velocities. 
In two-dimensional settings ($W_\infty = 0$ or $\beta = 0$) this
reduces to $\alpha_0 = \omega$, as used in the hypersonic calculations
of \S\ref{hypersonic}.

When the acoustic continuous spectrum carries significant energy, such
as in receptivity, resolvent or optimal disturbance calculations,
$\alpha_{1}$ may under-resolve acoustic content.  Setting
$\alpha_0 = d_2$ (Eq.~\ref{eqn:ds}), the acoustic branch midpoint,
centres resolution at the acoustic behaviour.  When both the acoustic and vorticity/entropy continuous
spectra carry significant energy, $\alpha_0 = (\alpha_1 + d_2)/2$ can
balance the two.

Unfactored OWNS is the limiting case $\alpha_0 = 0$, with no spectral
preference.

\paragraph{Summary}

In summary, iterating the Rayleigh quotient is the natural choice when a
single eigenfunction dominates, where it provides maximum resolution benefit.
When several eigenfunctions carry comparable energy and the excited
eigenfunctions are strongly non-orthogonal, the fixed carrier is preferred. It avoids the difficulties
associated with cross-term contamination of the Rayleigh quotient while still
centring the accuracy disc over the physically critical spectrum.  The choice
of carrier is application-dependent: $\alpha_1$ is
suited when discrete modes govern the instability
growth; $d_2$ is suited to acoustic-dominated problems; the midpoint of $d_2$ and $\alpha_1$ can balance acoustic and vortical/entropic contributions.
All results in this paper use the iterating variant; in the swept-cylinder case (\S\ref{hiemy_flow}) and two hypersonic test cases (Mode-S forcing \S\ref{sec:no_iter}, randomised inlet forcing \S\ref{sec:random}), the fixed carrier $\alpha_0=\alpha_1$ is additionally computed and compared against the iterating variant.

\section{OWNS and M-OWNS}\label{sec:projection}

Equation~(\ref{eqn:two-way2}) represents a two-way evolution equation, where the semi-discrete operator ${\mathbf{M}}$ accommodates eigenfunctions that propagate in both downstream and upstream directions. Through application of Briggs' criterion \citep{briggsElectronStreamInteractionPlasmas1964a}, the finite spectrum of ${\mathbf{M}}$ can be partitioned into two subsets:
\begin{equation}\label{eqn:twoway_spec}
    \sigma\left({\mathbf{M}}\right) = 
    \begin{cases}
        \text{downstream-propagating eigenvalues} \\
        \text{upstream-propagating eigenvalues} 
    \end{cases}
\end{equation}

Stable downstream marching of Eq.~(\ref{eqn:two-way2}) requires removing the upstream-propagating modes from ${\mathbf{M}}$; we therefore seek a parabolised operator ${\mathbf{M}}^+$ whose finite spectrum satisfies
\begin{equation}\label{eqn:oneway_spec}
    \sigma\left({\mathbf{M}}^+\right) = 
    \begin{cases}
        \text{downstream-propagating eigenvalues} \\
        \text{zero eigenvalues}.
    \end{cases}
\end{equation}
We call ${\mathbf{M}}^+$ parabolised if it satisfies Eq.~(\ref{eqn:oneway_spec}). The zero eigenvalues lie in the nullspace of ${\mathbf{M}}$ and are unaffected by the projection: they contribute nothing to the spatial dynamics regardless of whether they are formally assigned to ${\mathbf{M}}^+$.

\subsection{The Projection Operator}\label{sec:proj_operator}

Following \citet{towneAdvancementsJetTurbulence2016a,towneEfficientGlobalResolvent2021}, we construct a projection operator~$\mathbf{P}$ from the eigen-decomposition Eq.~\eqref{eqn:two-way4} such that ${\mathbf{M}}^+ = \mathbf{P}{\mathbf{M}}$ retains only the downstream-propagating eigenvalues (\citet{rudelNumericalFactorizationPropagation2022a} give the corresponding continuous theory).  On the finite-eigenvalue subspace,
\begin{equation}\label{eqn:projection}
    \mathbf{P} = \mathbf{V}\, \mathbf{E}\, {\mathbf{U}},
\end{equation}
where $\mathbf{E}$ is a diagonal selector,
\begin{equation}\label{eqn:ed}
    \left(\mathbf{E}\right)_j = \begin{cases}
        1 \quad &\text{if}\, (\mathbf{\Delta})_j \, \text{is downstream-propagating},\\
        0 \quad &\text{otherwise}.
    \end{cases}
\end{equation}
The one-way operator is then
\begin{equation}
    {\mathbf{M}}^+ = \mathbf{V}\, i\mathbf{E\Delta}\, {\mathbf{U}},
\end{equation}
whose spectrum satisfies Eq.~\eqref{eqn:oneway_spec}.  Since $\mathbf{E}$ and $\mathbf{\Delta}$ are both diagonal, $\mathbf{P}$ and ${\mathbf{M}}$ commute:
\begin{equation}
\mathbf{PM} = \mathbf{V}\, i\mathbf{E\Delta}\, {\mathbf{U}}= \mathbf{V}\, i\mathbf{\Delta E}\, {\mathbf{U}} = \mathbf{MP}.
\end{equation}

The projection operator $\mathbf{P}$ is constructed from the first-order pencil $(\mathbf{B},\mathbf{A})$ alone.  The second-order $\mathbf{C}$ terms are retained in the marching discretisation via backward differences but do not enter the projection construction.  Since $\mathbf{P}$ does not in general commute with $\mathbf{C}$, the upstream-downstream splitting of the full second-order system may in principle differ from that of the first-order pencil.  This question has not, to the authors' knowledge, been addressed in the literature.  The $\mathbf{C}$ terms contain only the second order streamwise viscous derivatives, which are small (scaling with the reciprocal of the Reynolds number) relative to $\mathbf{A}$ and $\mathbf{B}$; the results and conclusions of \citet{zhuRecursiveOnewayNavierStokes2023} and our own numerical tests across subsonic and hypersonic configurations show no evidence of error attributable to this omission.

\paragraph{Projection of the carrier-wave factored system}
We now show that the projection operator $\mathbf{P}$ parabolises the carrier-wave factored operator $\tilde{{\mathbf{M}}}$, provided it parabolises ${\mathbf{M}}$. Since $\mathbf{P}$ parabolises ${\mathbf{M}}$, the matrix $\mathbf{E}$ satisfies Eq.~\eqref{eqn:ed}. Expanding the factored operator $\tilde{{\mathbf{M}}}$ from Eq.~(\ref{eqn:Mtilde}) gives
\begin{equation}
    \tilde{{\mathbf{M}}} =  \mathbf{V}\left( i\mathbf{\Delta} - i\alpha_0 \mathbf{\mathds{I}} \right){\mathbf{U}}.
\end{equation}
Applying $\mathbf{P}$ and using the biorthogonality condition Eq.~\eqref{eqn:biorthog} yields
\begin{align}
\begin{split}
   \mathbf{P} \tilde{{\mathbf{M}}} =&\;   \mathbf{V}\mathbf{E}\mathbf{U} \mathbf{V}\left( i\mathbf{\Delta} - i\alpha_0 \mathbf{\mathds{I}} \right){\mathbf{U}} 
   = \mathbf{V}\, i\mathbf{E}\left( \mathbf{\Delta} - \alpha_0\mathbf{\mathds{I}}  \right){\mathbf{U}}.
\end{split}
\end{align}
From Eq.~\eqref{eqn:ed}, the operator $\mathbf{E}\left(\mathbf{\Delta} - \alpha_0\mathds{I}\right)$ has the property
\begin{equation}\label{eqn:ed_modal}
    \left[\mathbf{E}\left(\mathbf{\Delta} - \alpha_0\mathds{I}\right)\right]_j = \begin{cases}
        \left(\mathbf{\Delta} - \alpha_0\mathds{I}\right)_j \quad &\text{if}\, (\mathbf{\Delta})_j \, \text{is downstream-propagating},\\
        0 \quad &\text{otherwise},
    \end{cases}
\end{equation}
since $\mathbf{E}$ selects the same entries regardless of the shift by $\alpha_0$. Therefore $\tilde{{\mathbf{M}}}^+ = \mathbf{P} \tilde{{\mathbf{M}}}$ is a one-way operator: the same projection that parabolises the unfactored system also parabolises the carrier-wave factored system. No modification of $\mathbf{P}$ or the recursion parameter selection is required.

\subsection{The Approximate Projection Operator: OWNS-R}\label{sec:owns_r}

The eigenvalue decomposition in the projection operator $\mathbf{P}$, Eq.~(\ref{eqn:projection}), is prohibitively expensive to evaluate at each spatial marching step. \citet{zhuRecursiveOnewayNavierStokes2023} proposed the OWNS-R technique, which approximates $\mathbf{P}$ via an operator $\mathbf{P}_{N_\beta}$ that depends on a sequence of complex recursion parameters
\begin{equation}
    \beta^\pm = (\beta^+_j,\beta^-_j)_{j=1,\dots,N_\beta} \subset \mathbb{C}^2, \quad N_\beta\in\mathbb{N}.
\end{equation}
The sets $\beta^+$ and $\beta^-$ are called the positive and negative recursion parameters respectively, and are placed near the downstream- and upstream-propagating eigenvalues of $\mathbf{M}$. With appropriately placed $\beta^\pm$, the approximation satisfies $\lim_{N_\beta\rightarrow\infty} \mathbf{P}_{N_\beta} = \mathbf{P}$. 

One form of the approximate projection operator is
\begin{equation}\label{eqn:proj_1}
    \mathbf{P}_{N_\beta}= \left(\mathds{I} + c\prod^{N_\beta}_{k=1}\left(\mathbf{M}-i\beta_k^+\,\mathds{I} \right)\left(\mathbf{M}-i\beta_k^-\,\mathds{I} \right)^{-1} \right)^{-1},
\end{equation}
where $c>0$ is a tuning parameter that balances retention of downstream modes against removal of upstream modes; typical values are $c = 0.1$--$10$. Clearing the inner inverses and refactoring yields
\begin{align}\label{eqn:proj_2}
    \mathbf{P}_{N_\beta} =  \frac{1}{c+1}\prod^{N_\beta}_{k=1}\left(\mathbf{M}-i\beta_k^*\,\mathds{I} \right)^{-1}\left(\mathbf{M}-i\beta_k^-\,\mathds{I} \right),
\end{align}
where $\beta^*_j\in\mathbb{C}$ are the roots of the $N_\beta$-th order polynomial
\begin{equation}\label{eqn:beta_star}
    c\prod^{N_\beta}_{k=1}\left(\lambda-\beta_k^+ \right) +\prod^{N_\beta}_{k=1} \left(\lambda-\beta_k^- \right) = 
    (c+1)\prod^{N_\beta}_{k=1}\left(\lambda-\beta_k^* \right).
\end{equation}

In practice, $\mathbf{M}$ is never formed; all operations are expressed in terms of the pencil $(\mathbf{B},\mathbf{A})$. Equation~(\ref{eqn:proj_2}) takes the form
\begin{equation}\label{eqn:approx_project}
  \mathbf{P}_{N_\beta}  =\frac{1}{c+1}\prod^{N_\beta}_{k=1}\left(\mathbf{B}-i\beta_k^*\, \mathbf{A} \right)^{-1}\left(\mathbf{B}-i\beta_k^-\,\mathbf{A} \right),
\end{equation}
and the action of $\mathbf{P}_{N_\beta}$ on a state vector $\phi$ is computed via the recursion
\begin{align}
    \label{eqn:recursion_step_1} \phi_0 &= \frac{1}{c+1}\phi, \\ 
    \label{eqn:recursion_step_2} \left(\mathbf{B}-i\beta_{k}^*\, \mathbf{A} \right)\phi_{k} &=  \left(\mathbf{B}-i\beta_{k}^-\, \mathbf{A} \right)\phi_{k-1}, \quad k=1,\dots, N_{\beta}, \\ 
    \label{eqn:recursion_step_3} \phi^+ &=  \phi_{N_\beta}.
\end{align}
Since each solve involves the shifted pencil $(\mathbf{B}-i\beta_k^*\,\mathbf{A})$, and the $\beta_k^*$ are finite and distinct from the eigenvalues of $(\mathbf{B},\mathbf{A})$, the infinite eigenvalues associated with singular $\mathbf{A}$ do not enter the recursion and require no special treatment.

Each recursion step requires one banded linear solve of order $5n_y$, so the cost of applying $\mathbf{P}_{N_\beta}$ is $N_\beta$ solves.

\subsection{The M-OWNS Marching Algorithm}\label{sec:algorithm}

We now describe the complete marching algorithm. Since $\mathbf{P}$ and $\mathbf{M}$ commute, the two-way system can be solved first and the result projected, rather than forming and marching with $\mathbf{M}^+$ directly. The carrier-wave factored system Eq.~(\ref{eqn:factored_C}) provides the most general framework; unfactored OWNS is recovered by setting $\alpha_0 = 0$.

At each streamwise station, the algorithm proceeds as follows:
\begin{enumerate}
    \item \textbf{Solve.} Advance the carrier-wave factored two-way system one step using backward differencing to obtain $\tilde{\phi}$.
    \item \textbf{Iterate} (optional). Update the carrier wavenumber $\alpha_0$ via the Rayleigh quotient Eq.~(\ref{eqn:adaptive}) and re-solve step~1 with the updated $\alpha_0$. Repeat until $|\alpha_0^{(n+1)} - \alpha_0^{(n)}| < \mathcal{E}$.
    \item \textbf{Project.} Apply the approximate projection $\tilde{\phi}^+ = \mathbf{P}_{N_\beta}\tilde{\phi}$ using the recursion Eqs.~(\ref{eqn:recursion_step_1})--(\ref{eqn:recursion_step_3}).
\end{enumerate}

Three cases arise as parameter choices within this single framework:
\begin{itemize}
    \item \emph{Iterated $\alpha_0$} (iterated M-OWNS): the carrier wavenumber is updated at each station via the Rayleigh quotient. 
    \item \emph{Fixed $\alpha_0 = \alpha_{1}$} (non-iterating M-OWNS): the carrier wavenumber is set to the vorticity/entropy branch endpoint $(\omega - \beta\,{W}_\infty)/{U}_\infty$ from Eq.~(\ref{eqn:continuous_spectra}), which is known \emph{a priori}. No iterations are required.
    \item $\alpha_0 = 0$ (unfactored OWNS): the carrier-wave factoring is absent and the unfactored two-way system is solved directly. No iterations are required.
\end{itemize}

\paragraph{Computational cost}
At each streamwise station, the marching solve (one banded linear system of order $5n_y$) is performed $(N_{\text{it}} + 1)$ times: the initial solve plus $N_{\text{it}}$ Rayleigh quotient iterations. A single projection requiring $N_\beta$ recursion solves follows. The iteration and projection are sequential, not nested: each iteration repeats only the marching solve, and the projection is applied once after the iteration has converged. Table~\ref{tab:cost} summarises the per-step cost for each method.
\begin{table}[ht]
\centering
\caption{Number of linear solves per streamwise step for each method.
$N_{\text{it}}$ denotes the number of Rayleigh quotient iterations
(typically $2$--$3$ for $\mathcal{E}=10^{-9}$) and $N_\beta$ denotes the
number of recursion parameters (typically $20$--$60$).}\label{tab:cost}
\begin{tabular}{lcc}
\toprule
Method & Solves per step & Typical range \\
\midrule
PSE                      & $1 + N_{\text{it}}$                 & $3$--$4$ \\
OWNS                     & $1 + N_\beta$                       & $21$--$61$ \\
M-OWNS (iterating)       & $1 + N_{\text{it}} + N_\beta$      & $23$--$64$ \\
M-OWNS (non-iterating)   & $1 + N_\beta$                       & $21$--$61$ \\
\bottomrule
\end{tabular}
\end{table}
Non-iterating M-OWNS has identical per-step cost to unfactored OWNS. The iteration cost $N_{\text{it}}$ is additive at each step, not multiplicative on the projection cost $N_\beta$; the iterating variant therefore incurs an overhead of $N_{\text{it}}/(1+N_\beta)$ relative to OWNS, which is under $15\%$ across the typical parameter range. PSE is cheapest per step but is subject to a minimum step-size restriction that limits its applicability~\citep{towneCriticalParabolized2019}. The total cost additionally depends on the number of streamwise steps $n_x$ required for a given accuracy; quantitative comparisons are given in~\S\ref{sec:total_cost}.

\paragraph{Projection robustness}
The coarser step size permitted by carrier-wave factoring also eases the demands on the
approximate projection.  At a small step size the integrator's spectral window
is wide, so~$\mathbf{P}_{N_\beta}$ must suppress a broad portion of the
upstream-propagating spectrum.  Since the OWNS-R recursion has a practical upper
bound on~$N_\beta$ (\S\ref{sec:param_selection}), a wide spectral window
requires more carefully positioned recursion parameters.  M-OWNS resolves the
cluster of physically relevant modes near~$\alpha_{1}$ at a larger
step size, narrowing the spectral window so that fewer upstream modes fall
within it and the same~$N_\beta$ provides a more faithful projection. {{By the same leading-order argument, M-OWNS should not require a better projection than unfactored OWNS for the same
physical accuracy, provided $\alpha_0$ is never closer to the upstream spectrum than the origin.}}

\subsection{Recursion Parameter Selection}\label{sec:param_selection}
 
The recursion parameters $\beta^\pm$ introduced in \S\ref{sec:owns_r} must be chosen so that $\mathbf{P}_{N_\beta}$ faithfully approximates the exact projection $\mathbf{P}$: the positive parameters $\beta^+$ are placed near downstream-propagating eigenvalues to ensure their retention, and the negative parameters $\beta^-$ are placed near upstream-propagating eigenvalues to ensure their removal. The tuning parameter $c$ balances the relative strength of retention and removal.
 
The general strategy follows the geometric framework of
\citet{towneOnewaySpatialIntegration2015}. Each recursion parameter pair
$(\beta^+_k, \beta^-_k)$ defines a rational factor that drives the scalar
projection function $\mathcal{P}_{N_\beta}$ (the scalar version of
Eq.~\eqref{eqn:proj_1}) toward one near $\beta^+_k$ and toward zero near
$\beta^-_k$, with exponential convergence as $N_\beta$ increases. This convergence is limited in practice by the accumulation of numerical error in the recursive inversions, as first observed by \citet{sleeman2025greedy}. The parameter placement must reflect the spectral structure of the flow, which varies with Mach number, spanwise wavenumber and streamwise location. The wall-normal grid stretching produces a non-uniform sampling of the continuous spectral branches, with eigenvalues concentrated near the branch endpoints and increasingly sparse further along each branch. The recursion parameters must follow this distribution: clustered near the branch endpoints and more widely spaced away from them. 

\citet{sleeman2025greedy} proposed a greedy algorithm for the automatic selection of recursion parameters. Their method uses knowledge 
of the spectrum and applies the greedy selection to determine which 
eigenvalues to use as recursion parameters.
 
The parameters are recalculated at each streamwise station as the base flow evolves and the spectrum migrates. This is standard practice in OWNS implementations. The explicit parameter expressions for the subsonic and supersonic configurations studied in this work, together with their derivation and validation, are given in Appendix~\ref{app:recursion}.

\section{Numerical Methods}\label{sec:numerics}

The stability calculations are performed using an in-house compressible spatial marching code incorporating LST and PSE \citep{mughalactive}. The OWNS-R and M-OWNS results reported here are from an independently developed solver.

All spatial quantities are non-dimensionalised by $\delta_0$, the boundary-layer thickness at the inlet $\bar{x}_0$. The local Reynolds number based on boundary-layer thickness is
$
R_\delta = \sqrt{\bar{x}\, R_\infty},
$
where $R_\infty$ is the freestream Reynolds number and $\bar{x}$ is the dimensional distance from the origin. The inlet Reynolds number is $R = R_\delta(\bar{x}_0)$. The base flow is non-dimensionalised by freestream values at the inlet.

The non-dimensional frequency and spanwise wavenumber used throughout this work are
\begin{equation}\label{eqn:Fb}
F = \frac{\omega \times 10^{6}}{R}, \qquad b = \frac{\beta \times 10^{3}}{R},
\end{equation}
where $\omega$ and $\beta$ are the {{non-}}dimensional temporal frequency and spanwise wavenumber introduced in Eq.~\eqref{eqn:dist_decomp}.

The wall-normal domain $0 < \bar{y} < L_{\bar{y}}$ is discretised with $n_y$ points using fourth-order finite differences on the stretched grid of \citet{malikNumericalMethodsHypersonic1990}:
\begin{equation}\label{eqn:malik1}
y_{\mathrm{Malik}}\!\left(\eta,\, y_{\mathrm{max}},\, y_{\mathrm{half}}\right) = \frac{a\,\eta}{b - \eta},
\end{equation}
where $0 \le \eta \le 1$ is the uniform computational coordinate, $y_{\mathrm{max}}$ is the far-field truncation height, and more than half the grid points lie between $0$ and $y_{\mathrm{half}}$, with
\begin{equation}\label{eqn:malik3}
a = \frac{y_{\mathrm{max}}\, y_{\mathrm{half}}}{y_{\mathrm{max}} - 2\,y_{\mathrm{half}}}, \qquad
b = 1 + \frac{a}{y_{\mathrm{max}}}.
\end{equation}
This stretching function is also used to distribute recursion parameters non-uniformly in the complex plane (Appendix~\ref{app:recursion}).

The streamwise domain $0 < \bar{x} < L_{\bar{x}}$ is discretised with $n_x$
points using second-order backward differencing (BFD2), unless stated
otherwise.  Higher-order backward-difference schemes have smaller stability
regions for stiff systems and are not pursued here.  
Grid extents and
resolutions are specified for each flow configuration in the relevant
subsection.

At the wall, no-slip conditions are imposed on velocity disturbances with either a zero-disturbance or adiabatic condition for temperature. In the freestream, Thompson characteristic conditions suppress numerical reflections \citep{thompsonTimeDependentBoundary1987a}. Initial conditions are provided by one of three methods: a locally parallel eigenfunction from LST, a summation of different freestream waves, or random noise. Disturbances can also be generated via wall suction and blowing. The specific choice is stated for each case.

For iterated M-OWNS and PSE, the convergence criterion $\mathcal{E}$, {non-dimensionalised by the boundary-layer thickness at the inlet,} denotes the absolute difference between successive wavenumbers at each station. We use $\mathcal{E} = 10^{-9}$ unless otherwise stated, yielding $N_{\mathrm{it}} = 2$--$3$ iterations per station. 

\section{Results}\label{sec:results}

We assess M-OWNS across three flow configurations: subsonic flat-plate boundary-layers (\S\ref{flat_plate_subsonics}), incompressible flow over a swept cylinder (\S\ref{hiemy_flow}), and a Mach~$4.5$ hypersonic boundary-layer (\S\ref{hypersonic}). In every case, the spectral resolution comparison (\S\ref{sec:circle}) correctly predicts the resolution requirements of the two methods.

\subsection{Subsonic Flat-Plate Boundary Layers}\label{flat_plate_subsonics}

We consider isothermal flat-plate flows at freestream Reynolds number ${R}_\infty = 10^6 \, \text{m}^{-1}$ and temperature ${T}_\infty = 298$\,K, with Mach numbers $M=0.02$, $0.2$, $0.5$ and $0.8$.  The computational domain is initialised at $R_\delta=400$ using a locally parallel eigenfunction (LST) corresponding to the most unstable T-S mode. The analysis considers frequency $F = 86$ and spanwise wavenumbers $b = 0$, $0.1$, $0.2$ and $0.3$. All calculations employ $n_y=221$ wall-normal points. For this flat-plate geometry the curvature $\kappa=0$, so the metric term $\chi = 1$ in Eq.~\eqref{eqn:metric}. For every two-dimensional subsonic case, we use $N_\beta=30$ recursion parameters; exact parameters are given in \S\ref{app:subsonic}.

\subsubsection{Validation Across Mach Number and Obliqueness}

Figure~\ref{fig:vary_m} presents Rayleigh quotient and growth-rate comparisons across Mach number for a two-dimensional disturbance ($b = 0$). M-OWNS at both low resolution ($n_x = 110$) and high resolution ($n_x = 1000$) agrees with PSE ($n_x = 110$) and OWNS ($n_x = 1000$) throughout the domain.  Figure~\ref{fig:vary_b} demonstrates equivalent agreement across spanwise wavenumbers $b = 0$--$0.3$ in an incompressible boundary-layer ($M = 0.02$), confirming that M-OWNS accurately models three-dimensional oblique disturbances.

\definecolor{customgreen}{RGB}{34,139,34}
\definecolor{customblue}{RGB}{0,100,200}
\definecolor{customorange}{RGB}{220,120,0}
\definecolor{custompurple}{RGB}{150,0,150}

\begin{figure}[htbp]
    \centering
    \begin{subfigure}[b]{0.49\textwidth}
        \centering
        \includegraphics[width=\textwidth]{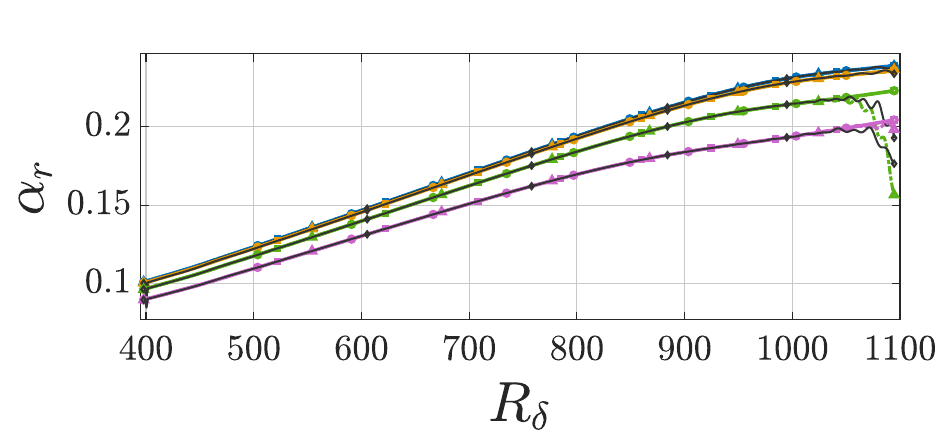}
        \caption{Rayleigh quotient}
        \label{fig:vary_m1}
    \end{subfigure}
    \begin{subfigure}[b]{0.49\textwidth}
        \centering
        \includegraphics[width=\textwidth]{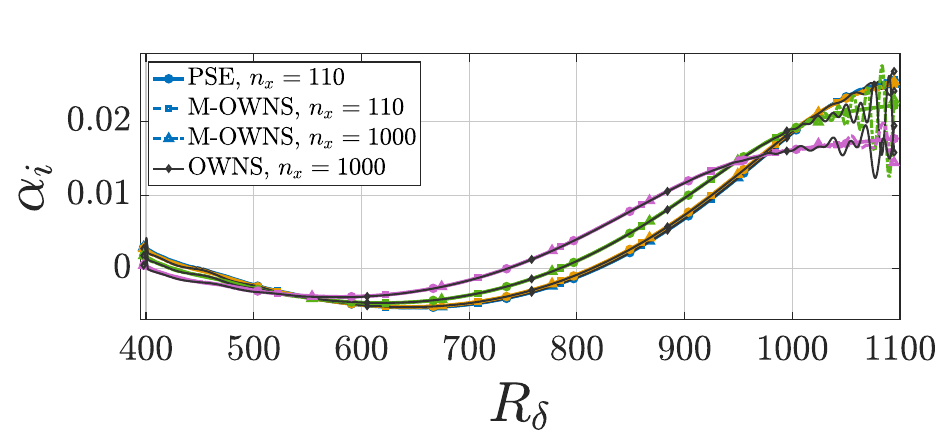}
        \caption{Growth rates}
        \label{fig:vary_m2}
    \end{subfigure}
\caption{(a) Rayleigh quotient and (b) growth-rate comparisons for a 2D-disturbance ($b=0$) with $F=86$ in a subsonic boundary-layer with variable Mach number.
{(\textcolor{customblue}{\raisebox{0.5ex}{\rule{1em}{1.2pt}}})}~$M=0.02$, (\textcolor{customorange}{\raisebox{0.5ex}{\rule{1em}{1.2pt}}})~$M=0.2$, (\textcolor{customgreen}{\raisebox{0.5ex}{\rule{1em}{1.2pt}}})~$M=0.5$, (\textcolor{custompurple}{\raisebox{0.5ex}{\rule{1em}{1.2pt}}})~$M=0.8$. All calculations include PSE with $n_x=110$, M-OWNS with $n_x=110$ and $n_x=1000$, and OWNS at $n_x=1000$. }
        \label{fig:vary_m}
    \end{figure}

Downstream, the Rayleigh quotient transitions from the T-S value to that of the vorticity wave, $\alpha_{1}=\omega$ in Eq.~\eqref{eqn:continuous_spectra}.  Non-parallel base flow evolution transfers energy from the decaying T-S wave to the more slowly decaying vorticity mode, which eventually dominates. This {transition} is resolved by OWNS and M-OWNS at sufficient $n_x$. 

    \begin{figure}[htbp]
    \centering
    \begin{subfigure}[b]{0.49\textwidth}
        \centering
        \includegraphics[width=\textwidth]{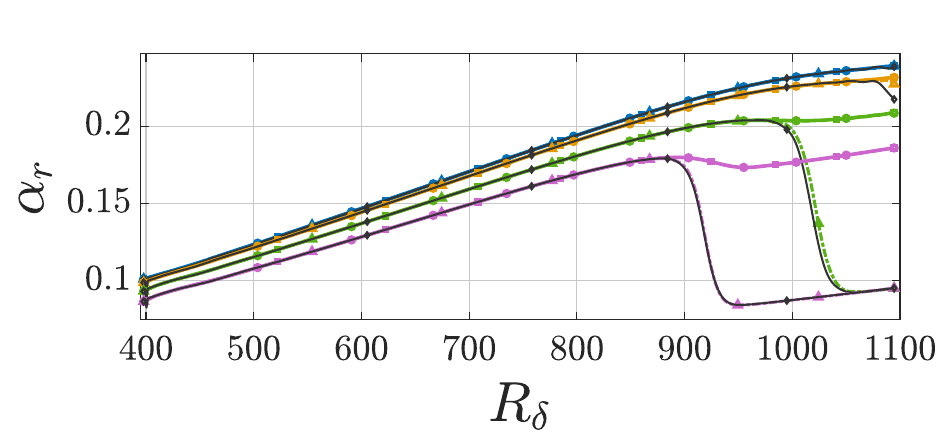}
        \caption{Rayleigh quotient}
        \label{fig:vary_b1}
    \end{subfigure}
    \begin{subfigure}[b]{0.49\textwidth}
        \centering
        \includegraphics[width=\textwidth]{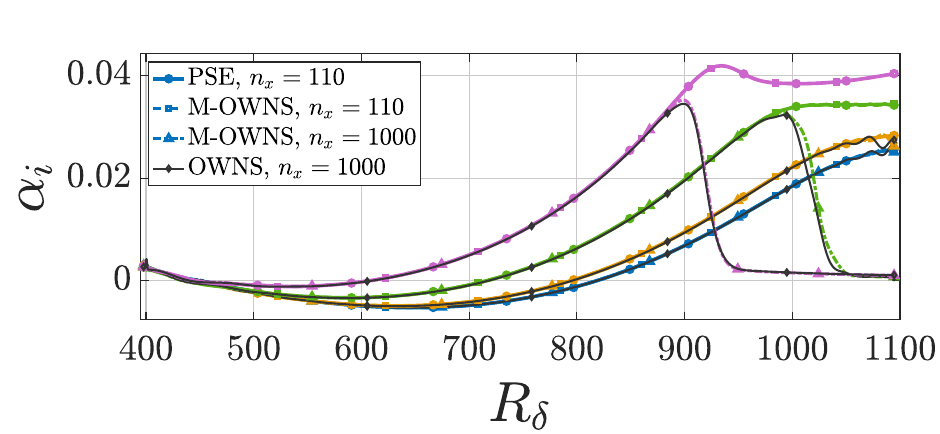}
        \caption{Growth rates}
        \label{fig:vary_b2}
    \end{subfigure}
\caption{(a) Rayleigh quotient and (b) growth-rate comparisons for a 3D-disturbance with $F=86$ and variable spanwise wavenumber in an incompressible boundary-layer $(M=0.02)$.
(\textcolor{customblue}{\raisebox{0.5ex}{\rule{1em}{1.2pt}}})~$b=0$, (\textcolor{customorange}{\raisebox{0.5ex}{\rule{1em}{1.2pt}}})~$b=0.1$, (\textcolor{customgreen}{\raisebox{0.5ex}{\rule{1em}{1.2pt}}})~$b=0.2$, (\textcolor{custompurple}{\raisebox{0.5ex}{\rule{1em}{1.2pt}}})~$b=0.3$. All calculations include PSE with $n_x=110$, M-OWNS with $n_x=110$ and $n_x=1000$, and OWNS at $n_x=1000$.}\label{fig:vary_b}
    \end{figure}

\subsubsection{M-OWNS in the Intermediate Resolution Regime}

Figure~\ref{fig:200} demonstrates a critical advantage of M-OWNS at intermediate streamwise resolution. At $n_x = 200$, both PSE and unfactored OWNS fail: PSE violates its step-size constraint, while OWNS exhibits oscillations leading to breakdown. M-OWNS maintains numerical stability and produces results consistent with the converged high-resolution calculations.  This intermediate regime, too coarse for OWNS and too fine for PSE, is precisely where the resolution benefit predicted by the spectral resolution comparison (\S\ref{sec:circle}) is most consequential.

\begin{figure}[htbp]
    \centering
    \begin{subfigure}[b]{0.75\textwidth}
        \centering
        \includegraphics[width=\textwidth]{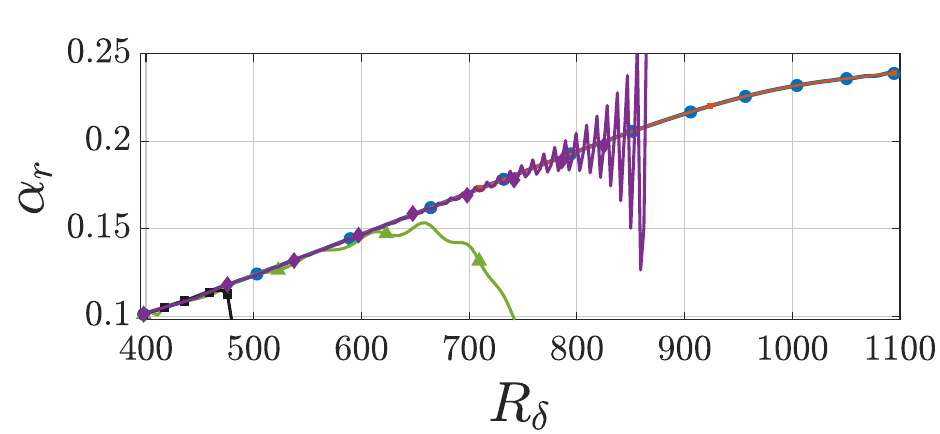}
        \label{fig:200a}
    \end{subfigure}
    \vspace{-1.2cm}
    
    \begin{subfigure}[b]{0.75\textwidth}
        \centering
        \includegraphics[width=\textwidth]{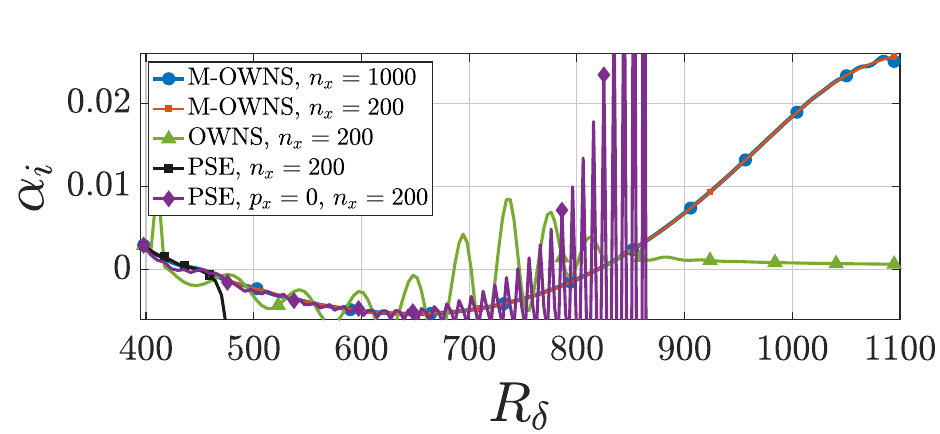}
        \label{fig:200b}
    \end{subfigure}

   \caption{(Top) Rayleigh quotient and (bottom) growth-rate comparisons for a disturbance with $F=86$ in an incompressible boundary-layer $(M=0.02)$ with zero spanwise wavenumber $(b=0)$. Only M-OWNS gives consistent and converged results with $n_{x}=200,\; 1000$; both OWNS and PSE with removed streamwise pressure gradient ($p_{x}=0$) fail on using $n_{x}=200$ points, while PSE with $p_{x}$ retained has the worst behaviour.}
        \label{fig:200}
    \end{figure}

 \begin{figure}[htbp]
    \centering
    \begin{subfigure}[b]{0.75\textwidth}
        \centering
        \includegraphics[width=\textwidth]{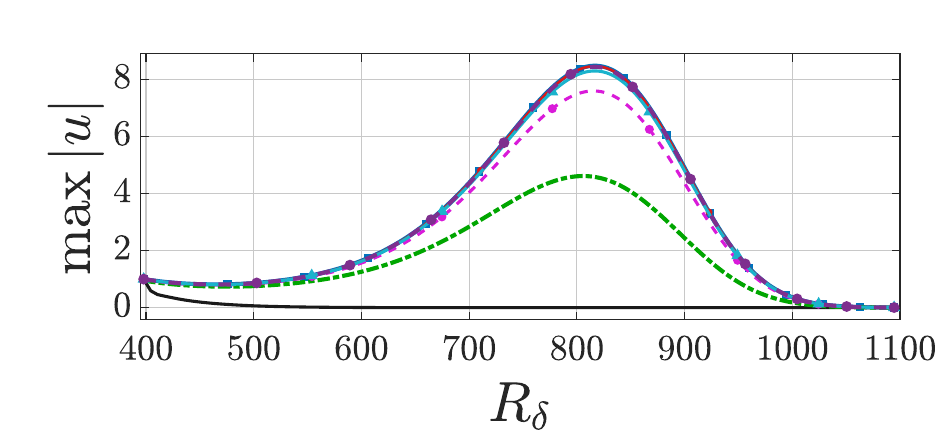}
        \label{fig:owns_norm1}
    \end{subfigure}     \vspace{-1.2cm}

    \begin{subfigure}[b]{0.75\textwidth}
        \centering
        \includegraphics[width=\textwidth]{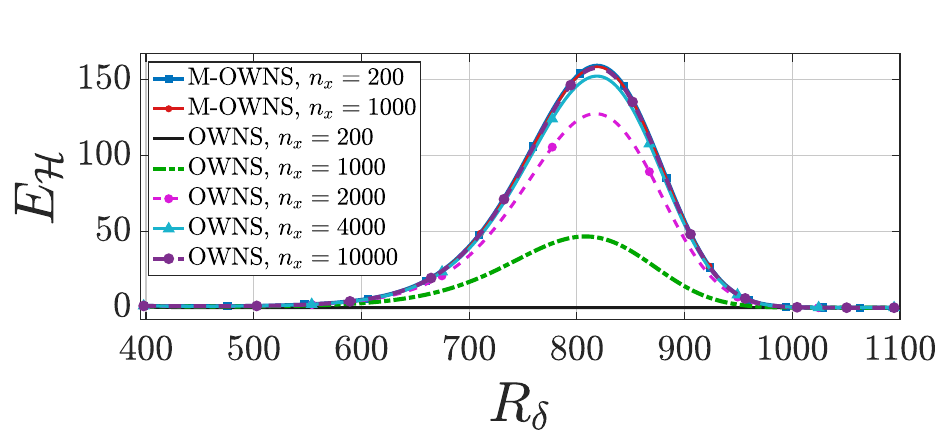}
        \label{fig:owns_norm2}
    \end{subfigure}

\caption{(Top) Maximum streamwise velocity magnitude and (bottom) disturbance energy for a disturbance with $F=86$ in an incompressible 2D boundary-layer $(M=0.02)$ with zero spanwise wavenumber $(b=0)$.}
        \label{fig:owns_norm}
    \end{figure}   

\subsubsection{Convergence of Disturbance Amplitude and Energy}

While normalised quantities such as Rayleigh quotient require relatively modest resolution, accurate prediction of disturbance amplitude and energy demands considerably finer $x$-discretisation with unfactored OWNS. Fig.~\ref{fig:owns_norm} shows that OWNS requires $n_x > 8000$ for converged maximum velocity and kinetic energy in an incompressible two-dimensional boundary-layer with $F = 86$. M-OWNS achieves converged predictions with $n_x \geq 150$. 

Similar resolution requirements emerge in compressible flows. Fig.~\ref{fig:wave_error} compares energy predictions for a Mach~$0.8$ boundary-layer with $b = 0.1$ and $F = 150$, using implicit Euler (BFD1) and BFD2 discretisations with $\mathcal{E} = 10^{-9}$.  OWNS energy predictions increase gradually with $n_x$, achieving agreement with M-OWNS only at extreme discretisations ($n_x > 4000$). First-order differencing performs worst, as expected.  M-OWNS demonstrates convergent behaviour across the entire resolution range.

\begin{figure}[ht!]
\centering
\begin{tikzpicture}
    \node[anchor=south west,inner sep=0] (image) at (0,0) {\includegraphics[width=0.95\linewidth]{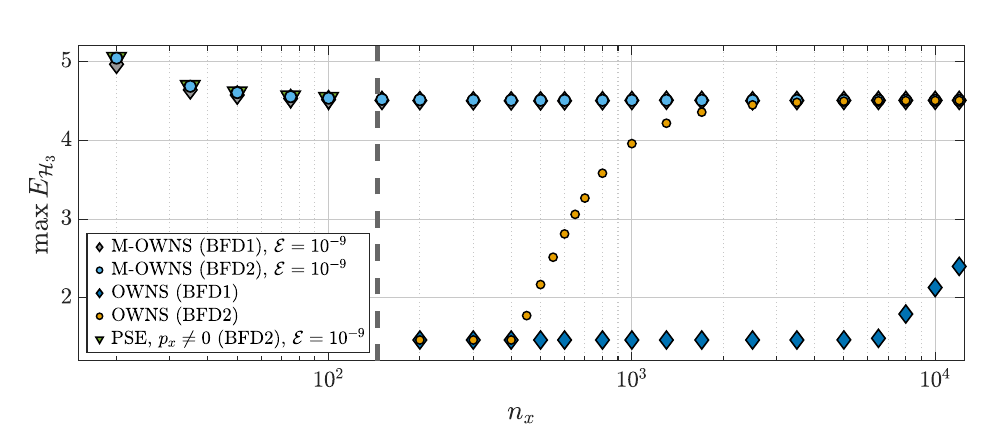}};
    
    \begin{scope}[x={(image.south east)},y={(image.north west)}]
        
        \fill[white] (0.0,0.63) rectangle (0.06,0.9);
        
        
    \end{scope}
\end{tikzpicture}
\caption{Maximum density-scaled kinetic energy ($E_{\!\mathcal{H}}$) versus streamwise resolution ($n_x$) for a $M = 0.8$ flat-plate boundary-layer, with $b = 0.1$ and frequency $F = 150$.}\label{fig:wave_error}
\end{figure}

\subsection{Swept Cylinder}\label{hiemy_flow}

To test M-OWNS in a three-dimensional flow with evolving spectra, we consider the effectively incompressible $(M = 0.0001)$ boundary layer over a swept cylinder with radius $0.5$\,m and sweep angle $65^\circ$, following \citet{malikCrossflowDisturbancesThreedimensional1994}. We set $T_\infty = 300$\,K and with $n_y = 301$ wall-normal points over the domain $132 \le R_\delta \le 1460$ and $N_\beta = 20$ recursion parameters. The curvature $\bar{\kappa} = 2/\text{m}$ is included via Eq.~\eqref{eqn:metric}. Two crossflow disturbances are examined at $\beta = 0.4$ ($b=3.030$): a stationary mode ($F = 0$) and a travelling mode ($F = 75$).

This flow introduces a computational challenge absent in flat-plate cases -- the vorticity/entropy branches migrate toward the acoustic spectrum as the crossflow ratio $W_\infty/U_\infty$ decreases downstream (Fig.~\ref{fig:cyl_spec}).  The recursion parameters must track this evolution; details of the adaptive strategy are given in Appendix~\ref{app:recursion}.

\begin{figure}[htbp]
    \centering
    \begin{subfigure}[b]{0.49\textwidth}
        \centering
        \includegraphics[width=\textwidth]{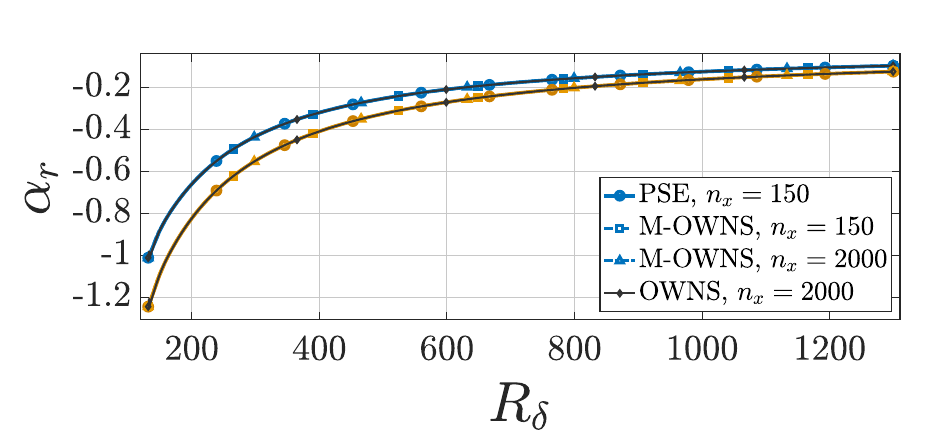}
        \caption{Rayleigh quotient}
        \label{fig:cyl_wave}
    \end{subfigure}
    \begin{subfigure}[b]{0.49\textwidth}
        \centering
        \includegraphics[width=\textwidth]{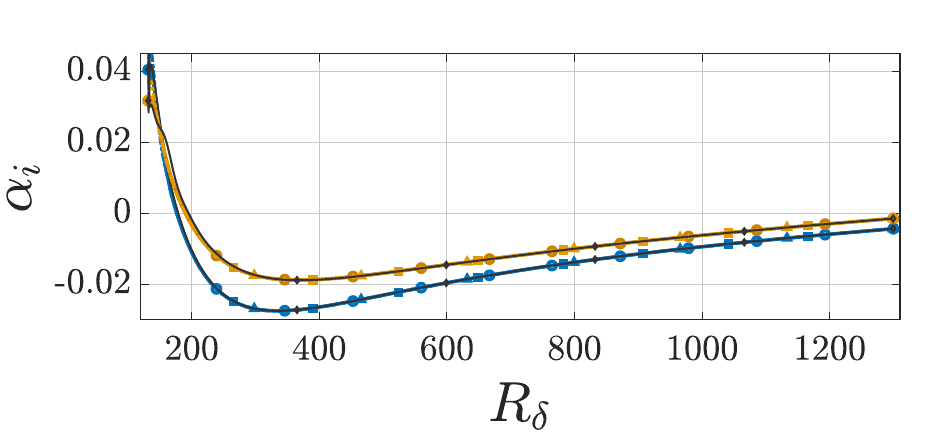}
        \caption{Growth rates}
        \label{fig:cyl_growth}
    \end{subfigure}

\caption{(a) Rayleigh quotient and (b) growth-rate comparisons for crossflow disturbances ($\beta=0.4$) in the swept cylinder analysis. Results shown for (\textcolor{customblue}{\raisebox{0.5ex}{\rule{1em}{1.2pt}}})~travelling ($F = 75$) and (\textcolor{customorange}{\raisebox{0.5ex}{\rule{1em}{1.2pt}}})~stationary ($F = 0$) crossflows. Computations performed using PSE ($n_x=150$), M-OWNS ($n_x=150, 2000$), and OWNS ($n_x=2000$).}
        \label{fig:cyl_Rayleigh_quotient}
\end{figure}

Figure~\ref{fig:cyl_Rayleigh_quotient} demonstrates agreement between PSE,
M-OWNS, and OWNS-R for both stationary and travelling crossflow modes.
Fig.~\ref{fig:cyl_ke} compares convergence with streamwise resolution for
three formulations: standard OWNS, M-OWNS with the iterated Rayleigh quotient
($\alpha_0 = \alpha$), and M-OWNS with the fixed carrier
$\alpha_0 = \alpha_1 = -\beta\, W_\infty / U_\infty$. 

The ordering reflects
spectral distance. At the inlet, the crossflow wave and the vorticity/entropy branches are far away from the origin/acoustic branches. As the flow develops downstream, the crossflow $W_\infty/U_\infty$ and the crossflow wave and the vorticity/entropy branches move towards the origin.
The iterated carrier sits on and tracks the crossflow eigenvalue at
every station, so the residual wavenumber is zero and convergence requires only
$n_x \sim 150$.  The fixed carrier $\alpha_1$ mode lies close to but not on the
eigenvalue, leaving a nonzero residual wavenumber throughout the march;
convergence requires $n_x \sim 1000$.  Standard OWNS measures all wavenumbers
from the origin, the largest spectral distance, and requires $n_x \sim 8000$.
The three panels therefore illustrate the resolution criterion of
\S\ref{sec:circle} in quantitative terms: convergence rate is governed by the
distance from the carrier to the dominant eigenvalue, not by the absolute
wavenumber of the mode.
 
   \begin{figure}[htbp]
    \centering
    \begin{subfigure}[b]{0.99\textwidth}
        \centering
        \includegraphics[width=\textwidth]{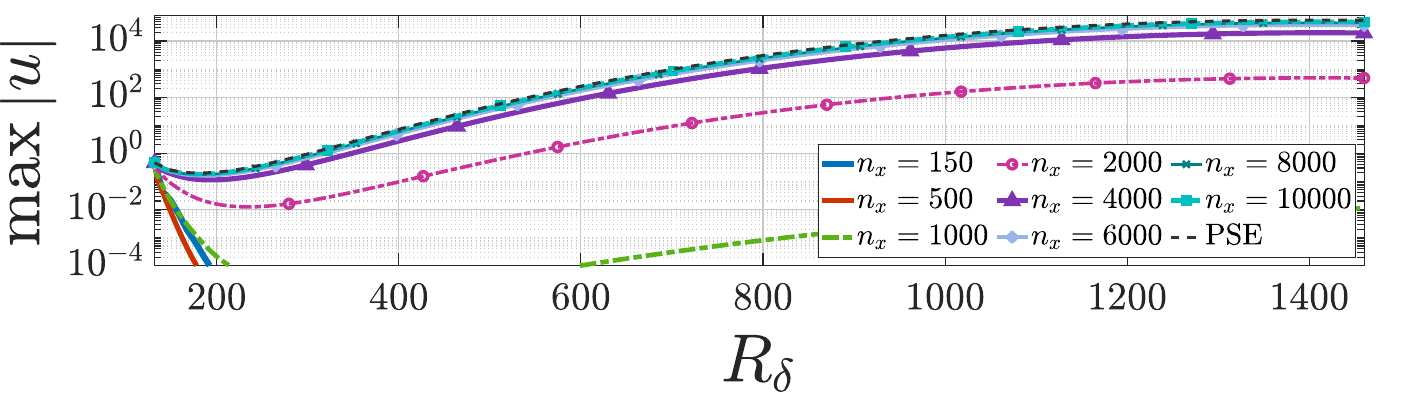}
    \end{subfigure}        \vspace{-1.1cm}

    \begin{subfigure}[b]{0.99\textwidth}
        \centering
        \includegraphics[width=\textwidth]{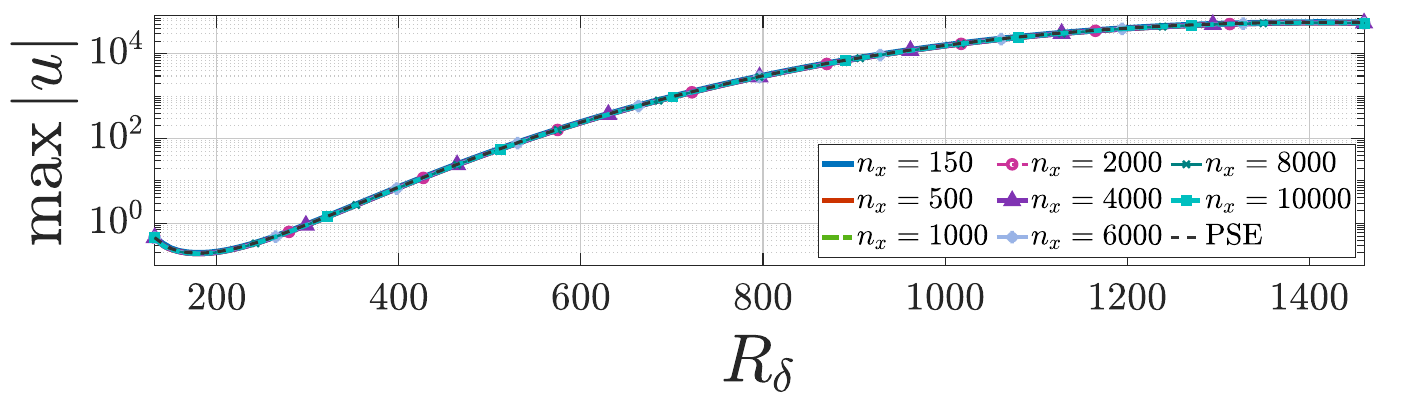}
    \end{subfigure}        \vspace{-1.1cm}

    \begin{subfigure}[b]{0.99\textwidth}
        \centering
        \includegraphics[width=\textwidth]{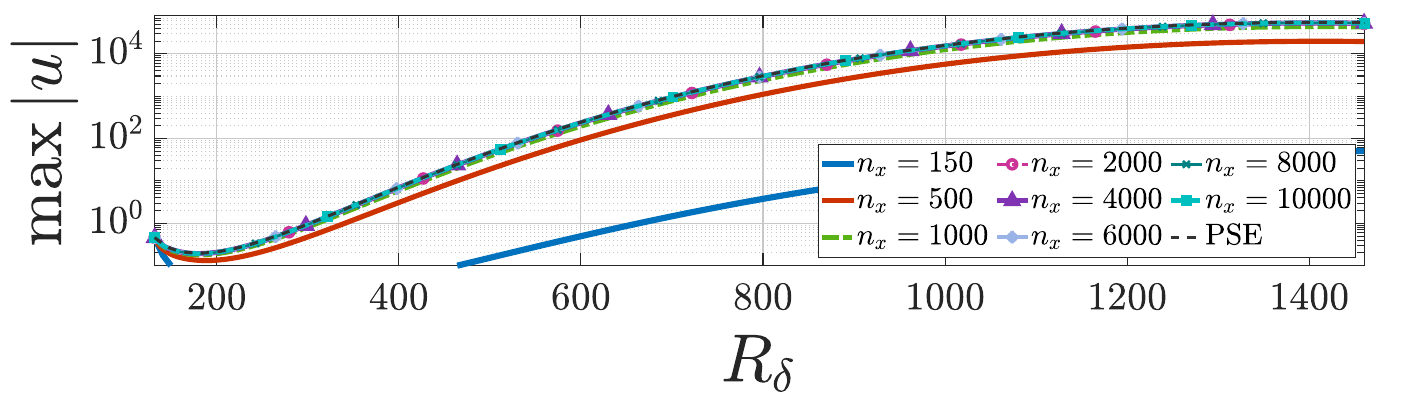}
    \end{subfigure}    
    
           \caption{Maximum streamwise velocity for stationary crossflow disturbances ($\beta=0.4$, $F=0$) in the swept cylinder analysis. Top: OWNS; middle: M-OWNS with iterated $\alpha_0=\alpha$; bottom: M-OWNS with fixed carrier-wave with $\alpha_0=\alpha_1=-{\beta\, W_\infty}/{U_\infty}$.  }
        \label{fig:cyl_ke}
    \end{figure} 

\clearpage

\subsection{Hypersonic boundary-layer}\label{hypersonic}

We examine a Mach~$4.5$ boundary layer following
\citet{mackBoundaryLayerLinearStability1984a}, with $R_\infty = 7.2 \times
10^6\,\text{m}^{-1}$, $T_\infty = 65.15$\,K, $p_\infty = 728.44$\,Pa,
Prandtl number $\sigma = 0.72$, frequency $F = 220$ and $b = 0$.  The
adiabatic wall condition $T_y = 0$ is imposed.  This configuration supports
multiple discrete modes: the slow mode~(S) and fast mode~(F) of
\citet{fedorovReceptivityHighspeedBoundary2003}, which synchronise far
downstream and generate acoustic radiation
\citep{fedorovBranchingDiscreteModes2010}.  We denote the associated
continuous spectral branch end-points as slow acoustic~(SA), fast
acoustic~(FA), and vorticity~(V).

Four forcing scenarios probe different aspects of M-OWNS capability:
(i)~Mode~S inlet forcing to study synchronisation,
(ii)~wall suction/blowing for multimodal excitation,
(iii)~simultaneous freestream wave forcing with SA, FA and V waves to test
the broadband, disparate-eigenvalue case, and
(iv)~randomised inlet forcing, where the absence of a dominant upstream
mode provides a direct test of the leading-order spectral resolution condition and a comparison of the
iterated and fixed-carrier closures.

Supersonic parameter placement is distinct from subsonic and is discussed in \S\ref{app:supersonic}. All calculations use a BFD2 scheme.

\subsubsection{Mode S Forcing}\label{sec:slow_mode}

The computational domain spans $R_\delta = 400$ to $1410$ with $n_y = 600$ and $N_\beta=40$.
The inlet is forced with a Mode~S eigenfunction from LST.

The wall pressure $|p_\mathrm{wall}|$ tracks the modal content at the
surface.  Fig.~\ref{fig:hyper_pwall} shows that OWNS converges between
$n_x = 4000$ and~$8000$; M-OWNS is converged at all resolutions tested,
down to $n_x = 650$.

The pressure field $\mathrm{Re}(p)$ reveals the acoustic radiation
generated by S--F synchronisation in the freestream.
Fig.~\ref{fig:superU} compares OWNS and M-OWNS at $n_x = 12\,000$
and~$2000$.  At the finer resolution, both methods capture the near-wall
second mode and its freestream radiation.  At $n_x = 2000$, OWNS fails to
resolve the radiation pattern while M-OWNS maintains both near-wall
accuracy and radiation field structure.  At $n_x = 1000$ (not shown),
M-OWNS still resolves the near wall Mode~S growth but not the freestream
radiation; OWNS captures neither.

\begin{figure}[htbp]
    \centering
    \begin{subfigure}[b]{0.75\textwidth}
        \centering
        \includegraphics[width=\textwidth]{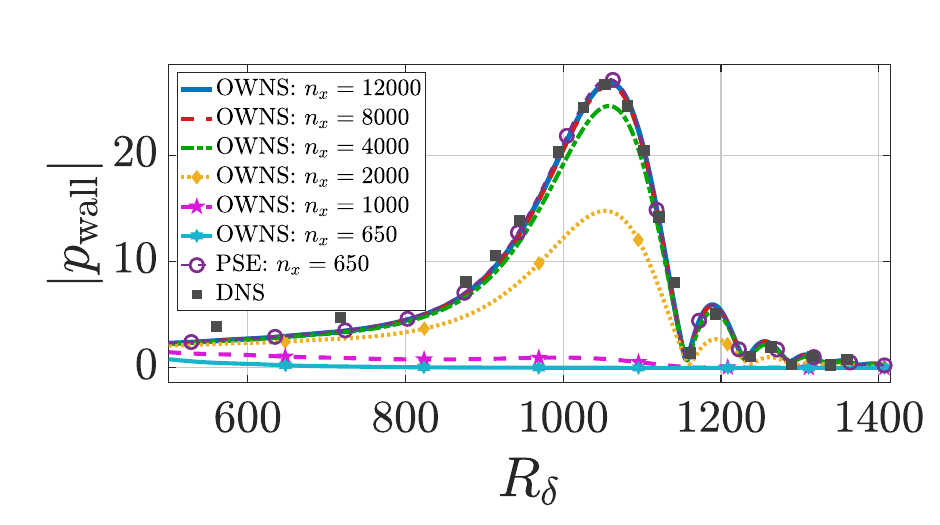}
    \end{subfigure}
    \vspace{-1.5cm}
    
    \begin{subfigure}[b]{0.75\textwidth}
        \centering
        \includegraphics[width=\textwidth]{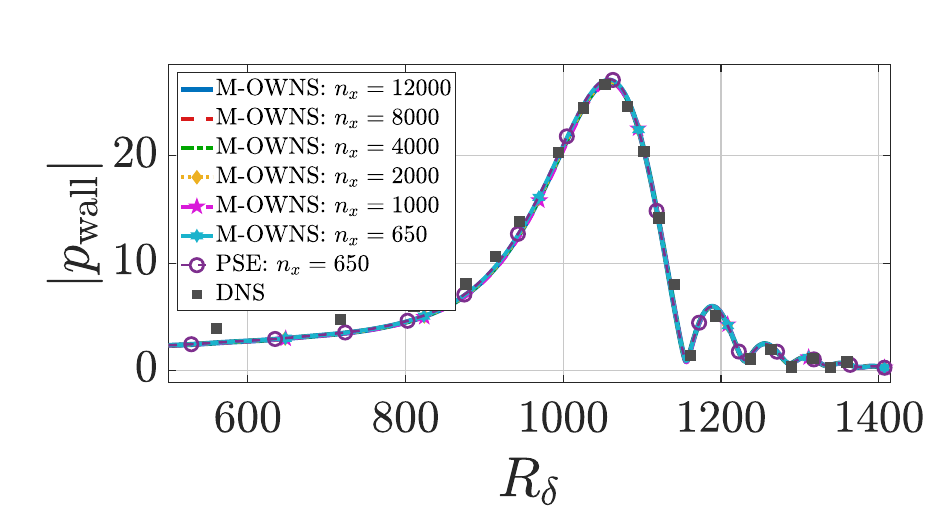}
    \end{subfigure}
    \caption{Wall pressure amplitude $|p_\mathrm{wall}|$ for the Mach~4.5
    boundary layer at $F = 220$.  Top: OWNS; bottom: M-OWNS\@.  DNS
    reference from \citet{maReceptivitySupersonicBoundary2003}.  OWNS
    converges between $n_x = 4000$ and~$8000$; M-OWNS is converged at all
    resolutions tested.}\label{fig:hyper_pwall}
\end{figure}

\begin{figure}[ht!]
\centering
\begin{subfigure}{0.49\textwidth}
    \includegraphics[width=\linewidth]{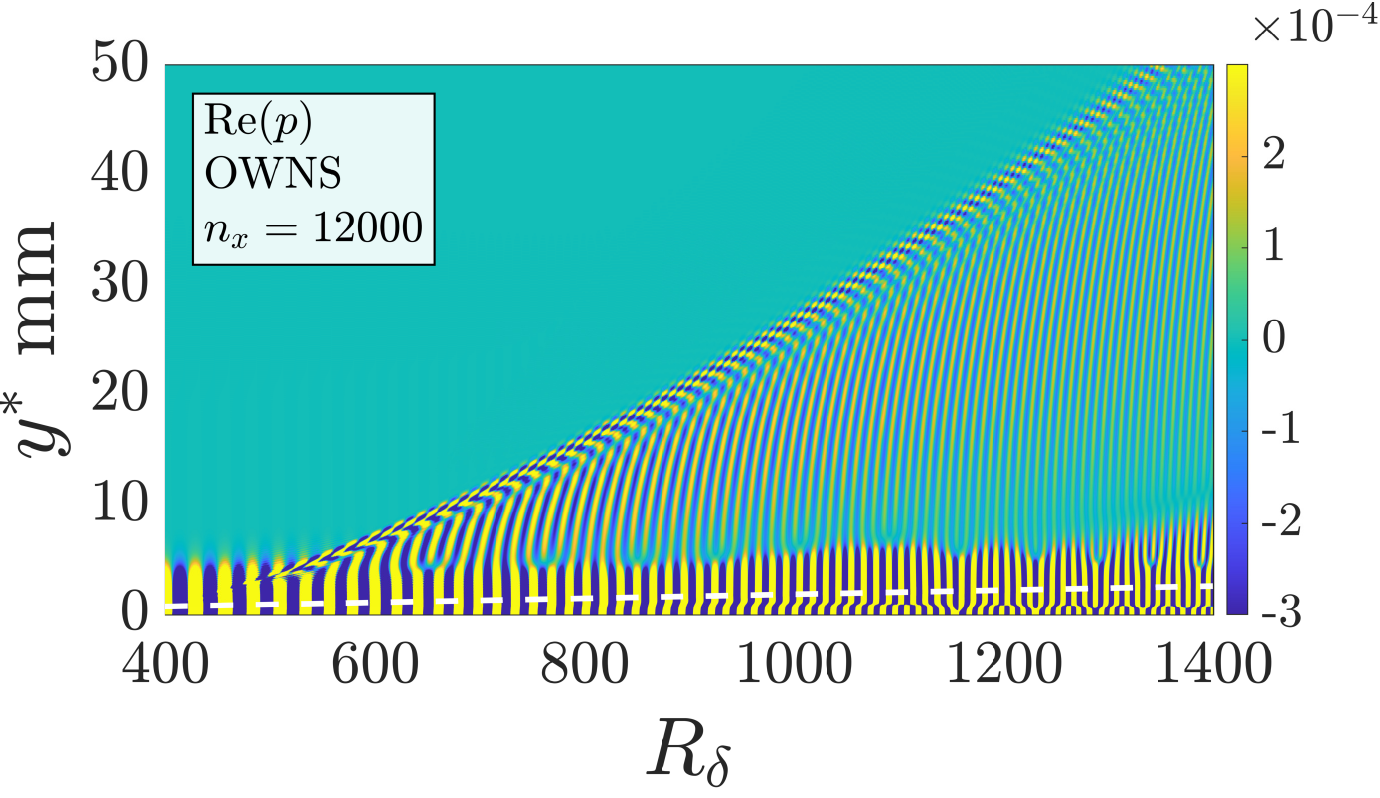}
\end{subfigure}%
\begin{subfigure}{0.49\textwidth}
    \includegraphics[width=\linewidth]{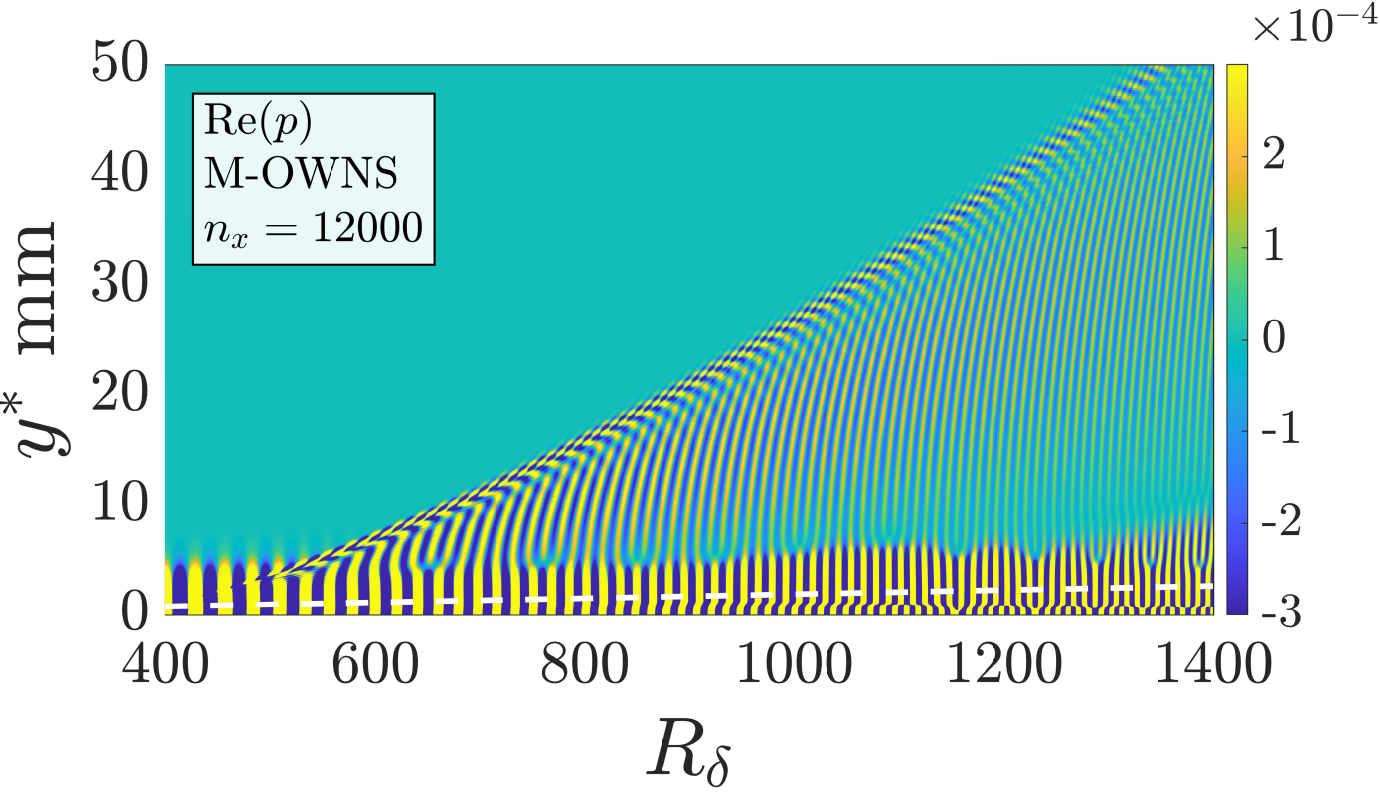}
\end{subfigure}
    \vspace{-0.5cm}
    
\begin{subfigure}{0.49\textwidth}
    \includegraphics[width=\linewidth]{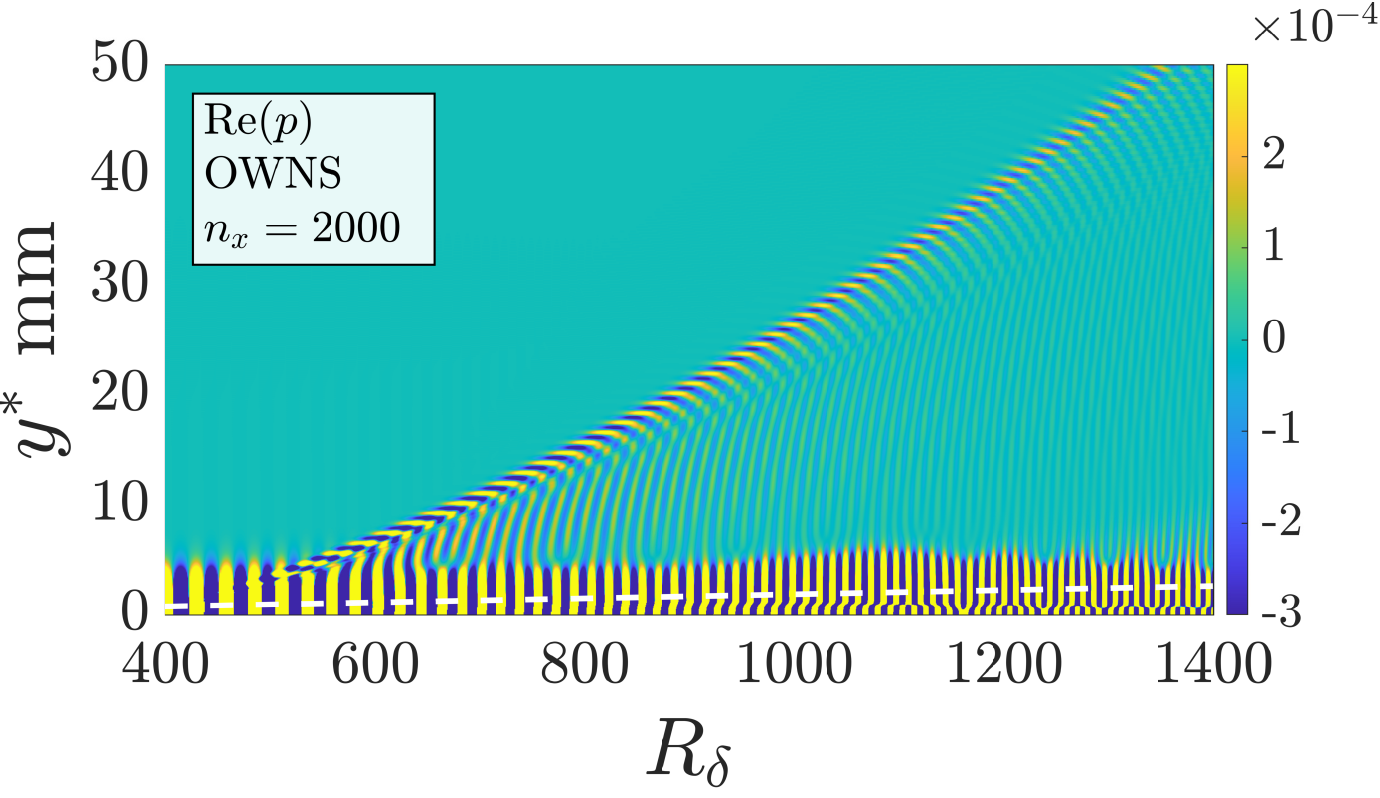}
\end{subfigure}%
\begin{subfigure}{0.49\textwidth}
    \includegraphics[width=\linewidth]{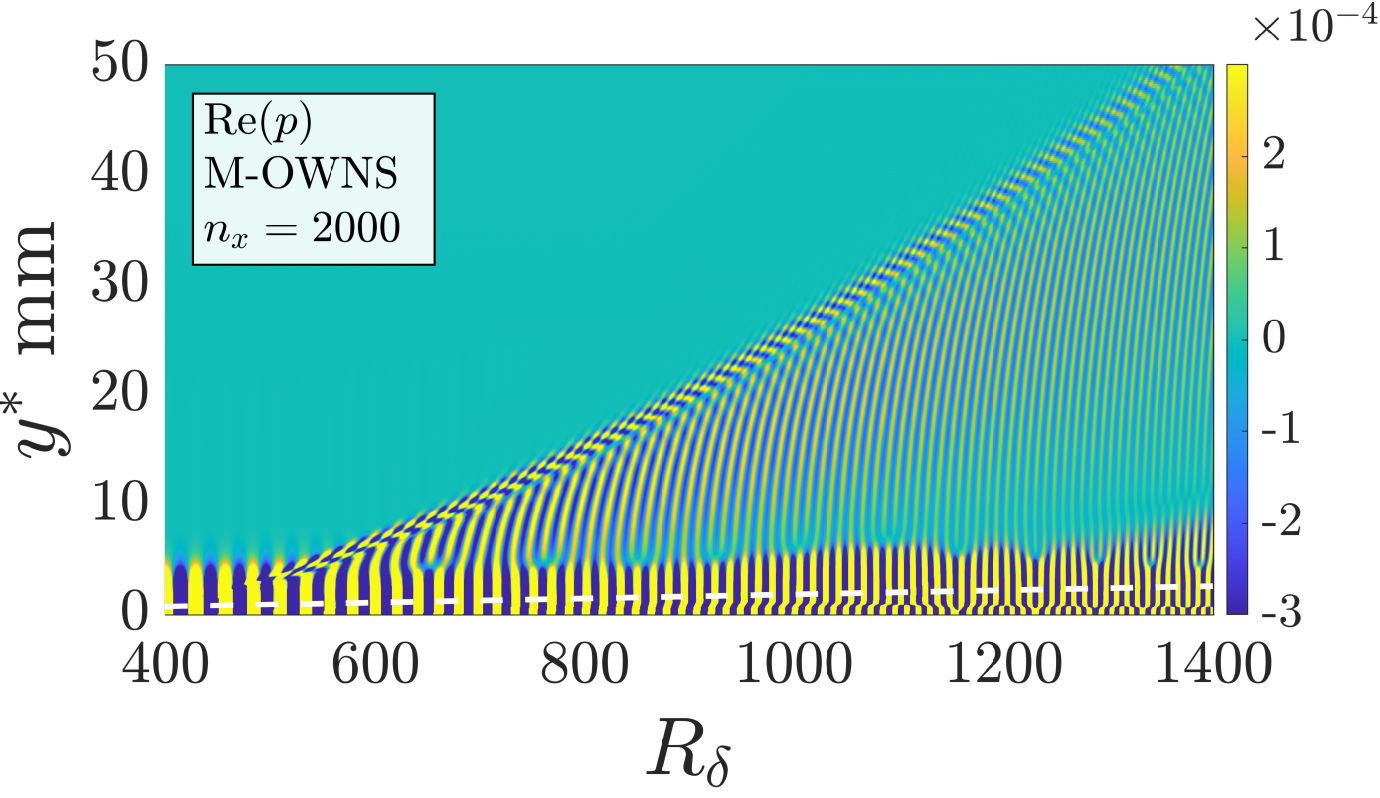}
\end{subfigure}
    \caption{Pressure field $\mathrm{Re}(p)$ for Mode~S forcing in the
    Mach~4.5 boundary layer.  Left: OWNS; right: M-OWNS\@.  Top row:
    $n_x = 12\,000$; bottom row: $n_x = 2000$.  At the coarser
    resolution, OWNS fails to resolve the acoustic radiation field that
    M-OWNS captures.  The dashed white line denotes the boundary-layer
    edge.}\label{fig:superU}
\end{figure}

\paragraph{Non-iterating M-OWNS}\label{sec:no_iter}

We repeat the Mode~S calculation with the fixed carrier
$\alpha_0 = \alpha_{1} = \omega$ (\S\ref{sec:closure}), so that
$N_\mathrm{it} = 0$ and the per-step cost is identical to OWNS.
Fig.~\ref{fig:fixed_b} compares iterated M-OWNS
($\mathcal{E} = 10^{-14}$), non-iterating M-OWNS, and OWNS, all at
$n_x = 1000$.  The two M-OWNS variants are indistinguishable in both
Rayleigh quotient and wall pressure.  OWNS at the same~$n_x$ under-predicts
the amplitude by more than an order of magnitude.  

\begin{figure}[ht!]
\centering
\begin{subfigure}{0.49\textwidth}
    \includegraphics[width=\linewidth]{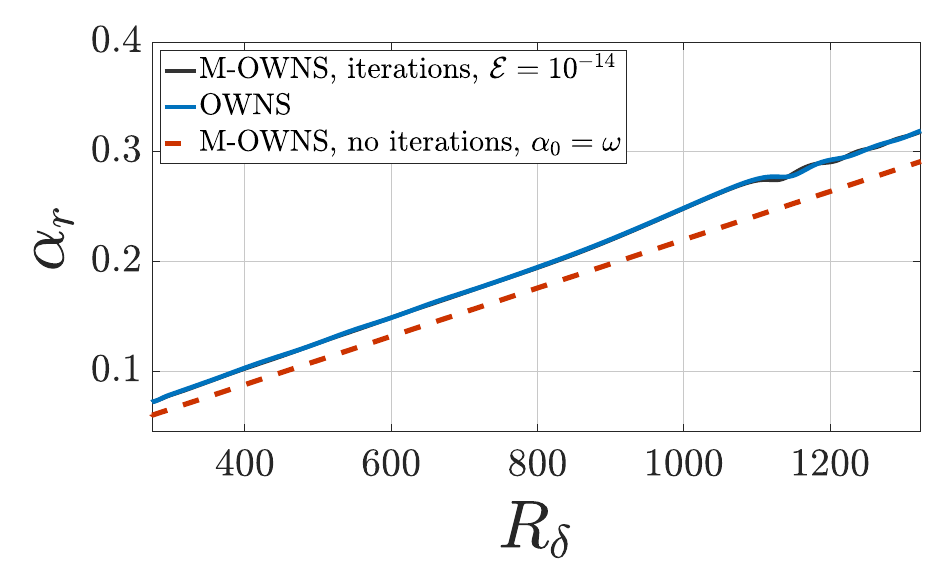}
\end{subfigure}
\hfill
\begin{subfigure}{0.49\textwidth}
    \includegraphics[width=\linewidth]{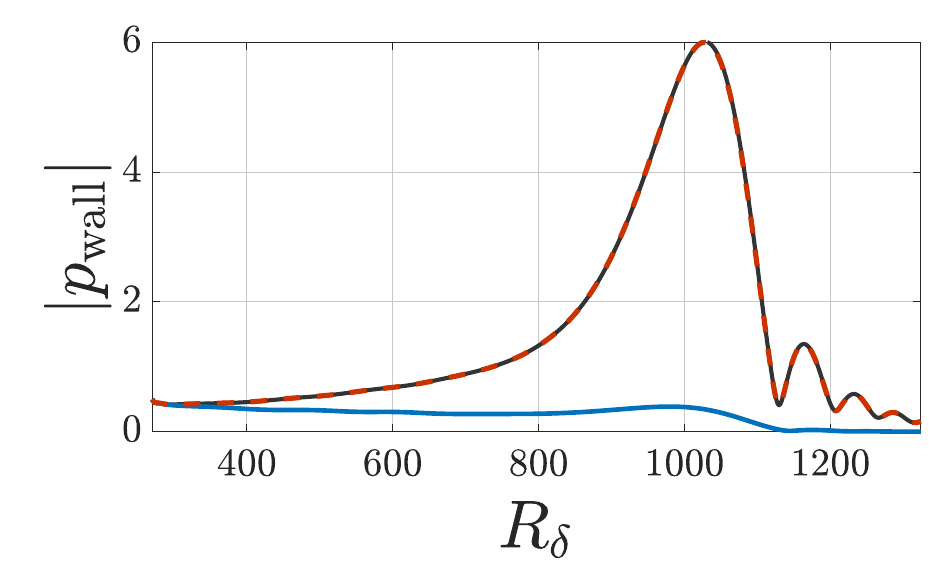}
\end{subfigure}
    \caption{Left: Rayleigh quotient; right: wall pressure, for Mode~S
    forcing in the Mach~4.5 boundary layer at $n_x = 1000$.  Three
    methods are compared: iterated M-OWNS
    ($\mathcal{E} = 10^{-14}$), non-iterating M-OWNS
    ($\alpha_0 = \omega$), and OWNS\@.  In the left panel the
    non-iterating curve shows the fixed carrier~$\alpha_0 = \omega$
    rather than a Rayleigh quotient; the Rayleigh quotient recovered
    from the non-iterating solution (not shown) is identical to the
    iterated one.  The non-iterating variant has identical per-step cost
    to OWNS\@.}\label{fig:fixed_b}
\end{figure}

This is a comparison at equal per-step cost and equal~$n_x$.  For all
resolutions tested, the iterated and non-iterating variants produce
identical results.  Non-iterating M-OWNS captures the S--F synchronisation
and second-mode amplification that OWNS at the same resolution does not. As
predicted by the spectral resolution comparison, the physically relevant
eigenvalues cluster near~$\alpha_{1}$, so the fixed carrier centres the
accuracy disc over the excited spectrum without iteration.

\subsubsection{Wall Forcing via Suction and Blowing}\label{sec:wall_forcing}

The wall forcing of
\citet{maNumericalSimulationReceptivity2001,maReceptivitySupersonicBoundary2003a}
is applied:
\begin{equation}\label{m4p5mazhong}
    v_{\mathrm{wall}}(\bar{x}) = \begin{cases}
    a_{0} \sin \left( 2\pi(\bar{x}-\bar{x}_{0})/\bar{x}_{l} \right),
      & \text{if } |\bar{x}-\bar{x}_{0}| \leq \bar{x}_{l}/2, \\
    0, & \text{otherwise},
    \end{cases}
\end{equation}
where $\bar{x}_{0} = 0.0125\,\mathrm{m}$ ($R_{\delta} = 300$),
$\bar{x}_{l} = 0.0029127\,\mathrm{m}$, and $a_{0}$ is arbitrary.  The
forcing extends over $282 \leq R_{\delta} \leq 317$.  The spatial scales of
the forcing are smaller than PSE's minimum stable step size, precluding its
use. The domain is $275 \leq R_{\delta} \leq 1275$ and $n_y=721$ points in the wall-normal direction, with $N_\beta=40$.

Figure~\ref{fig:forced} shows the normalised wall pressure.  M-OWNS at
$n_x = 1000$ agrees with LHNS at $n_x = 8949$ throughout the domain.  OWNS
at $n_x = 1000$ under-predicts the amplitude.  The pressure and temperature
fields (Fig.~\ref{fig:force}) confirm that M-OWNS and LHNS produce nearly
identical distributions.  LHNS is used as the reference here because this is the only case where upstream-propagating content could differ materially between the one-way and elliptic treatments. Minor differences in the pressure field near the
forcing region are attributable to the projection operator removing
upstream-propagating content in the subsonic sublayer, which the LHNS
retains. 

The normalised wall-normal profiles at $R_\delta \approx 1050$
(Fig.~\ref{fig:pselhnscf}) show agreement in eigenfunction shape across all
three methods.  OWNS captures the shape correctly but not the streamwise amplitude.

\begin{figure}[ht!]
\centering
    \includegraphics[width=0.7\linewidth]{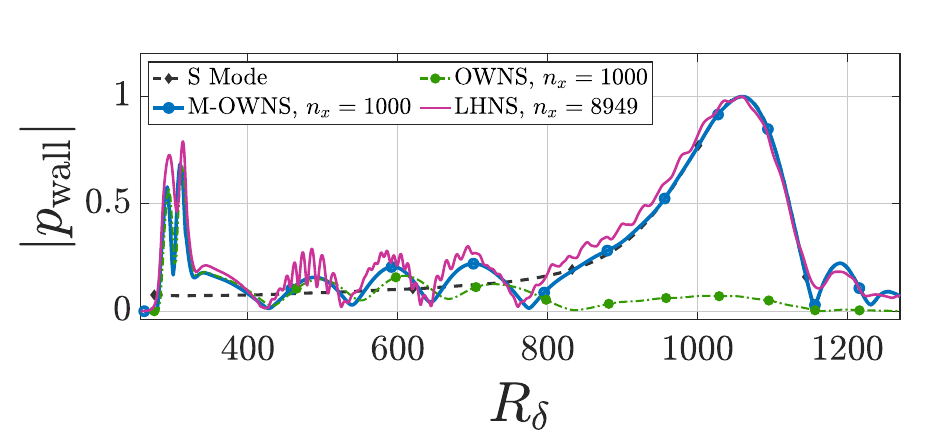}
    \caption{Normalised wall pressure $|p_\mathrm{wall}|$ for the $M = 4.5$
    boundary layer forced by surface suction/blowing
    (Eq.~\ref{m4p5mazhong}).  The dashed black curve is M-OWNS forced
    with a Mode~S eigenfunction at the inlet, shown for
    reference.}\label{fig:forced}
\end{figure}

\begin{figure}[ht!]
\centering
\begin{subfigure}{0.49\textwidth}
    \includegraphics[width=\linewidth]{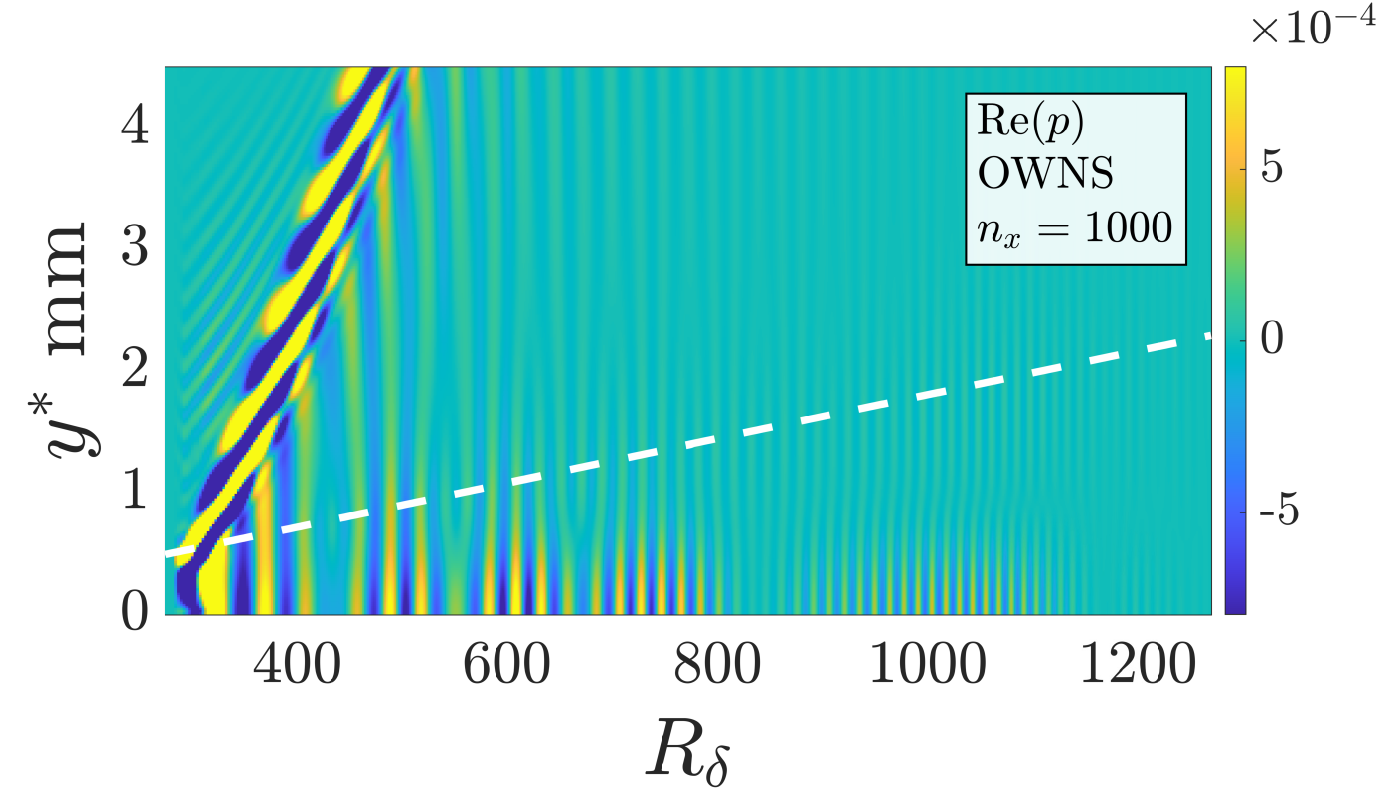}
    \caption{OWNS pressure}\label{fig:force1}
\end{subfigure}%
\begin{subfigure}{0.49\textwidth}
    \includegraphics[width=\linewidth]{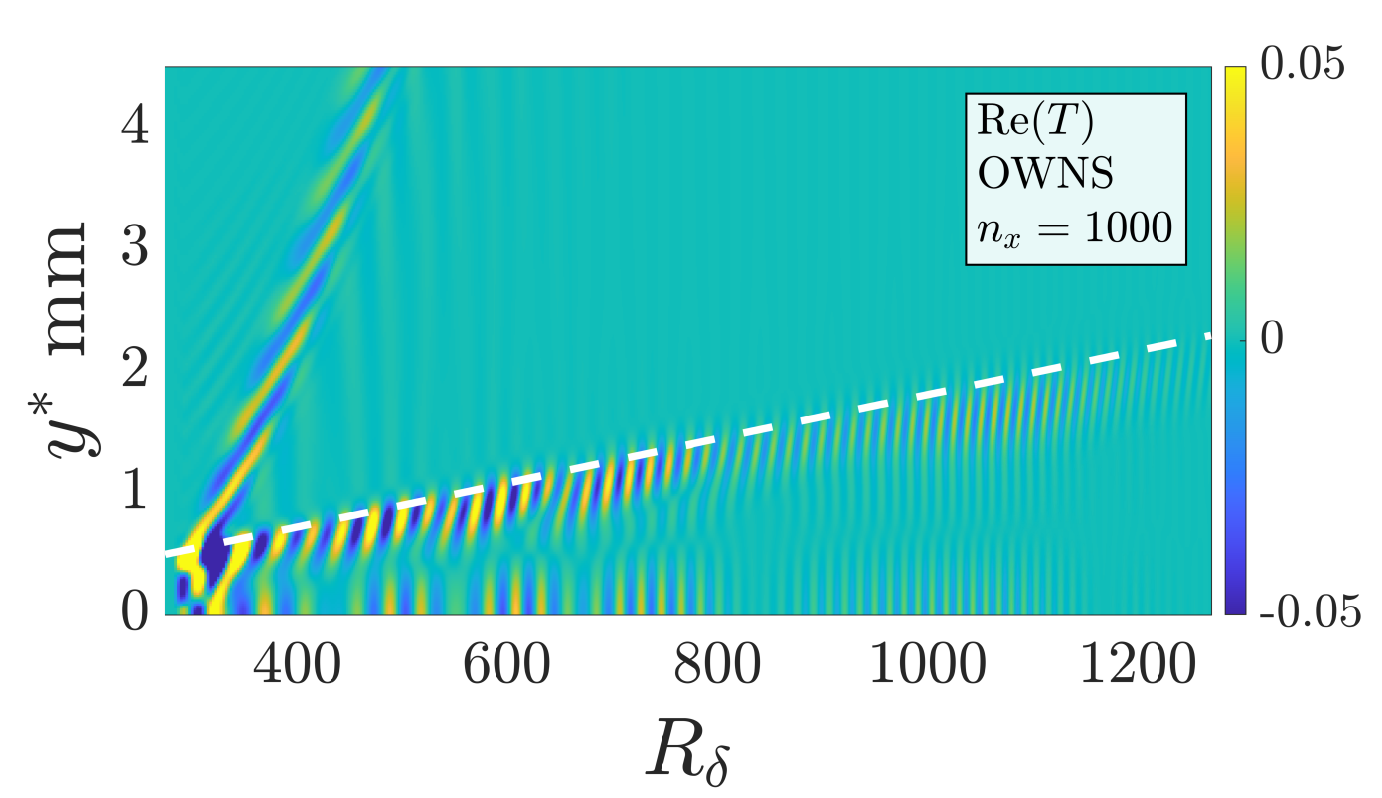}
    \caption{OWNS temperature}\label{fig:force2}
\end{subfigure}

\begin{subfigure}{0.49\textwidth}
    \includegraphics[width=\linewidth]{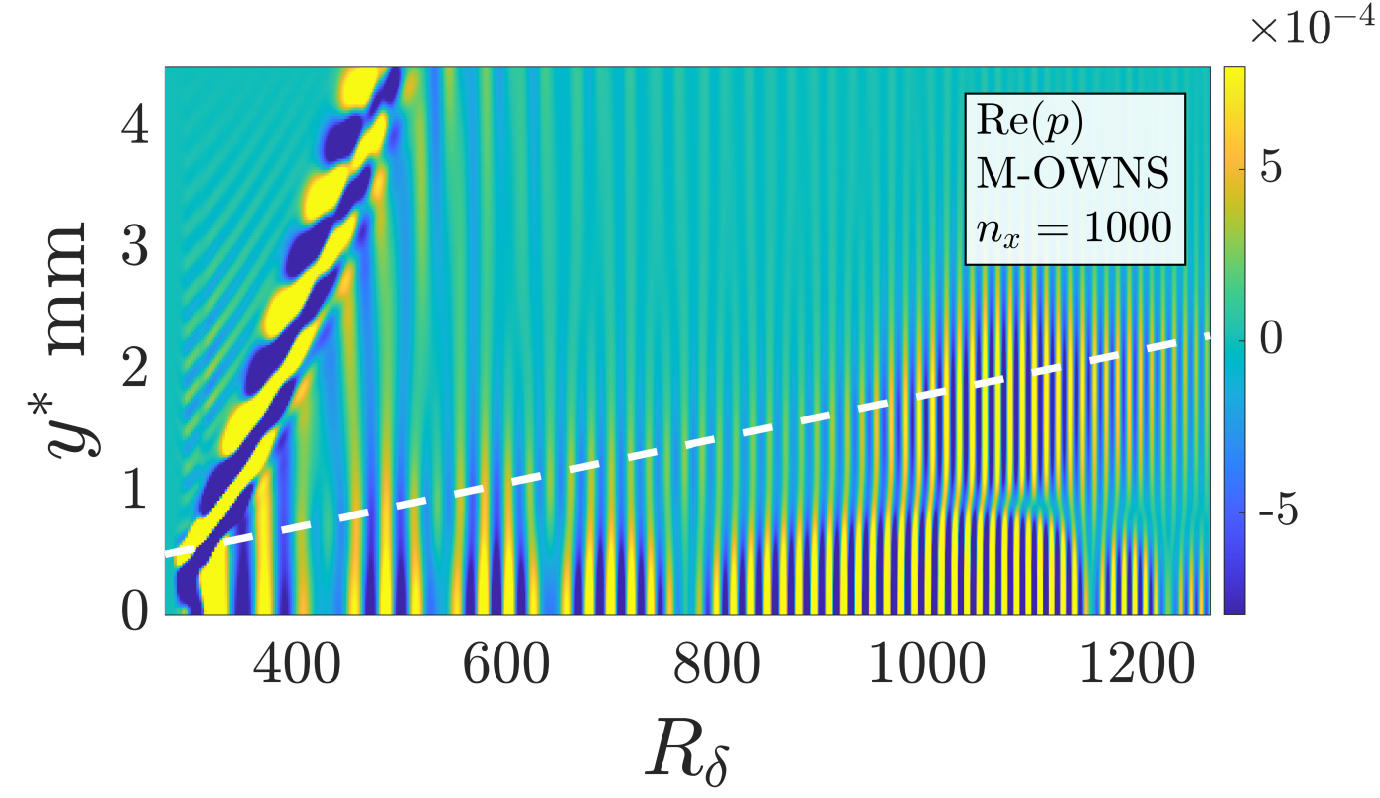}
    \caption{M-OWNS pressure}\label{fig:force3}
\end{subfigure}%
\begin{subfigure}{0.49\textwidth}
    \includegraphics[width=\linewidth]{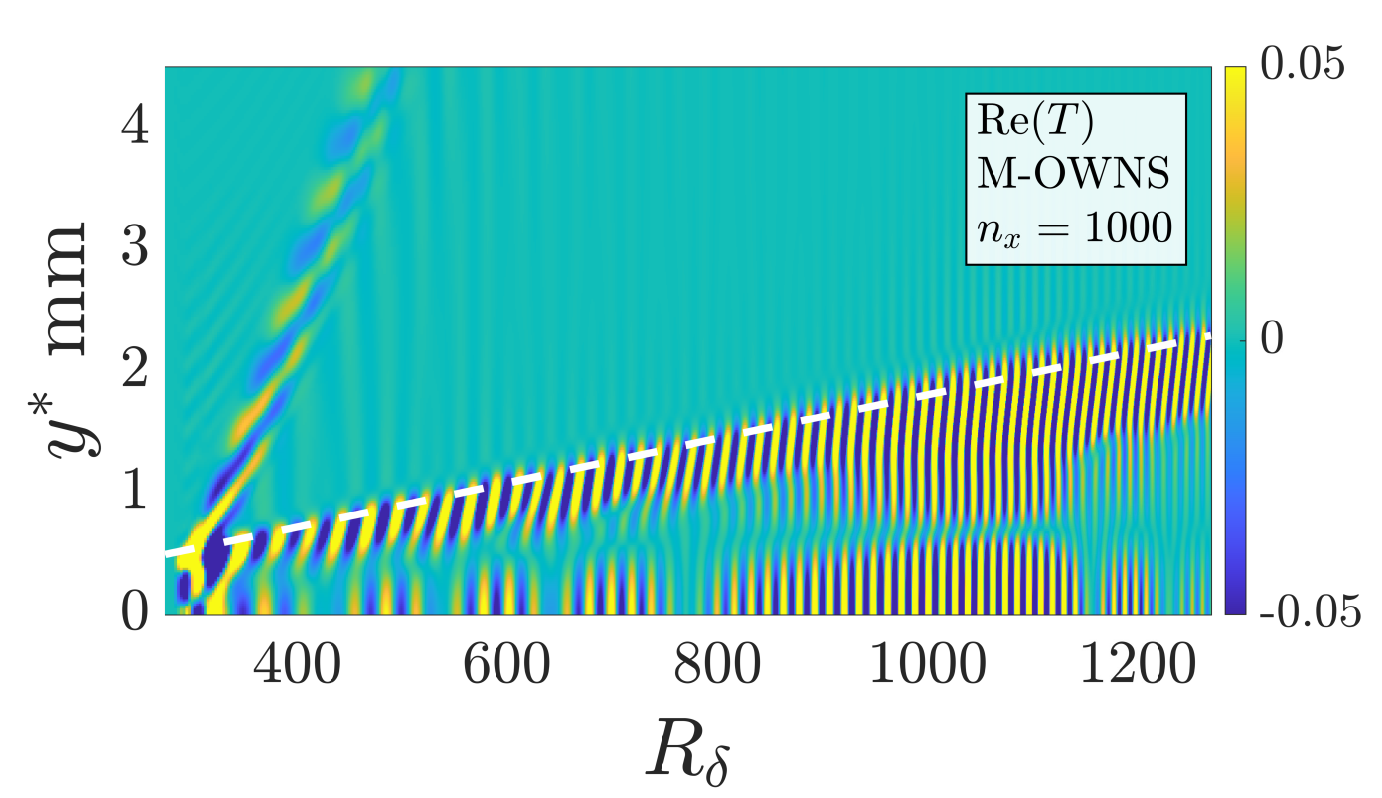}
    \caption{M-OWNS temperature}\label{fig:force4}
\end{subfigure}

\begin{subfigure}{0.49\textwidth}
    \includegraphics[width=\linewidth]{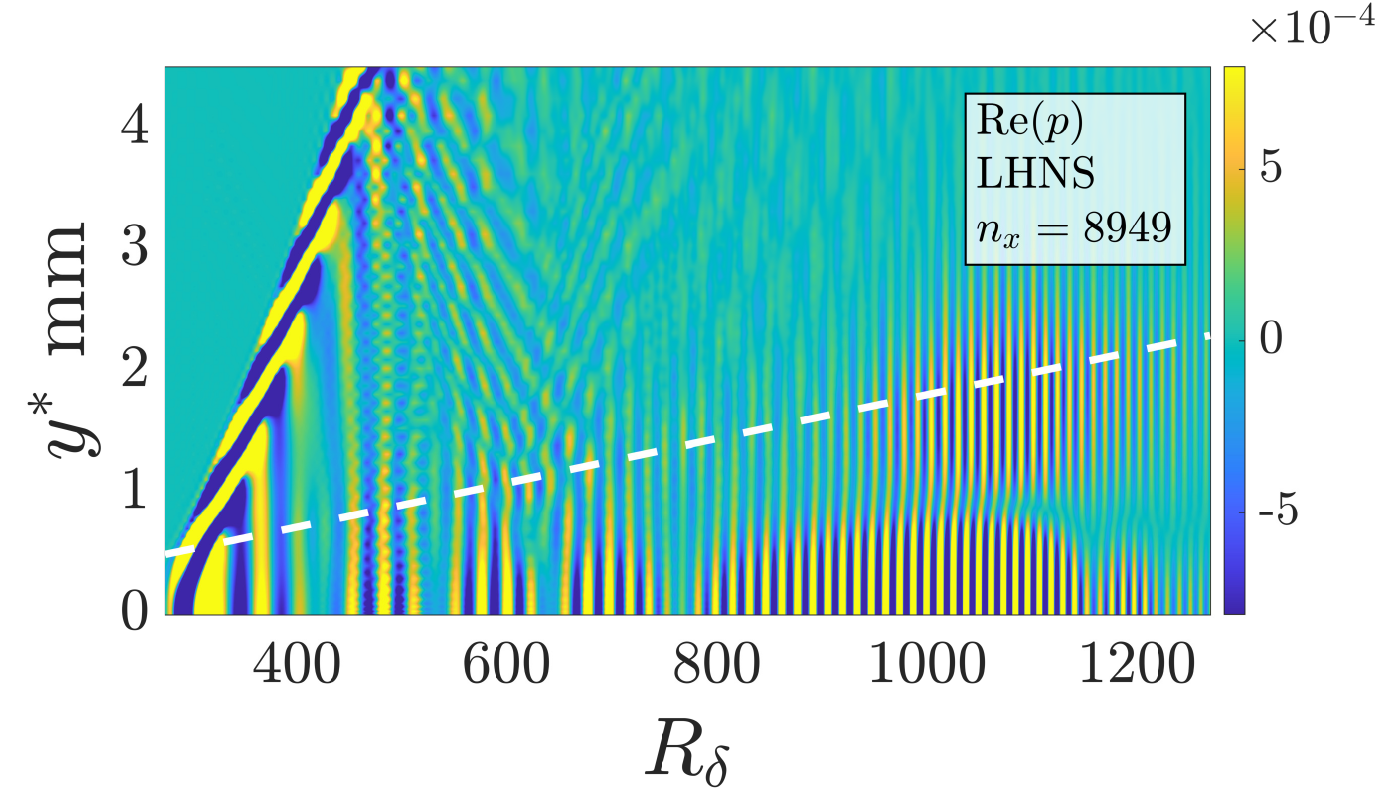}
    \caption{LHNS pressure}\label{fig:force5}
\end{subfigure}%
\begin{subfigure}{0.49\textwidth}
    \includegraphics[width=\linewidth]{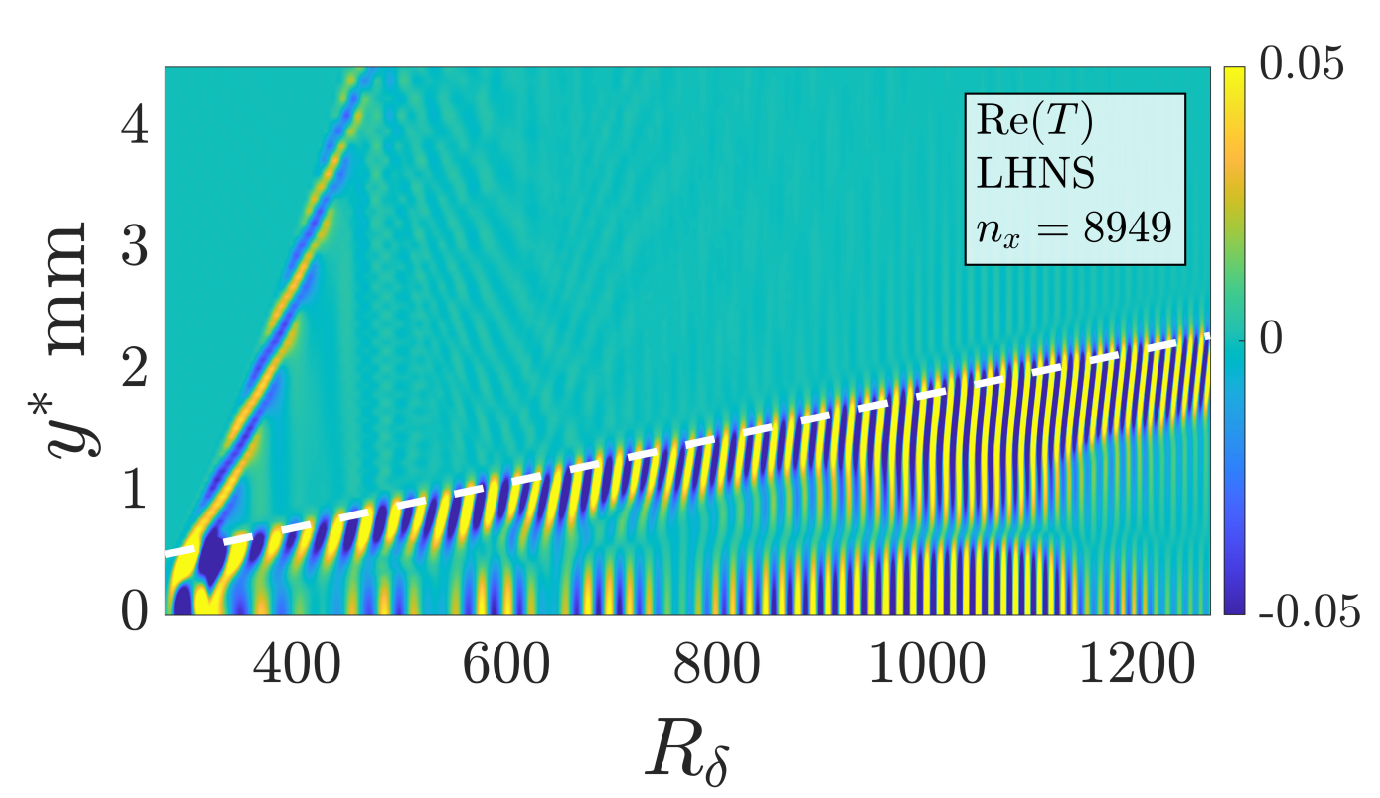}
    \caption{LHNS temperature}\label{fig:force6}
\end{subfigure}
    \caption{Disturbance pressure $\mathrm{Re}(p)$ (left) and temperature
    $\mathrm{Re}(T)$ (right) for the $M = 4.5$ boundary layer with
    surface suction/blowing.  Top: OWNS, $n_x = 1000$; middle: M-OWNS,
    $n_x = 1000$; bottom: LHNS, $n_x = 8949$.  The white dashed line
    denotes the boundary-layer edge.}\label{fig:force}
\end{figure}

\begin{figure}[ht!]
\centering
\begin{subfigure}{0.45\textwidth}
    \includegraphics[width=\linewidth]{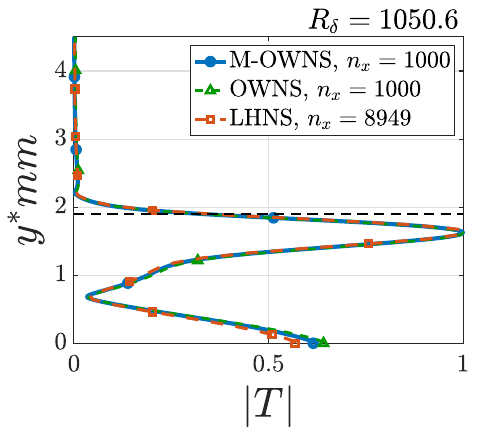}
\end{subfigure}
\hfill
\begin{subfigure}{0.45\textwidth}
    \includegraphics[width=\linewidth]{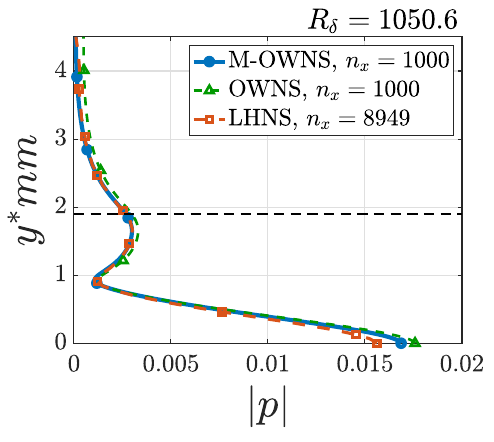}
\end{subfigure}
    \caption{Wall-normal profiles of temperature $|T|$ (left) and pressure
    $|p|$ (right) at $R_{\delta} = 1050.6$.  M-OWNS and OWNS at
    $n_x = 1000$; LHNS at $n_x = 8949$.  Temperature is normalised by
    its maximum.  All three methods agree in eigenfunction shape; the
    unnormalised maximum temperature for OWNS is $0.0241$, compared with
    $0.4981$ for M-OWNS, a factor of~20.  The slight offset at
    $y^{\ast} = 0$ is due to the effect of the one-way projection. Dashed line: boundary-layer
    edge.}\label{fig:pselhnscf}
\end{figure}

\subsubsection{Multi-Mode Freestream Forcing}\label{sec:multi_mode}

The preceding test cases produced responses dominated by Mode~S\@.  To
probe M-OWNS performance when multiple spectrally distinct wave types
contribute simultaneously, we initialise
the computation with a superposition of three freestream wave types: slow
acoustic (SA, $\alpha = \omega M/(M-1)$), fast acoustic (FA,
$\alpha = \omega M/(M+1)$), and vorticity (V, $\alpha = \omega$), each with
equal magnitude and the total disturbance energy normalised to unity.  No
LST eigenfunctions are used.  The computation uses $n_y = 721$, $N_\beta=60$ and marches
from $R_\delta = 270$ to~$1321$.  This configuration simultaneously excites
modes spanning a wide range of Rayleigh quotients, as shown by the
individual wave evolutions in Fig.~\ref{fig:multi_a}.

Despite this spread, the spectral resolution comparison
(\S\ref{sec:circle}) predicts that M-OWNS should retain its resolution
advantage.  The physically relevant criterion is not eigenvalue proximity to
one another, but proximity to the carrier~$\alpha_0$.  All three wave types
are excited at the forcing frequency~$\omega$, so their eigenvalues cluster
near~$\omega$. 

Figure~\ref{fig:multi_b} confirms this prediction.  OWNS at $n_x = 1000$
exhibits large Rayleigh quotient oscillations and wall pressure that departs
from the converged solution.  M-OWNS at $n_x = 1000$ is indistinguishable
from the $n_x = 8000$ reference in both quantities.  The pressure fields
(Fig.~\ref{fig:multi_c}) show the same result: OWNS at $n_x = 1000$
produces unphysical structure absent from the $n_x = 8000$ reference; M-OWNS
at $n_x = 1000$ agrees with the reference throughout.

\begin{figure}[ht!]
\centering
\begin{subfigure}{0.49\textwidth}
    \includegraphics[width=\linewidth]{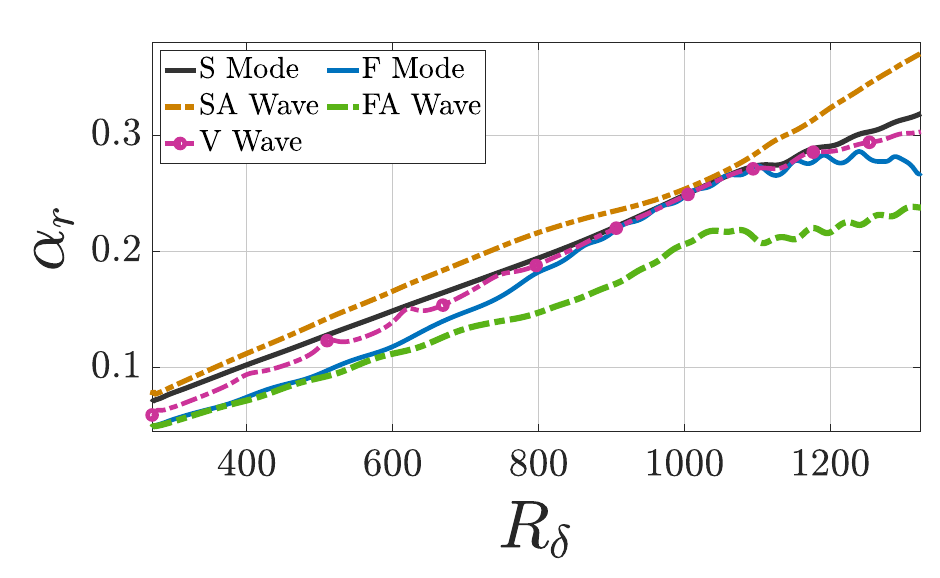}
\end{subfigure}%
\begin{subfigure}{0.49\textwidth}
    \includegraphics[width=\linewidth]{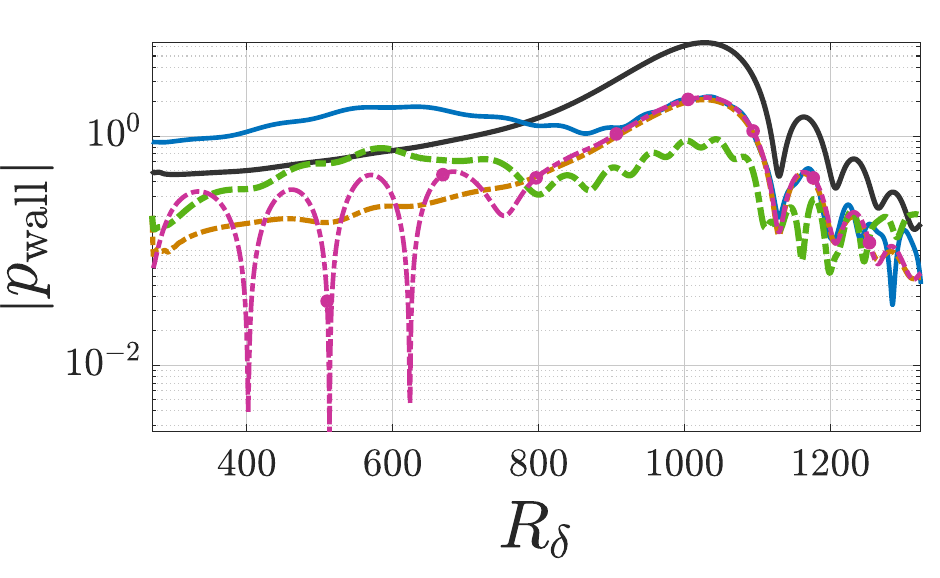}
\end{subfigure}
    \caption{Rayleigh quotient (left) and wall pressure (right) for
    individual freestream wave forcing in the Mach~4.5 boundary layer
    ($F = 220$, $b = 0$): Mode~S
(\textcolor{black}{\raisebox{0.5ex}{\rule{1.2em}{1.2pt}}}), Mode~F
(\textcolor{customblue}{\raisebox{0.5ex}{\rule{1.2em}{1.2pt}}}), SA~wave
(\textcolor{customorange}{\raisebox{0.5ex}{\rule{0.4em}{1.2pt}\kern0.2em\rule{0.4em}{1.2pt}\kern0.2em\rule{0.4em}{1.2pt}}}),
FA~wave
(\textcolor{customgreen}{\raisebox{0.5ex}{\rule{0.4em}{1.2pt}\kern0.2em\rule{0.4em}{1.2pt}\kern0.2em\rule{0.4em}{1.2pt}}}),
and V~wave
(\textcolor{custompurple}{\raisebox{0.5ex}{\rule{0.4em}{1.2pt}\kern0.2em\rule{0.4em}{1.2pt}\kern0.2em\rule{0.4em}{1.2pt}}}).
    Each wave is computed individually with M-OWNS at $n_x = 1000$.
    These evolutions serve as reference for the combined forcing in
    Fig.~\ref{fig:multi_b}.}\label{fig:multi_a}
\end{figure}

\begin{figure}[ht!]
\centering
\begin{subfigure}{0.49\textwidth}
    \includegraphics[width=\linewidth]{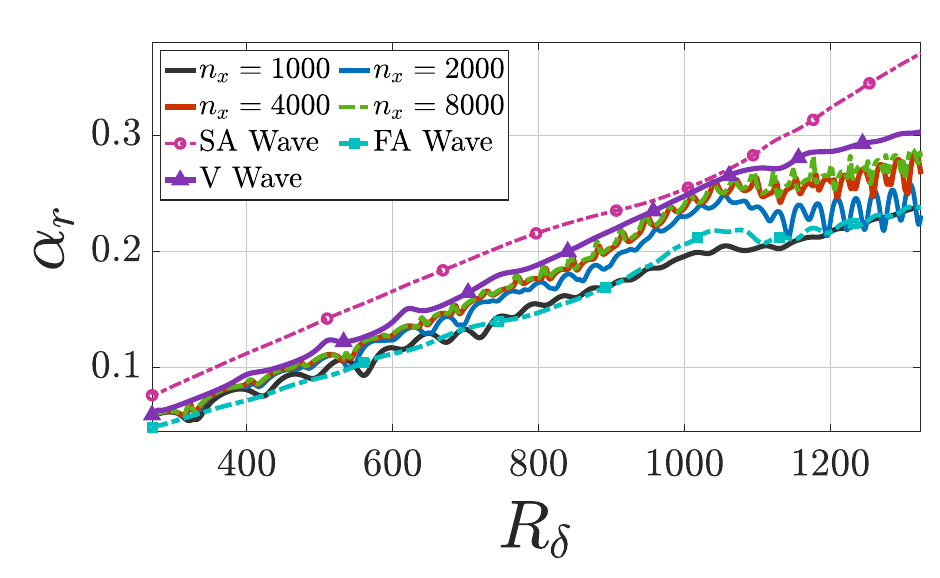}
\end{subfigure}
\hfill
\begin{subfigure}{0.49\textwidth}
    \includegraphics[width=\linewidth]{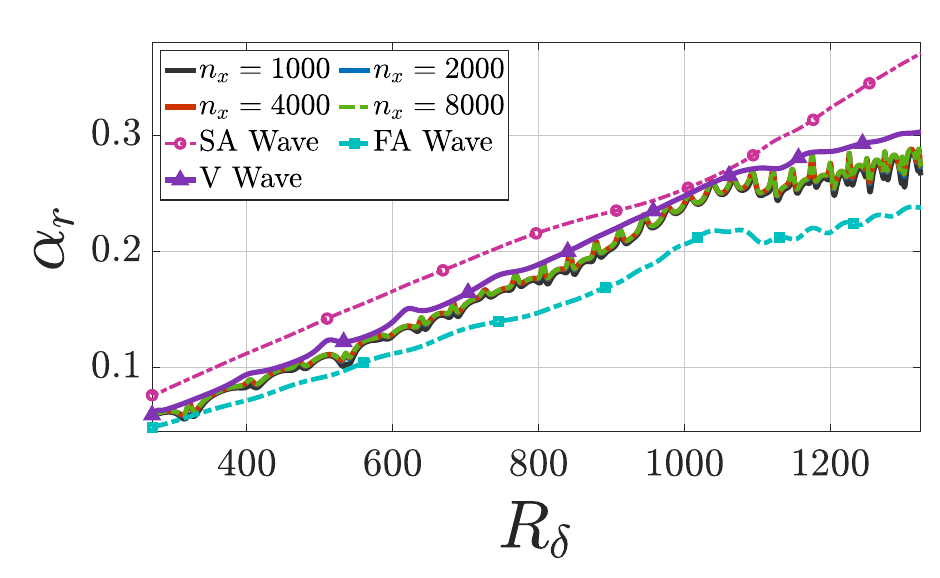}
\end{subfigure}

\begin{subfigure}{0.49\textwidth}
    \includegraphics[width=\linewidth]{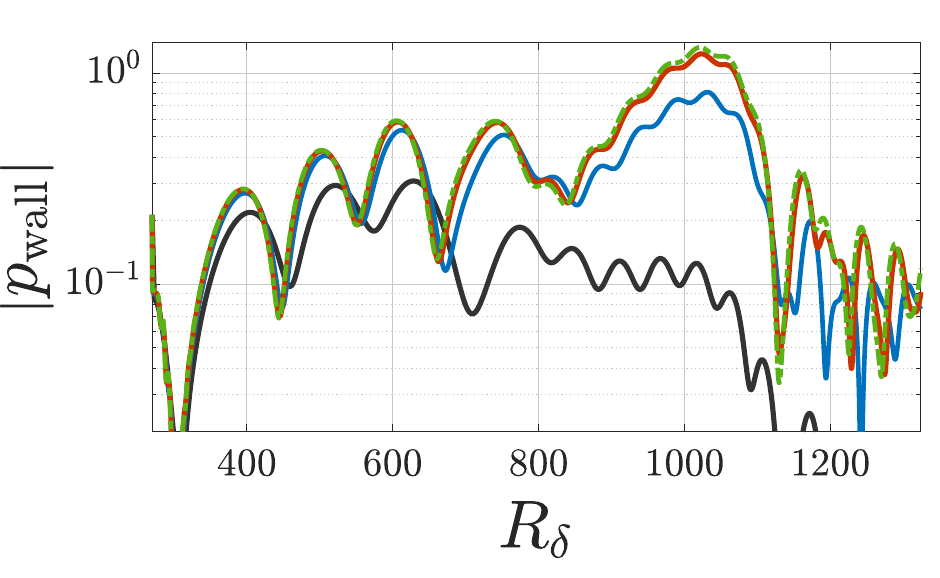}
\end{subfigure}
\hfill
\begin{subfigure}{0.49\textwidth}
    \includegraphics[width=\linewidth]{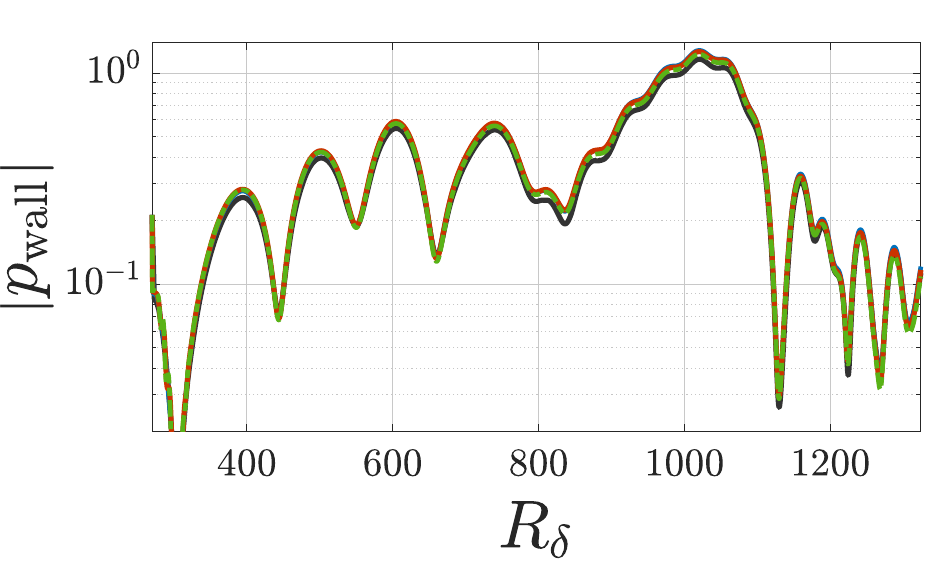}
\end{subfigure}
    \caption{Multi-mode freestream forcing (SA $+$ FA $+$ V) in the
    Mach~4.5 boundary layer.  Left: OWNS; right: M-OWNS\@.  Top row:
    Rayleigh quotient; bottom row: wall pressure.  Resolutions:
    $n_x = 1000$, $2000$, $4000$ and~$8000$.  The individual SA, FA and
    V Rayleigh quotients from Fig.~\ref{fig:multi_a} are shown for
    reference.  OWNS exhibits strong resolution dependence at
    $n_x \leq 2000$; M-OWNS is converged at all
    resolutions.}\label{fig:multi_b}
\end{figure}

\begin{figure}[ht!]
\centering
\begin{subfigure}{0.99\textwidth}
    \includegraphics[width=\linewidth]{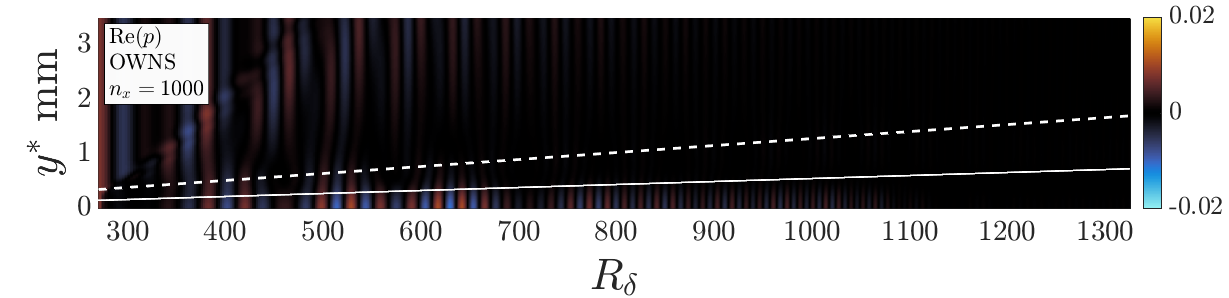}
\end{subfigure}\vspace{-0.5cm}

\begin{subfigure}{0.99\textwidth}
    \includegraphics[width=\linewidth]{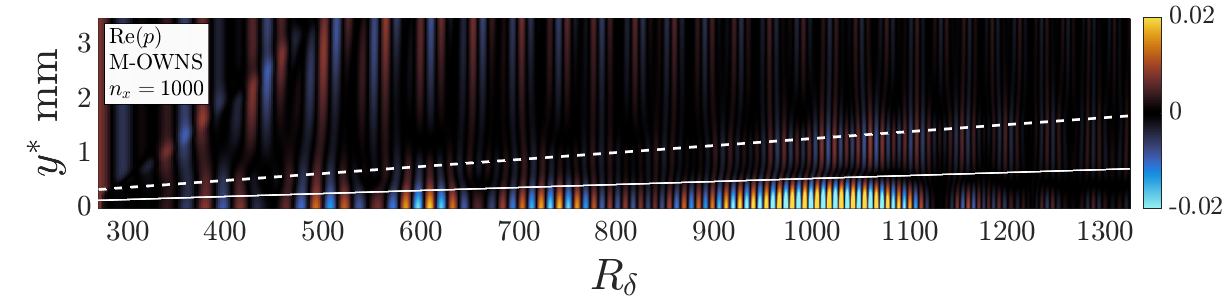}
\end{subfigure}\vspace{-0.5cm}

\begin{subfigure}{0.99\textwidth}
    \includegraphics[width=\linewidth]{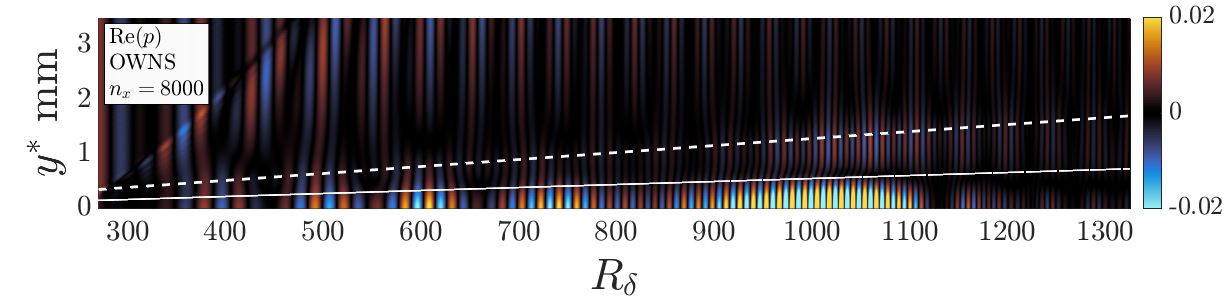}
\end{subfigure}\vspace{-0.5cm}

\begin{subfigure}{0.99\textwidth}
    \includegraphics[width=\linewidth]{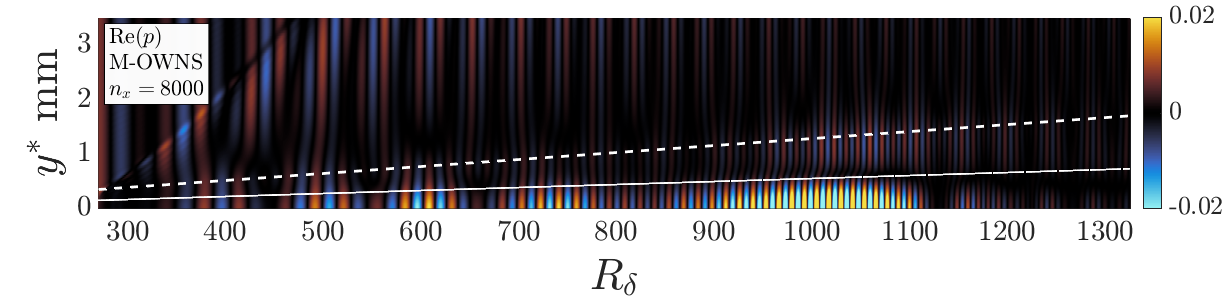}
\end{subfigure}
    \caption{Pressure field $\mathrm{Re}(p)$ for multi-mode freestream
    forcing (SA $+$ FA $+$ V) in the Mach~4.5 boundary layer.  From top
    to bottom: OWNS at $n_x = 1000$, M-OWNS at $n_x = 1000$, OWNS at
    $n_x = 8000$, M-OWNS at $n_x = 8000$.  White dashed line:
    boundary-layer edge; white solid line: sonic line.  At $n_x = 1000$,
    OWNS fails to resolve the near-wall second-mode structure that
    M-OWNS captures; at $n_x = 8000$ both methods
    agree.}\label{fig:multi_c}
\end{figure}

\subsubsection{Randomised Inlet Forcing}\label{sec:random}

The multi-mode forcing of \S\ref{sec:multi_mode} prescribes specific wave
types at the inlet.  To test robustness under less controlled conditions, we
replace the structured inlet with a randomised disturbance profile.  Each
component of the state vector at every wall-normal grid point is multiplied
by an independent random factor drawn uniformly from $[-1,\,1]$.  The
profile is truncated to zero above $y^{\ast} = 10$\,mm to preserve the
freestream boundary conditions (Fig.~\ref{fig:random0}).  The computation
uses $n_y = 721$, $N_\beta=60$ and marches from $R_\delta = 270$ to~$1321$.  Despite the
incoherent inlet, the field rapidly organises into a coherent wave structure
(Fig.~\ref{fig:random_field}). 

Figure~\ref{fig:random1} compares wall pressure convergence for OWNS,
iterated M-OWNS ($\alpha_0 = \alpha$), and fixed-carrier M-OWNS
($\alpha_0 = \omega$).  At $n_x = 1000$, OWNS and M-OWNS resolve
complementary parts of the spectrum.  In the upstream region
($R_\delta \lesssim 500$), OWNS captures small-scale oscillations in
$|p_{\mathrm{wall}}|$ but not the correct amplitude, while M-OWNS captures
the amplitude but not the oscillations.  This is consistent with the spectral resolution
condition (\S\ref{sec:circle}): eigenvalues nearer the origin than~$\omega$
satisfy $|\lambda_k| < |\lambda_k - \omega|$ and are better resolved by the
unfactored system, while eigenvalues near~$\omega$ are better resolved by
the factored system.  The Rayleigh quotient data
(Fig.~\ref{fig:random_alpha}) support this: $\alpha_r$ lies below~$\omega$
in the upstream region, indicating that the dominant content has wavenumber
closer to the origin than to~$\omega$. 

Downstream, Mode~S growth through S--F synchronisation produces content with
Rayleigh quotient near~$\omega$.  OWNS at $n_x = 1000$ loses the solution
by $R_\delta \approx 600$; $n_x = 3000$ begins to under-resolve Mode~S
growth by $R_\delta \approx 700$; convergence requires
$n_x \approx 6000$.  Both M-OWNS variants resolve the full domain by
$n_x = 3000$.

The fixed carrier $\alpha_0 = \omega$ performs comparably to the iterated
Rayleigh quotient throughout.  The iteration converges for this realisation at every streamwise step, despite the Rayleigh quotient being used as a recursion parameter (see \S\ref{app:supersonic}).
The fixed carrier requires no convergence and so avoids the cross-term
sensitivity of the Rayleigh quotient under strongly non-orthogonal
excitation~(\S\ref{sec:energy_wavenumber}); this non-orthogonality may cause the iterations to fail for different initial conditions~(§\ref{sec:closure}).


\begin{figure}[ht!]
\centering
    \includegraphics[width=0.4\linewidth]{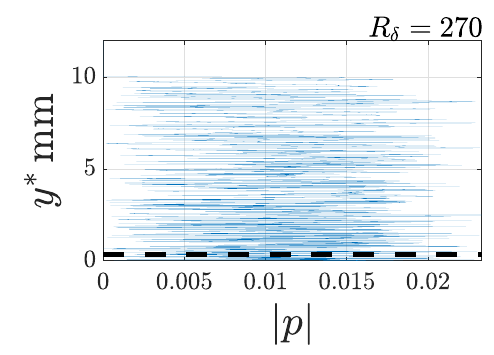}
    \caption{Randomised initial condition at $R_\delta = 270$.  The
    profile is set to zero above $y^{\ast} = 10$\,mm.  Dashed line:
    boundary-layer edge.}\label{fig:random0}
\end{figure}

\begin{figure}[ht!]
\centering
\begin{subfigure}{0.99\textwidth}
    \includegraphics[width=\linewidth]{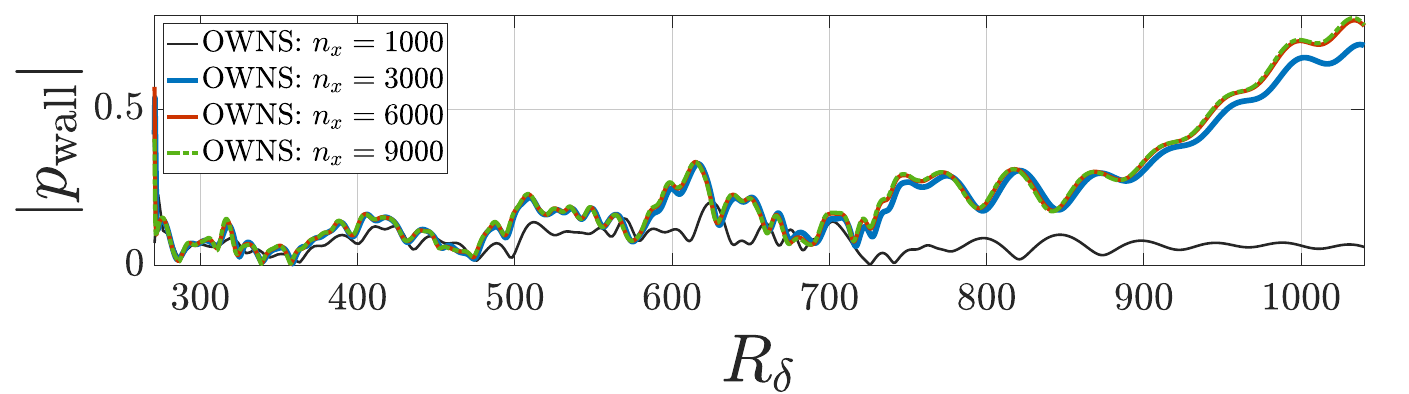}
\end{subfigure}\vspace{-1.1cm}

\begin{subfigure}{0.99\textwidth}
    \includegraphics[width=\linewidth]{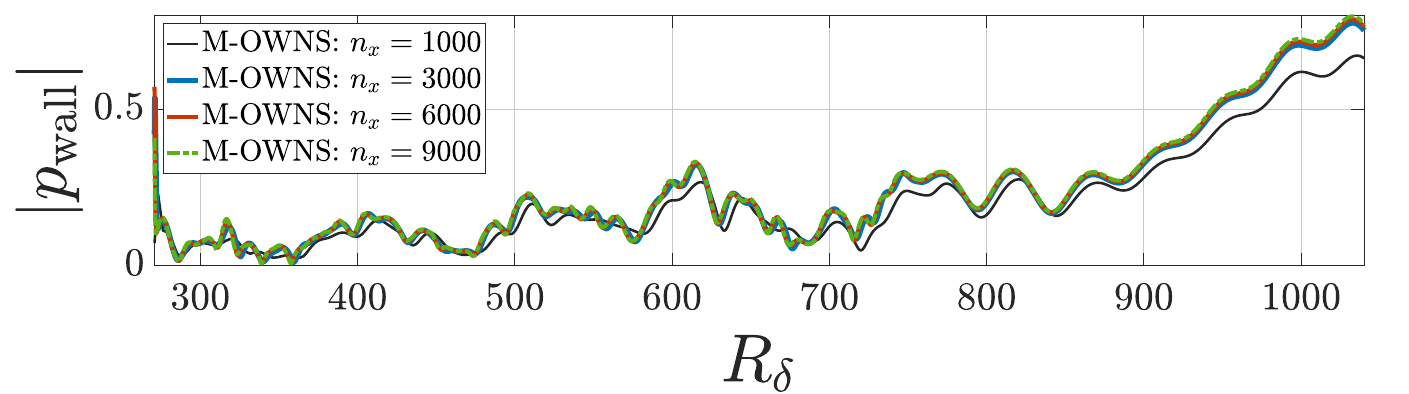}
\end{subfigure}\vspace{-1.1cm}

\begin{subfigure}{0.99\textwidth}
    \includegraphics[width=\linewidth]{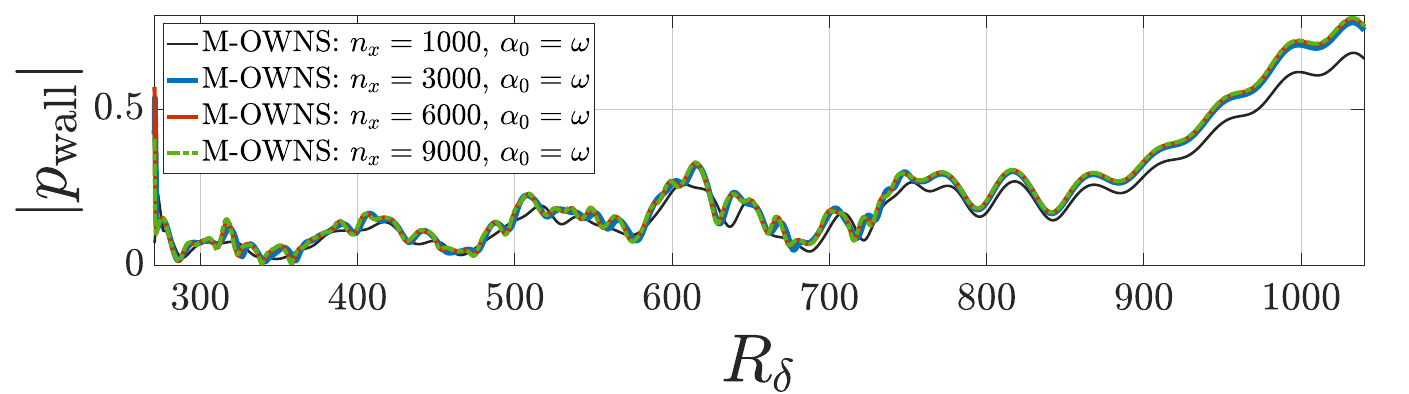}
\end{subfigure}
    \caption{Wall pressure $|p_{\mathrm{wall}}|$ for randomised inlet
    forcing.  Top: OWNS; middle: iterated M-OWNS ($\alpha_0 = \alpha$);
    bottom: fixed-carrier M-OWNS ($\alpha_0 = \omega$).  Each at
    $n_x = 1000$, $3000$, $6000$ and~$9000$.}\label{fig:random1}
\end{figure}

\begin{figure}[ht!]
\centering
    \includegraphics[width=0.8\linewidth]{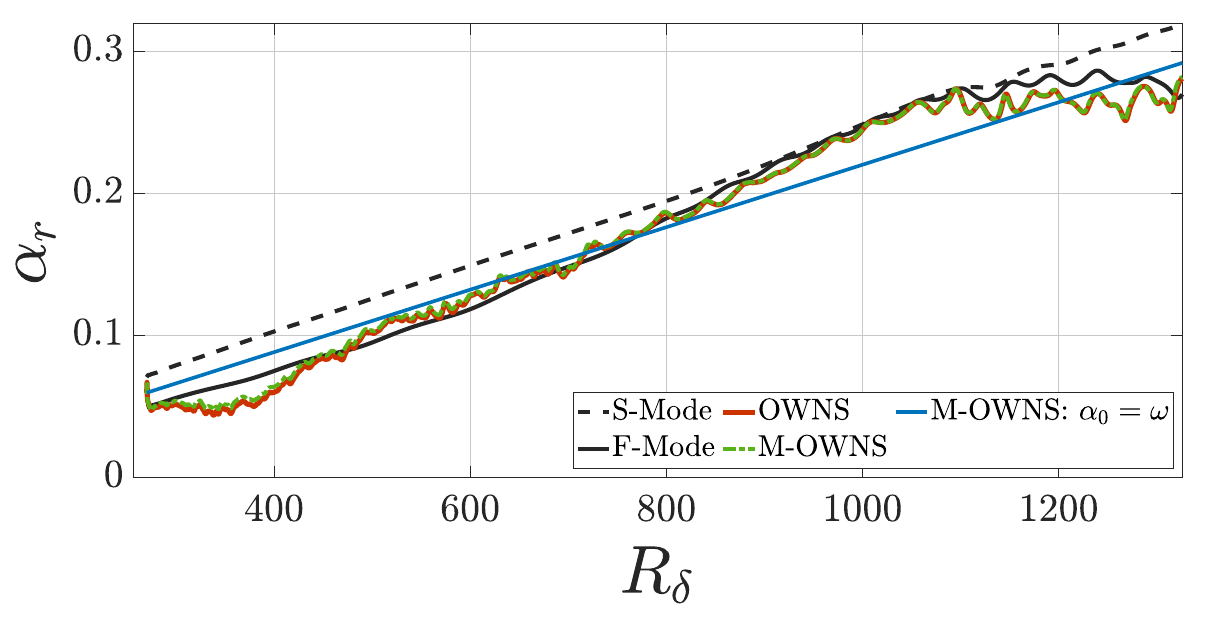}
    \caption{Rayleigh quotient $\alpha_r$ for the randomised inlet
    forcing, with Mode~S and Mode~F Rayleigh quotients from OWNS with
    LST initialisation.  The fixed-carrier M-OWNS curve shows
    $\alpha_0 = \omega$ to indicate that no iterations are used.  All
    randomised forcing calculations are converged at
    $n_x = 9000$.}\label{fig:random_alpha}
\end{figure}

\begin{figure}[ht!]
\centering
    \includegraphics[width=0.99\linewidth]{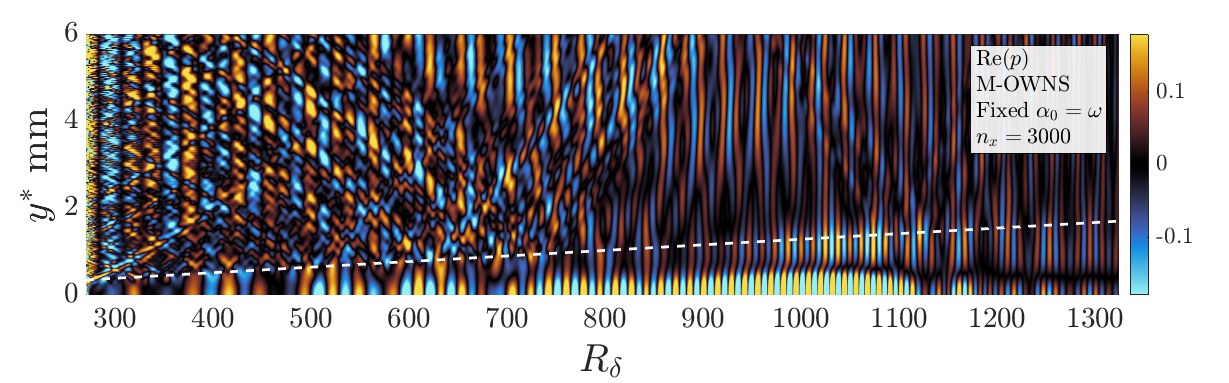}
    \caption{Pressure field $\mathrm{Re}(p)$ for fixed-carrier M-OWNS
    ($\alpha_0 = \omega$) at $n_x = 3000$.  White dashed line:
    boundary-layer edge.}\label{fig:random_field}
\end{figure}


\subsection{Total cost}\label{sec:total_cost}

The total cost of each method scales as $n_x$ times the per-step cost of
Table~\ref{tab:cost}. {{To leading order, the step count $n_x$ required for a
given accuracy is set by the spectral resolution comparison of~\S\ref{sec:circle}:
carrier-wave factoring reduces the residual wavenumbers of modes near $\alpha_0$,
permitting coarser step sizes for any consistent spatial discretisation. Beyond leading order the gain depends on the scheme and on the eigenvalue
phase}}; the convergence studies below report the $n_x$ each method requires in
practice.

The randomised-inlet case of~\S\ref{sec:random} is the most stringent test of M-OWNS. Both M-OWNS variants converge at $n_x = 3000$ while OWNS requires $n_x = 6000$. With $N_\beta = 60$, this corresponds to $3.7 \times 10^5$ solves for OWNS. For iterating M-OWNS with $N_{\text{it}} = 3$, the total solve count is $1.9 \times 10^5$, a reduction of $47.5\%$. For fixed-carrier M-OWNS the total solve count is $1.8 \times 10^5$, a reduction of $50.0\%$. {{Even}} larger reductions are obtained in less stringent cases. 

The swept-cylinder configuration of~\S\ref{hiemy_flow} uses $N_\beta = 20$ and requires $n_x = 8000$ for OWNS, corresponding to $1.7 \times 10^5$ solves. Iterating M-OWNS converges at $n_x = 150$ with $N_{\text{it}} = 3$, giving $3.6 \times 10^3$ solves, a factor of 47 below OWNS. Fixed-carrier M-OWNS converges at $n_x = 1000$, giving $2.1 \times 10^4$ solves, a factor of $8$ below OWNS. OWNS resolves the full inter-branch span with $\alpha_0 = 0$; both M-OWNS variants centre the accuracy disc near the crossflow wavenumber, with the iterating variant tracking it exactly.

Across all test cases in~\S\ref{sec:results}, the randomised-inlet study shows the smallest reduction in $n_x$ at equivalent accuracy; the other cases show larger reductions. All computations reported here were performed on standard desktop hardware.

\section{Conclusions}\label{sec:conclusions}
This paper has presented M-OWNS, a spatial marching method that combines
the carrier-wave factoring of the parabolised stability equations (PSE)
with the one-way projection of the recursive one-way Navier--Stokes (OWNS-R)
framework. We have shown that this additional step offers significant computational benefits, while still providing equivalent modelling fidelity to standard OWNS computations. The method is validated across subsonic and hypersonic
boundary-layer flows as well as for three-dimensional crossflow disturbances. In addition, 
four forcing scenarios (at inlet or the wall) in a Mach~4.5 boundary-layer, provide results consistent with conventional OWNS and LHNS models.

The resolution advantage is explained by the spectral resolution comparison
derived in~\S\ref{sec:circle}: the carrier-wave factoring translates the
integrator's accuracy region from the origin to the carrier wavenumber
$\alpha_0$ in spectral space, reducing the residual wavenumbers of all modes near $\alpha_0$.
At a single forcing frequency, the physically
relevant eigenvalues cluster near the vorticity/entropy branch
endpoint~$\alpha_{1}$, so the criterion is generically satisfied. 
{To leading order the spectral resolution condition is set by the eigenvalue geometry (\S\ref{sec:circle}); higher-order corrections depend on the scheme.} In
practice, wave-factoring {{allows}} the one-way projection to suppress the
upstream-propagating modes at coarser streamwise step size $\Delta x$, a role the
regularisation strategies of PSE cannot play without introducing characteristic
errors \citep{towneCriticalParabolized2019}.

For the fixed-carrier variant, M-OWNS reduces the total solve count by factors of two to eight in all test cases considered. The lower bound corresponds to the
randomised-inlet case~(\S\ref{sec:random}), the most stringent multi-modal
test: OWNS requires $n_x = 6000$ while fixed-carrier M-OWNS converges at
$n_x = 3000$ with $N_\beta = 60$. The upper bound corresponds to the
swept-cylinder configuration~(\S\ref{hiemy_flow}), where fixed-carrier
M-OWNS converges at $n_x = 1000$ against $n_x = 8000$ for OWNS. When a
single mode dominates, the carrier-wave approach achieves even further reductions. For the swept-cylinder {{study}}, iterating M-OWNS converges {{with as few as}}
$150\; (n_x)$ {{grid points}}: a $47$-fold reduction in computational cost relative to pure OWNS.

When the iterated Rayleigh quotient is used as a closure condition, a poorly chosen
or divergent~$\alpha_0$ can degrade accuracy by displacing the accuracy
region away from the excited spectrum~(\S\ref{sec:closure}). The
non-iterating variant, with $\alpha_0$ set to the vorticity/entropy branch
endpoint~$\alpha_{1}$ from Eq.~\eqref{eqn:continuous_spectra}, eliminates
this risk at identical per-step cost to unfactored OWNS. The factoring
provides no benefit when~$\alpha_{1}$ itself is near the origin; in this case $\alpha_0 = 0$ recovers unfactored OWNS with no penalty.

The choice $\alpha_0 = \alpha_1$ is appropriate when the excited spectrum
is dominated by the vorticity/entropy branch and nearby discrete modes,
which covers all test cases reported here. For regimes in which the
acoustic and vortical responses are of comparable amplitude (\emph{e.g.} optimal-disturbance and resolvent analyses in which both
branches contribute to the gain) a single branch-endpoint carrier need
not be optimal, and a minimax placement that balances the residual
wavenumbers across the relevant branch endpoints may be preferable. More extensive investigations using the M-OWNS approach to modelling forced
problems summarised in this paper may be found in~\citet{ejb2025}.

\section*{Funding}
This work was supported by the United Kingdom Defence Science and Technology Laboratory (DSTL), under contract DSTLX1000152974.

\section*{Declarations}
\textbf{Competing interests:} The authors declare no competing interests. \\

\noindent\textbf{Author contributions:} Both authors contributed to the study conception
and design. Material preparation, data collection and analysis were performed
by E. Badcock. The first draft of the manuscript was written by E. Badcock and
both authors commented on previous versions of the manuscript. Both authors
have read and approved the final manuscript.

\section*{Acknowledgements}
We thank the anonymous reviewers for the care and detail of their comments and suggestions.
Their comments improved the rigour of the analysis and the overall clarity of the paper.

\noindent
We also thank Dr Joe Coppin (DSTL) for his continued support of this research.

\begin{appendices}
\numberwithin{equation}{section}
\section{Body-fitted compressible Navier--Stokes equations}\label{apendeqnsbf}
We work in a body-fitted orthogonal coordinate system $(x,y,z)$, where $x$ is
the streamwise direction aligned normal to a wing's leading edge, $y$ is
wall-normal, and $z$ is spanwise. The metric coefficient
\begin{equation}\label{eqn:metric}
\chi(x,y) = \frac{1}{1 - y\,\kappa(x)},
\end{equation}
accounts for streamwise curvature $\kappa = \theta_x$, with $\theta$ the angle
between the Cartesian $x_1$ axis and $x$. Spatial dimensions are normalised by
the boundary-layer thickness at the initial streamwise station. Terms arising
from the streamwise variation of the curvature, $\kappa_x$, are neglected.

The equations below are written in terms of the total (instantaneous) fields
$(\rho, u, v, w, p, T, \mu)$, denoting density, velocity components,
pressure, temperature, and dynamic viscosity. The corresponding state vector is
\begin{equation}
    q^{\mathrm{tot}} = (p, u, v, w, T)^\top,
\end{equation}
with $\rho$ recovered from the equation of state. 

For an ideal gas with ratio of specific heats $\gamma$, freestream Mach number
$M$, Reynolds number $R$ (based on the same reference length), Prandtl number
$\sigma$, and bulk-to-shear viscosity ratio $m$, the non-dimensional equations
read:
\begin{align}\label{eqn:1}
  {\rho}_t + \chi {u}{\rho}_x + {v}{\rho}_y + {w}{\rho}_z
  + {\rho}\!\left(\chi {u}_x + {v}_y + {w}_z - \kappa\chi {v}\right) = 0,
\end{align}
\begin{align}
\begin{split}
{\rho}&\left({u}_t + \chi {u}{u}_x + {v}{u}_y + {w}{u}_z - \kappa\chi {u}{v}\right) = -\chi {p}_x \\
&+ \frac{1}{R}\bigg[
  {\mu}_x\left[(m+2)\chi^2({u}_x - \kappa{v}) + m\chi({v}_y + {w}_z)\right]
  \\&\quad+ {\mu}_y\left[{u}_y + \chi({v}_x + \kappa{u})\right]
  + {\mu}_z({u}_z + \chi{w}_x) \\
&\quad + {\mu}\Big[
  (m+2)\chi^2({u}_{xx} - \kappa{v}_x)
  + {u}_{yy} - \kappa\chi{u}_y + {u}_{zz}
  + (m+1)\chi({v}_{xy} + {w}_{xz}) \\
&\qquad - \kappa\chi^2({v}_x + \kappa{u})
\Big]\bigg],
\end{split}
\end{align}
\begin{align}
\begin{split}
{\rho}&\left({v}_t + \chi {u}{v}_x + {v}{v}_y + {w}{v}_z + \kappa\chi {u}^2\right) = -{p}_y \\
&+ \frac{1}{R}\bigg[
  {\mu}_x\left[\chi{u}_y + \chi^2({v}_x + \kappa{u})\right] \\
&\quad
  + {\mu}_y\left[(m+2){v}_y + m\chi({u}_x - \kappa{v}) + m{w}_z\right]
  + {\mu}_z({w}_y + {v}_z) \\
&\quad + {\mu}\Big[
  \chi^2{v}_{xx} + (m+2)({v}_{yy} - \kappa\chi{v}_y) + {v}_{zz}
  + (m+2)\kappa\chi^2({u}_x - \kappa{v}) + \kappa\chi^2{u}_x \\
&\qquad + (m+1)(\chi{u}_{xy} + {w}_{yz})
\Big]\bigg],
\end{split}
\end{align}
\begin{align}
\begin{split}
{\rho}&\left({w}_t + \chi {u}{w}_x + {v}{w}_y + {w}{w}_z\right) = -{p}_z \\
&+ \frac{1}{R}\bigg[
  {\mu}_x(\chi{u}_z + \chi^2{w}_x)
  + {\mu}_y({v}_z + {w}_y)
   + {\mu}_z\left[(m+2){w}_z + m{v}_y + m\chi({u}_x - \kappa{v})\right] \\
&\quad + {\mu}\Big[
  \chi^2{w}_{xx} + {w}_{yy} + (m+2){w}_{zz} - \kappa\chi{w}_y \\
&\qquad
  + (m+1){v}_{yz} + (m+1)\chi({u}_{xz} - \kappa{v}_z)
\Big]\bigg],
\end{split}
\end{align}
\begin{align}
\begin{split}
{\rho}&\left({T}_t + \chi {u}{T}_x + {v}{T}_y + {w}{T}_z\right)
  = (\gamma-1)M^2\!\left({p}_t + \chi {u}{p}_x + {v}{p}_y + {w}{p}_z\right) \\
  &\qquad + \frac{1}{\sigma R}\!\left(
  \chi^2{\mu}_x{T}_x + {\mu}_y{T}_y + {\mu}_z{T}_z
  + {\mu}\!\left(\chi^2{T}_{xx} + {T}_{yy} + {T}_{zz} - \kappa\chi{T}_y\right)
\right)\\
&\qquad + \frac{(\gamma-1)M^2}{R}\,{\mu}\,\mathcal{D} ,
\end{split}
\end{align}
where the viscous dissipation is
\begin{align}
\begin{split}
\mathcal{D} ={}&
  (m+2)\!\left({w}_z^2 + \left[{v}_y + \chi({u}_x - \kappa{v})\right]^2\right)
  + 2m\!\left({v}_y{w}_z + \chi({u}_x{w}_z - \kappa{v}{w}_z)\right) \\
&+ 2\!\left(\chi{u}_z{w}_x + \chi{u}_y{v}_x + {v}_z{w}_y\right)
  + \chi^2({v}_x^2 + {w}_x^2)
  + 4\chi(\kappa{v}{v}_y - {u}_x{v}_y) \\
&+ 2\kappa\chi{u}({u}_y + \chi{v}_x)
  + {u}_y^2 + {w}_y^2 + {u}_z^2 + {v}_z^2 + \kappa^2\chi^2{u}^2.
\end{split}
\end{align}
The system is closed by the ideal gas law
\begin{equation}\label{eqn:2}
{p} = \frac{{T}{\rho}}{\gamma M^2},
\end{equation}
and Sutherland's law ${\mu} = {\mu}({T})$.

The linearised equations used
in \S\ref{sec:two} follow from the decomposition $q^{\mathrm{tot}} = Q + q$,
where $Q$ is the base flow and $q = (p', u', v', w', T')^\top$ is the
perturbation. Primes are dropped in \S\ref{sec:two} onwards, so that
$q = (p, u, v, w, T)^\top$ denotes the perturbation throughout.
\section{Recursion Parameter Selection}\label{app:recursion}

Recursion parameter selection for OWNS-R is more demanding than for other
OWNS formulations.  \citet{sleeman2025greedy} showed that heuristic parameters
yielding a stable OWNS-Projection \citep{towneEfficientGlobalResolvent2021} march can produce an unstable OWNS-R \citep{zhuRecursiveOnewayNavierStokes2023} march.  Both retention and removal must therefore be
controlled independently, making explicit parameter expressions essential for
reproducibility.  This appendix provides the complete formulae used for the
subsonic and supersonic configurations in this work.
Here, the spectrum of ${\mathbf{M}}$ at each streamwise location is computed via the QZ algorithm.

\subsection{Spectral Structure}

The continuous spectral branches of Eq.~\eqref{eqn:LHNS} are obtained by
frozen-coefficient analysis with freestream values.  Substituting
$\partial/\partial x \to i\alpha$ and $\partial/\partial y \to i\eta$ into
Eq.~\eqref{eqn:LHNS} yields
\begin{align}\label{eqn:continuous_spectra}
\begin{split}
\alpha_1 = &\frac{\omega - \beta W_\infty}{U_\infty},\qquad \alpha_{2,3}(\eta)
= \frac{\omega - \beta W_\infty}{U_\infty}
  + \frac{i\,(\beta^2+\eta^2)}{U_\infty R},\\
&\alpha_{4,5}(\eta)
= \frac{ M^2 U_\infty\,(\omega - \beta W_\infty) \;\pm\; \sqrt{\mu(\eta)} }
       { M^2 U_\infty^2 - T_\infty }, 
\end{split}
\end{align}
where
\begin{equation}\label{eqn:mu}
\mu(\eta)
= T_\infty\!\left[\,\bigl(M^2 U_\infty^2 - T_\infty\bigr)\bigl(\beta^2+\eta^2\bigr)
              + M^2\,(\omega - \beta W_\infty)^2\,\right].
\end{equation}
and $U_\infty$, $W_\infty$ and $T_\infty$ denote the freestream streamwise velocity, crossflow velocity
and temperature respectively, each non-dimensionalised by the boundary-layer edge values at
the inlet.  The branches $\alpha_{1,2,3}$ represent vorticity/entropy modes;
$\alpha_{4,5}$ are acoustic.  In subsonic flows, $\alpha_{1\text{--}4}$
propagate downstream and $\alpha_5$ propagates upstream; in supersonic flows
all five propagate downstream
\citep{balakumarmalik1992,fedorovTransitionStabilityHighSpeed2011a}.  The
critical spectral locations are
\begin{equation}\label{eqn:ds}
    d_1  = \alpha_1,\quad
    d_{2}  = \frac{\alpha_4+\alpha_5}{2}, \quad
    d_{3,4}  = \alpha_{4,5}(0),
\end{equation}
where $d_1$ is the vorticity/entropy branch endpoint, $d_2$ is the acoustic
branch midpoint, and $d_{3,4}$ are the propagating acoustic endpoints.
The wall-normal grid stretching produces a non-uniform sampling of these
branches: eigenvalues cluster near the branch endpoints (small $|\eta|$) and
become sparse further along each branch (large $|\eta|$).  The recursion
parameters must follow this distribution.

\subsection{Subsonic Flows}\label{app:subsonic}

The acoustic branches exhibit rotational symmetry about $d_2$.  We exploit
this by placing $\beta^+$ to capture downstream modes and obtaining $\beta^-$
through rotation by $\pi$ about $d_2$:
\begin{equation}\label{eqn:recursion_parameters2}
\beta^-_k = -(\beta^+_k - d_2) + d_2.
\end{equation}
This ensures balanced filtering strength on both sides of the spectrum.

The $N_\beta$ parameters are partitioned as
\begin{equation}
N_{\mathrm{v}} = \lceil N_\beta/3 \rceil + 2, \quad N_{\mathrm{a}} = N_\beta - N_{\mathrm{v}},
\end{equation}
allocating $N_{\mathrm{v}}$ to the vorticity/entropy branches and $N_{\mathrm{a}}$
to the acoustics, with denser placement near the real axis where the
discrete eigenvalues concentrate.  The downstream parameters are
\begin{equation}\label{eqn:recursion_parameters1}
\beta^+_k =
\begin{cases}
d_1 + i|\alpha|\, y_{\mathrm{Malik}}\!\left(\dfrac{k-1}{N_{\mathrm{v}} - 3},1,0.4\right) & k=1,\ldots, N_{\mathrm{v}}-2, \\[6pt]
\alpha & k = N_{\mathrm{v}}-1, \\[4pt]
d_1 + 0.6(\alpha - d_1) & k = N_{\mathrm{v}}, \\[4pt]
d_2 + i\alpha_{\mathrm{max}}\, y_{\mathrm{Malik}}\!\left(\dfrac{k-N_{\mathrm{v}}-1}{N_{\mathrm{a}} - 1},1,0.12\right) + p & k=N_{\mathrm{v}} + 1,\ldots, N_\beta,
\end{cases}
\end{equation}
where $\alpha$ denotes the Rayleigh quotient (Eq.~\eqref{eqn:wavenumber_extract}) of the disturbance (or equally $|d_1|$ if in a strongly non-modal regime),
$y_{\mathrm{Malik}}$ implements the non-uniform spacing of
Eq.~\eqref{eqn:malik1}, and $\alpha_{\mathrm{max}} = 28$ bounds the acoustic
branch extent.  The adjustment parameter
\begin{equation}
p = -5 \,\mathrm{sign}(d_1) \times 10^{-2} + 7.5\, i\times 10^{-3}
\end{equation}
serves dual purposes: the imaginary component shifts $\beta^-$ away from
$\sigma({\mathbf{M}})$ to prevent singularities in $\mathbf{P}_{N_\beta}$,
while the real component clamps the contour $|\mathcal{P}_{N_\beta}|=0.5$ so
that it passes through the origin parallel to the real axis, separating the
competing spectral regions.

The parameters are recalculated at each streamwise station as the base flow
evolves and the spectrum migrates.  The discrete mode (T-S or crossflow wave) must be
tracked throughout the march; failure to do so leads to convergence
difficulties or spurious migration to alternative eigenfunctions within
$\sigma({\mathbf{M}})$.

\paragraph{Incompressible two-dimensional flow}
Figure~\ref{fig:spec1} illustrates the parameter placement for a flat-plate
boundary-layer ($M=0.02$, $R=400$, $F=86$, $b=0$).  This is the most
challenging configuration: the downstream and upstream acoustic branches
intersect near the origin, forcing $\beta^+$ and $\beta^-$ parameters into
direct competition.  The rotational symmetry about $d_2$ resolves this by
constructing fictitious upstream counterparts to the vorticity/entropy branches,
balancing the filtering strength and preventing any single parameter from
dominating.

\begin{figure}[ht!]
    \centering
    \includegraphics[width=0.8\linewidth]{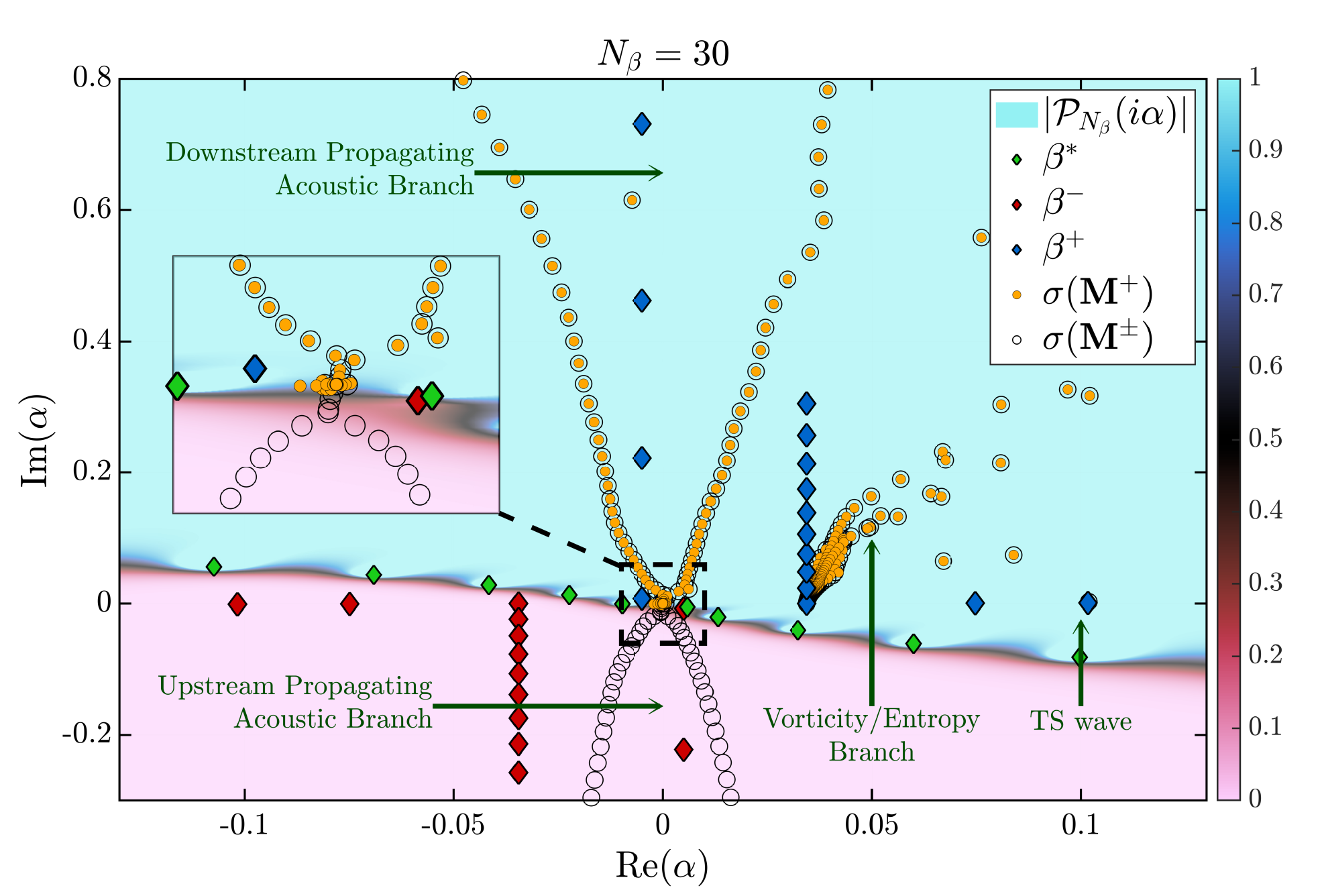}
\caption{Numerical spectrum $\sigma(\mathbf{M})$ and parabolised spectrum $\sigma(\mathbf{P}_{N_\beta}\mathbf{M})$ for a 2D flat-plate boundary layer
    ($M=0.02$, $R=400$, $F=86$, $b=0$, $N_\beta=30$, $c=1$).
    Blue and red diamonds mark the recursion parameters $\beta^+$ and $\beta^-$;
    green diamonds mark the roots $\beta^*$ of the characteristic polynomial,
    lying near the $|\mathcal{P}_{N_\beta}|=0.5$ contour.
    Teal regions ($|\mathcal{P}_{N_\beta}| \approx 1$) indicate retained modes;
    pink regions ($|\mathcal{P}_{N_\beta}| \approx 0$) indicate suppressed modes.}
    \label{fig:spec1}
\end{figure}

\paragraph{Variable Mach number}
Figure~\ref{fig:mach_spec} shows the effect of increasing Mach number at
fixed $b=0$.  As $M$ increases, the real parts of the acoustic branches
$\alpha_{4,5}$ become more pronounced, progressively separating the downstream
and upstream spectra and easing the parameter selection.

\begin{figure}[htbp]
    \centering
    \begin{subfigure}[b]{0.49\textwidth}
        \centering
        \includegraphics[width=\textwidth]{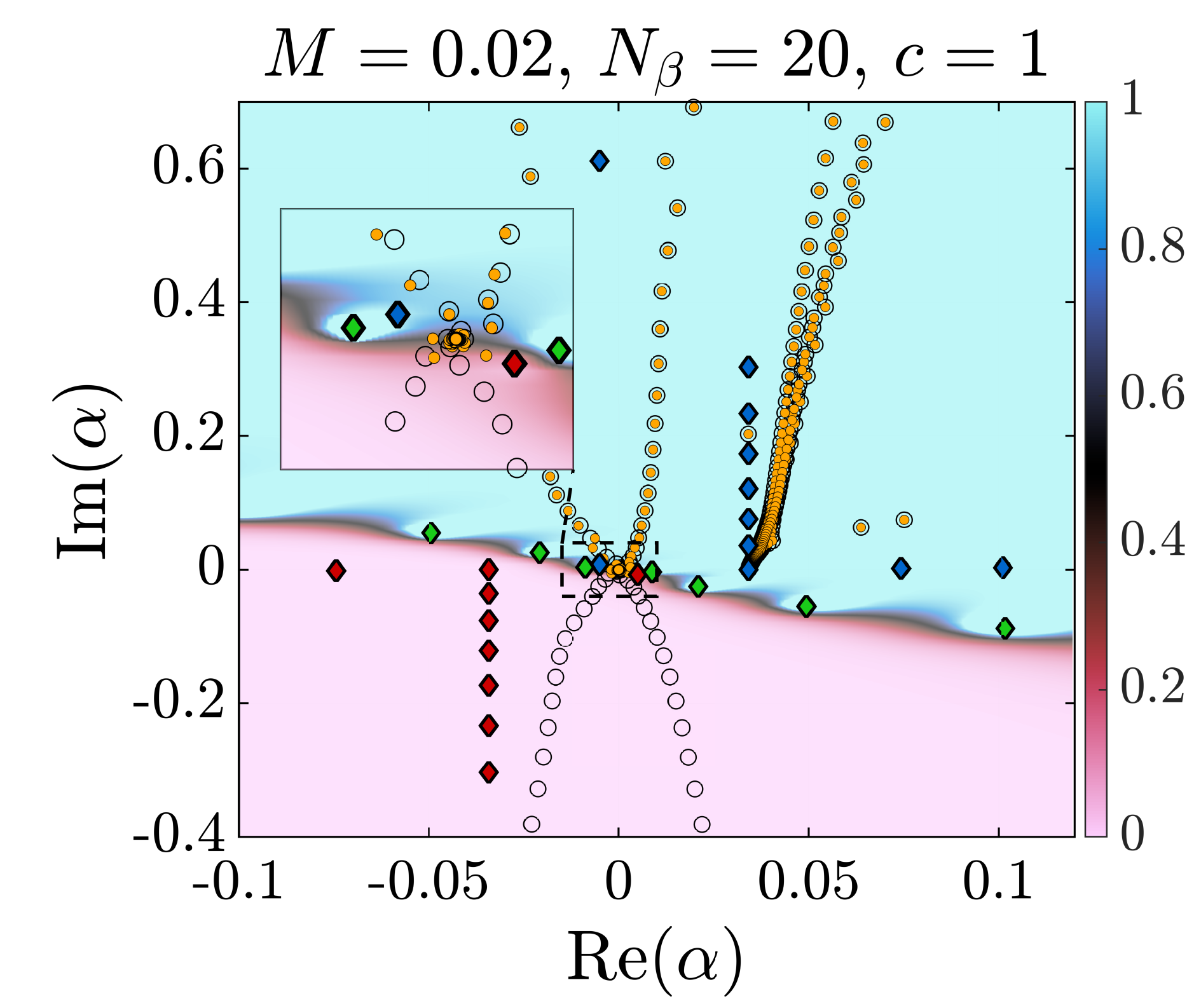}
        \caption{}
        \label{fig:mach_speca}
    \end{subfigure}
    \begin{subfigure}[b]{0.49\textwidth}
        \centering
        \includegraphics[width=\textwidth]{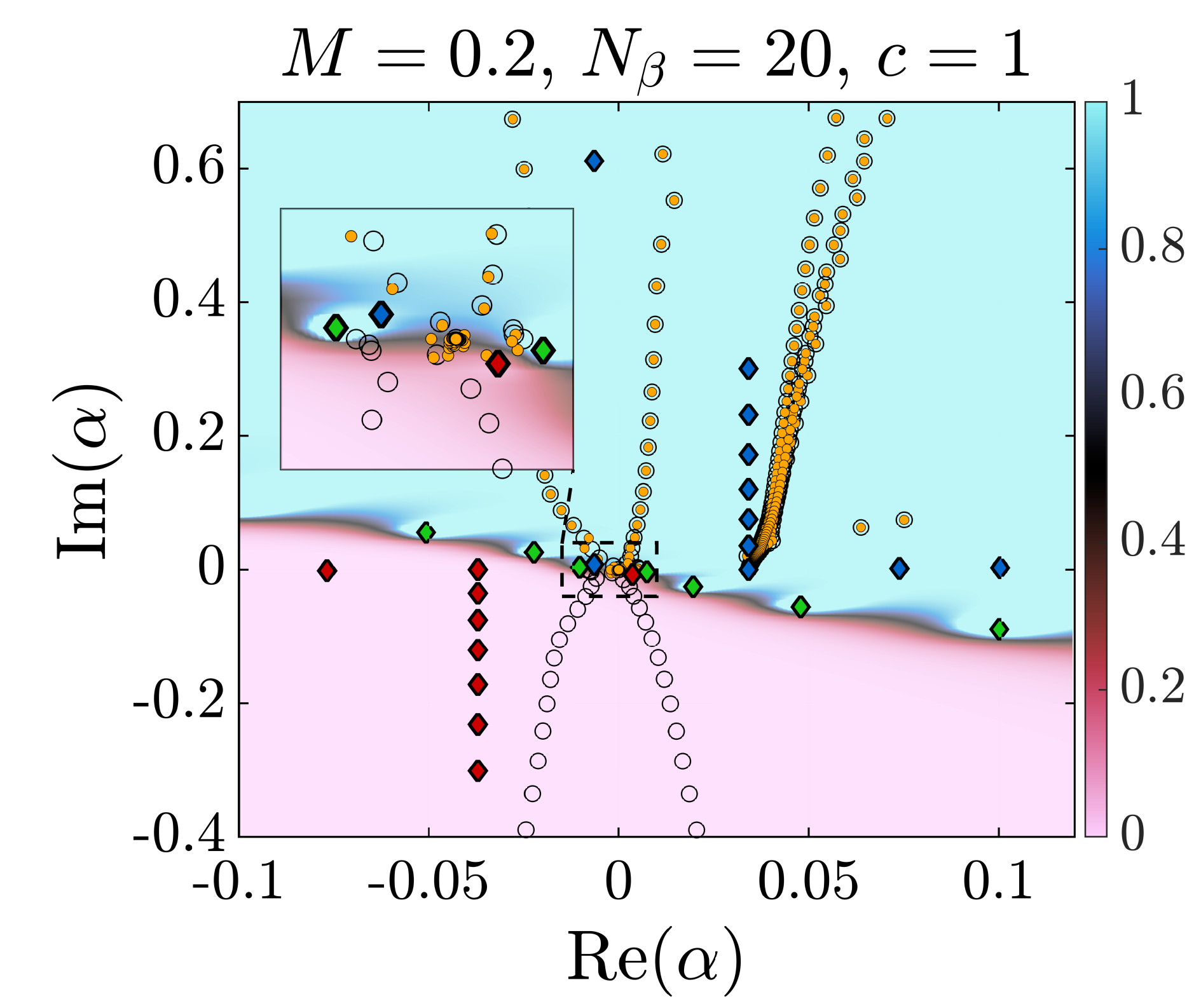}
        \caption{}
        \label{fig:mach_specb}
    \end{subfigure}
    \begin{subfigure}[b]{0.49\textwidth}
        \centering
        \includegraphics[width=\textwidth]{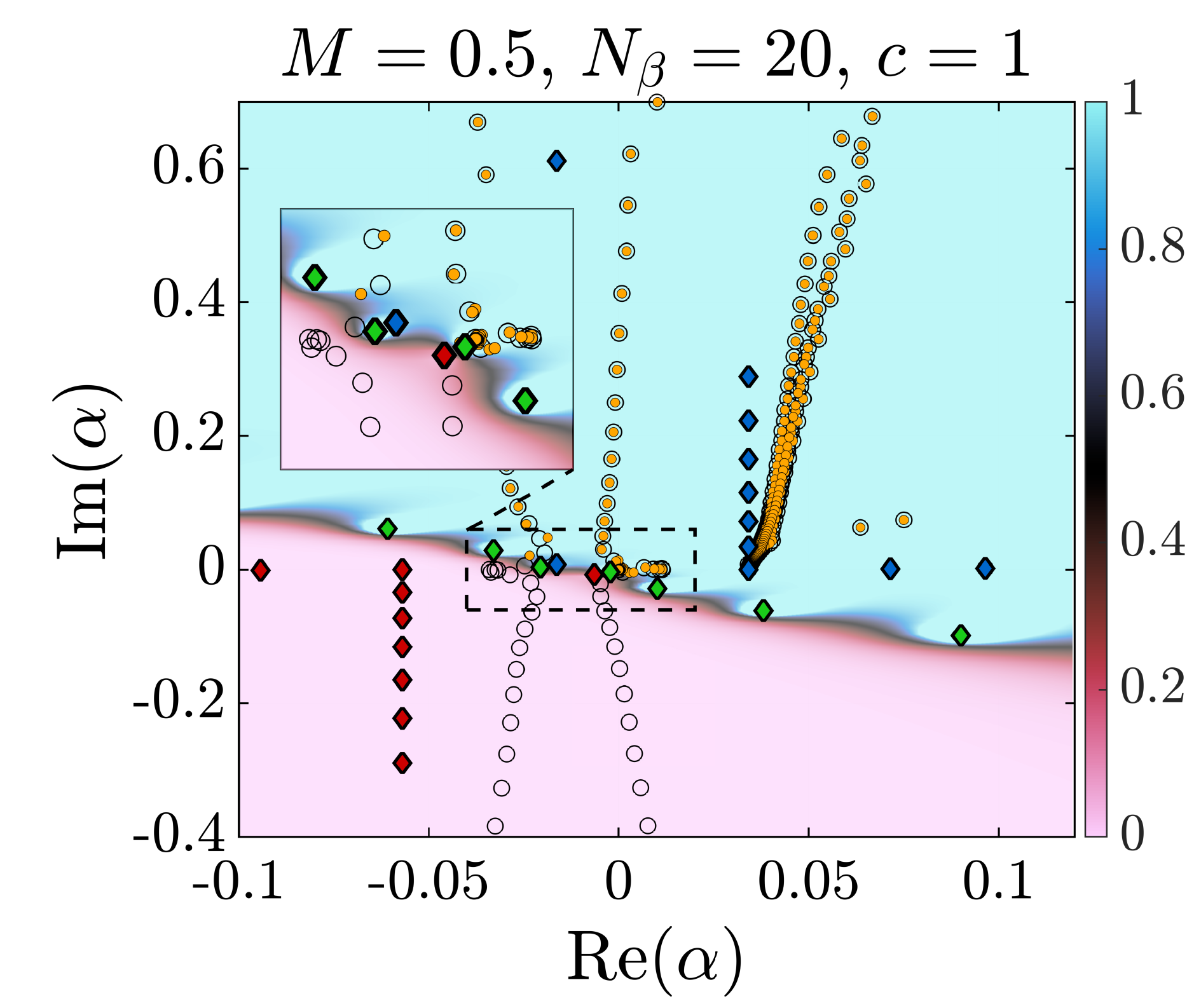}
        \caption{}
        \label{fig:mach_specc}
    \end{subfigure}
    \begin{subfigure}[b]{0.49\textwidth}
        \centering
        \includegraphics[width=\textwidth]{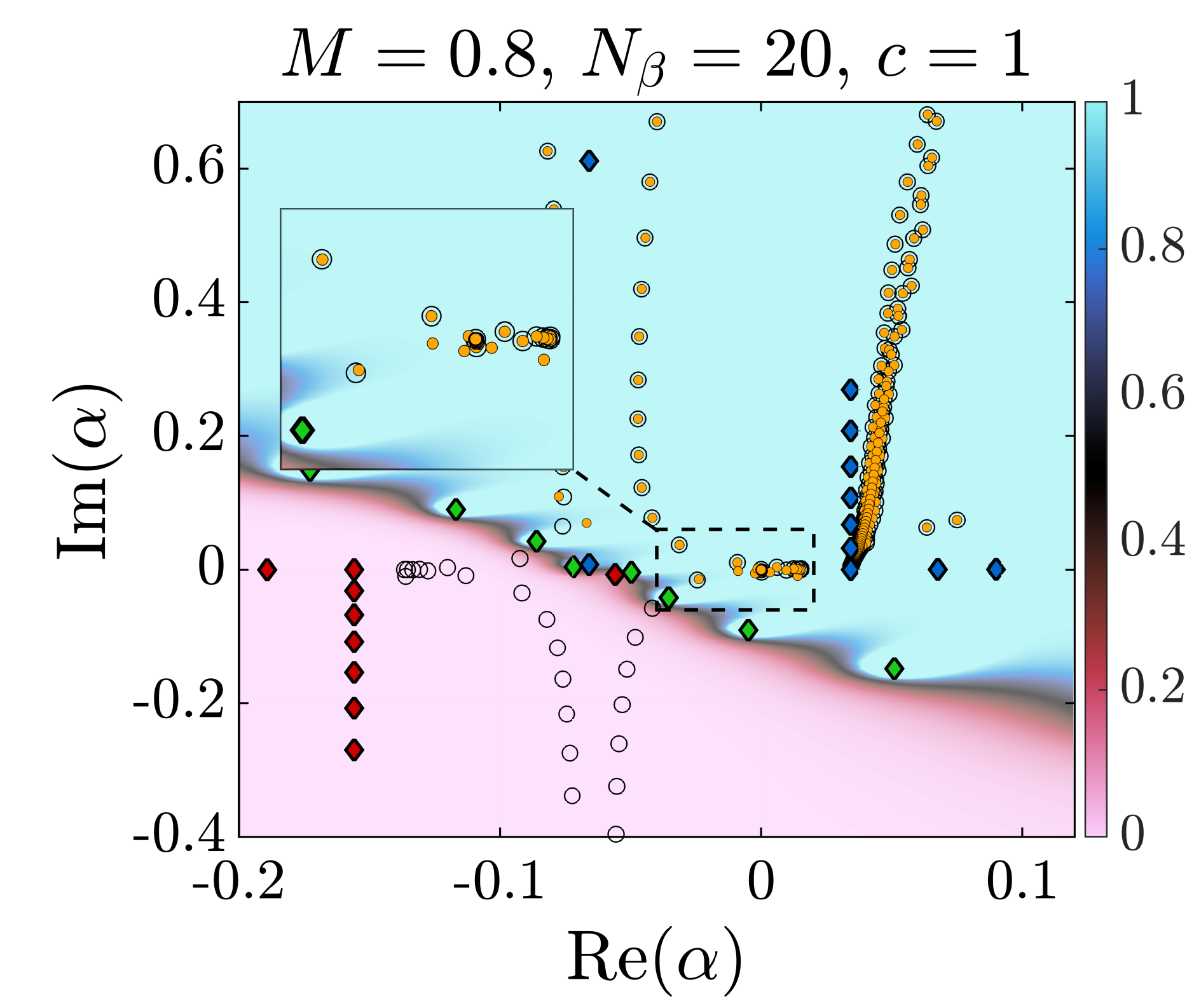}
        \caption{}
        \label{fig:mach_specd}
    \end{subfigure}
    \caption{Recursion parameter placement for subsonic 2D boundary-layers at
    $R_\delta=400$, $b=0$, $W_\infty=0$, $F=86$ with variable Mach number.}
    \label{fig:mach_spec}
\end{figure}

\paragraph{Variable spanwise wavenumber}
Figure~\ref{fig:beta_spec} shows that increasing $b$ separates the acoustic
branches along the imaginary axis, eliminating the competition between
$\beta^+$ and $\beta^-$ parameters entirely.  Parameter selection becomes
straightforward for $b \gtrsim 0.1$.

\begin{figure}[htbp]
    \centering
    \begin{subfigure}[b]{0.49\textwidth}
        \centering
        \includegraphics[width=\textwidth]{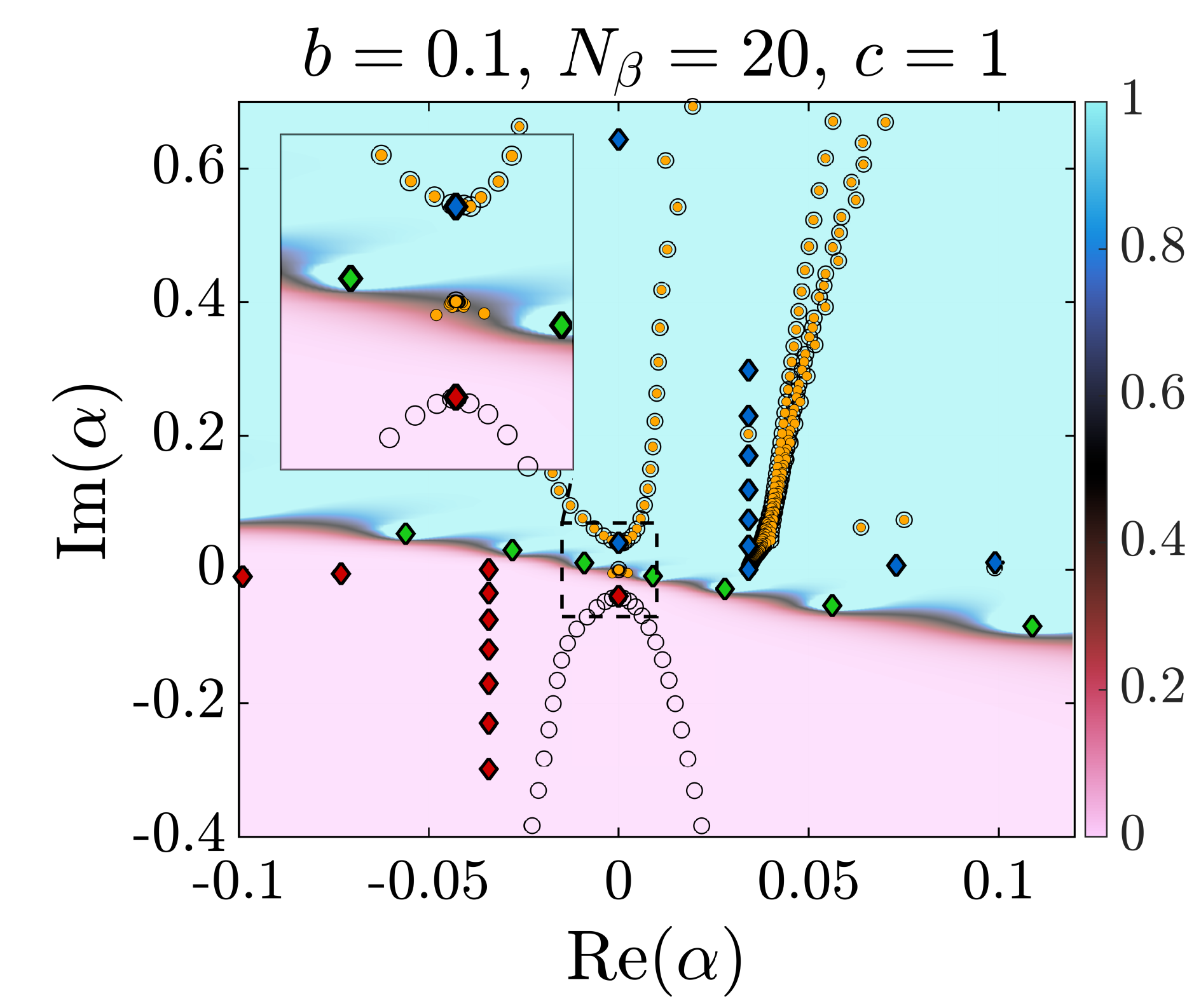}
        \caption{}
        \label{fig:beta_speca}
    \end{subfigure}
    \begin{subfigure}[b]{0.49\textwidth}
        \centering
        \includegraphics[width=\textwidth]{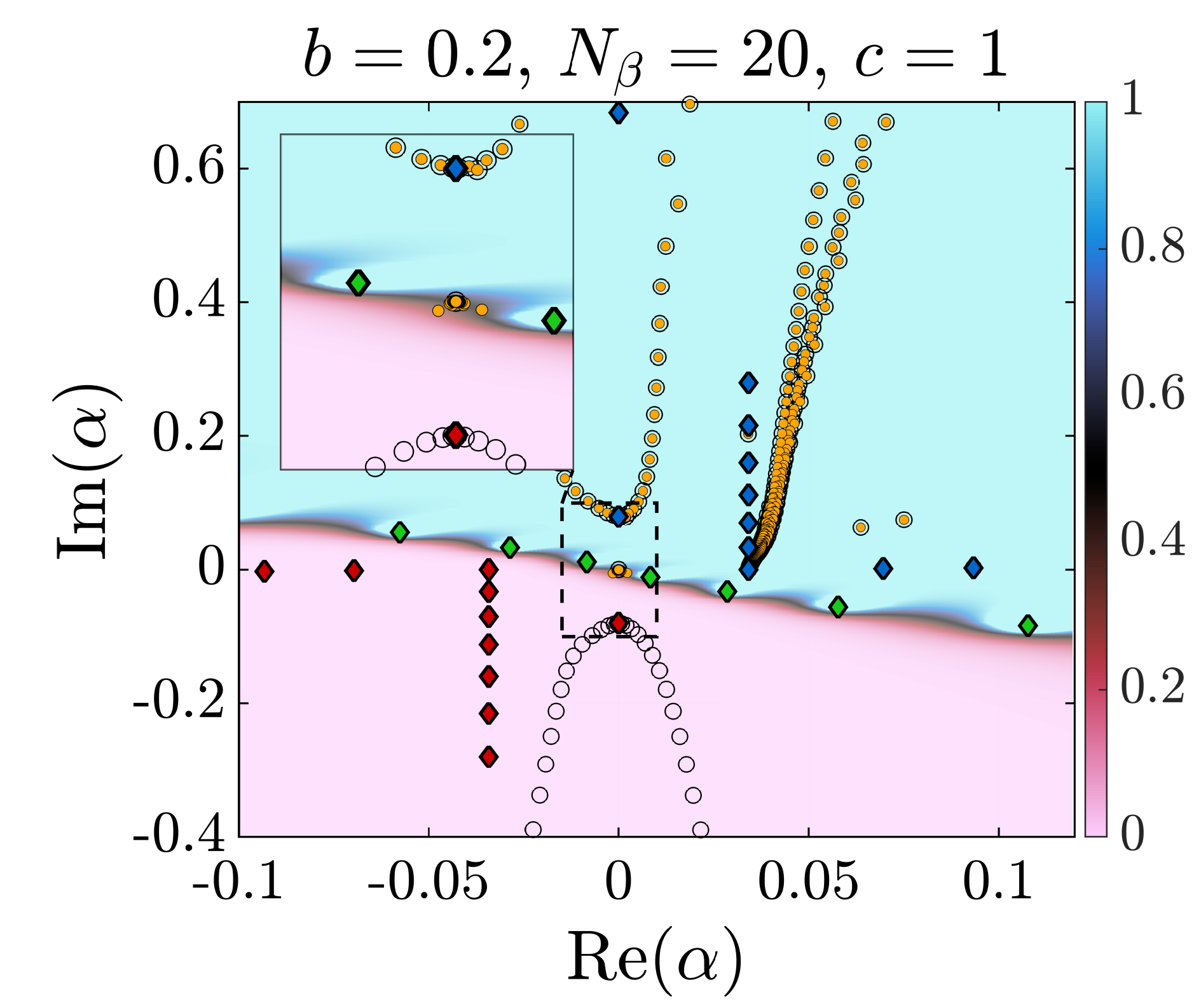}
        \caption{}
        \label{fig:beta_specb}
    \end{subfigure}
    \caption{Recursion parameter placement for subsonic boundary-layers at
    $R_\delta=400$, $M=0.02$, $W_\infty=0$, $F=86$ with $b=0.1$ and $b=0.2$.}
    \label{fig:beta_spec}
\end{figure}

\paragraph{Swept cylinder}
The swept cylinder flow ($R=132$, $\beta=0.4$) introduces an evolving
spectral structure absent in flat-plate cases: the crossflow ratio $W_\infty/U_\infty$
decreases downstream, causing the vorticity/entropy branches to migrate toward
the acoustic spectrum (Fig.~\ref{fig:cyl_spec}).
Equations~\eqref{eqn:recursion_parameters1}
and~\eqref{eqn:recursion_parameters2} accommodate this through two mechanisms.
The symmetric placement of $\beta^\pm$ about $d_2$ maintains balanced filtering
despite the migration, and the $|\alpha|$ scaling factor in
Eq.~\eqref{eqn:recursion_parameters1} contracts the vorticity/entropy parameter
extent as the branches shorten downstream.  The nonzero spanwise wavenumber
keeps the acoustic branches well separated throughout, so the parameter
selection remains robust despite the spectral evolution.

\begin{figure}[htbp]
    \centering
    \begin{subfigure}[b]{0.49\textwidth}
        \centering
        \includegraphics[width=\textwidth]{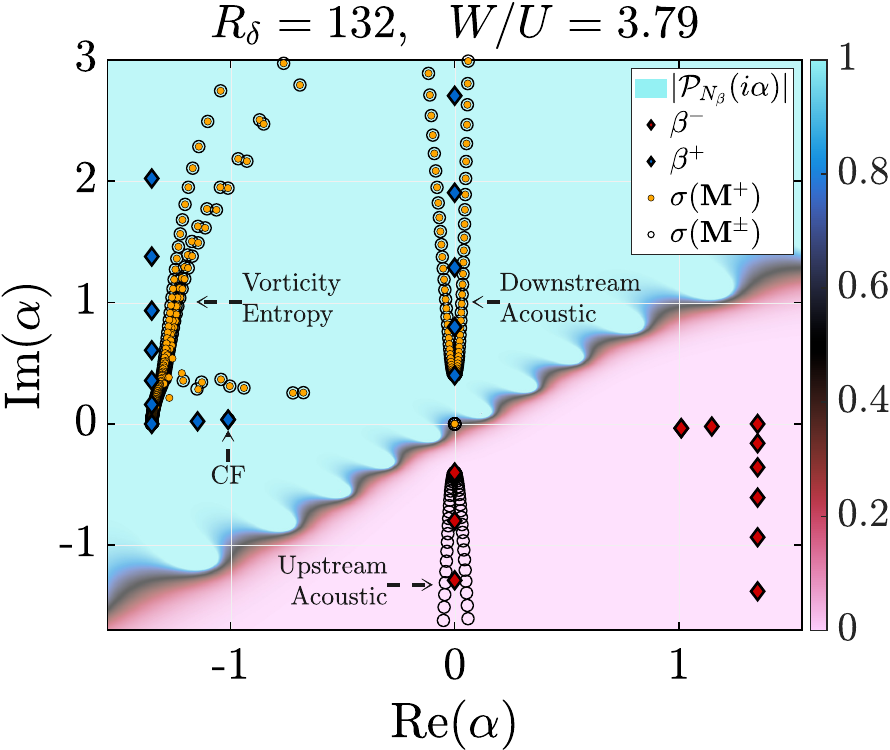}
        \caption{Initial location, $R_\delta=132$}
        \label{fig:cyl_speca}
    \end{subfigure}
    \begin{subfigure}[b]{0.49\textwidth}
        \centering
        \includegraphics[width=\textwidth]{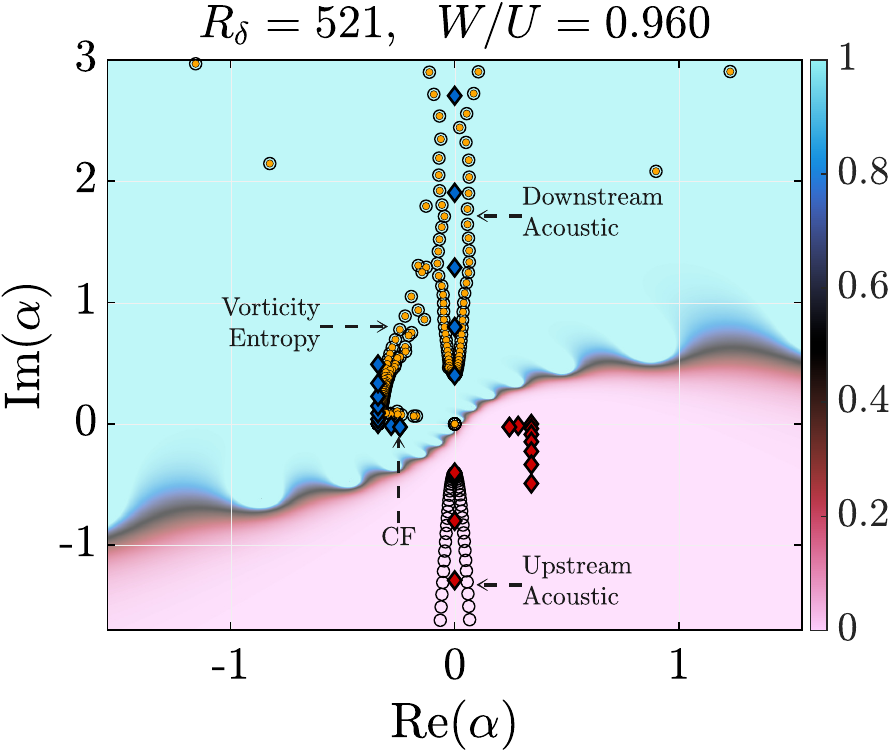}
        \caption{Further downstream, $R_\delta=521$}
        \label{fig:cyl_specb}
    \end{subfigure}
    \caption{Spectral evolution for the swept cylinder ($R=132$,
    $\beta=0.4$, $N_\beta=20$).  As $R_\delta$ increases, $W_\infty/U_\infty$ decreases from
    $3.79$ to $0.960$ and the vorticity/entropy branches migrate toward the
    acoustic spectrum.  The same subsonic parameter formulas
    (Eqs.~\ref{eqn:recursion_parameters1}--\ref{eqn:recursion_parameters2})
    handle this evolution without modification.}
    \label{fig:cyl_spec}
\end{figure}

These parameters have been validated across wall-normal discretisations
($n_y$ from 71 to 1201) and with the tuning parameter $c \in [0.2, 10]$.

\subsection{Hypersonic Flows}\label{app:supersonic}

In supersonic and hypersonic boundary-layers all continuous branches propagate downstream.
Upstream propagation occurs only through discrete acoustic modes, trapped
between the wall and sonic line, that populate the negative imaginary
half-plane with increasing density
\citep{fedorovTransitionStabilityHighSpeed2011a}.  The rotational symmetry
exploited above no longer applies, and the $\beta^+$ and $\beta^-$ parameters
require independent placement strategies.

The downstream parameters are positioned along the vorticity/entropy branches:
\begin{equation}
    \beta^+_j = d_1 + i\, y_{\mathrm{Malik}}\!\left(\frac{j-1}{N_\beta-1},5,0.2\right).
\end{equation}
The upstream parameters target the discrete modes through enhanced clustering
near the real axis:
\begin{equation}
    \beta^-_j = -0.15 - 0.01i - \left(10(-1)^j + 50i\right)y_{\mathrm{Malik}}\!\left(\frac{j-1}{N_\beta-1},1,0.01\right).
\end{equation}
The alternating term $(-1)^j$ provides lateral spreading while maintaining
dense coverage near $\mathrm{Im}(\alpha) = 0$.
A $\beta^+$ can additionally be placed on the Rayleigh quotient if the quantity is well-defined and inside the convex hull of the dominant spectrum.

Figure~\ref{fig:spec_super} confirms that this configuration ($N_\beta = 30$,
$c = 0.5$, optimised for $M = 4.5$) retains the complete downstream spectrum
while eliminating upstream propagation. Modified versions of these parameters have been applied to transonic, supersonic $(M=1.6)$ and an additional hypersonic ($M=6$) case. 

\begin{figure}[ht!]
\centering
    \includegraphics[width=0.8\linewidth]{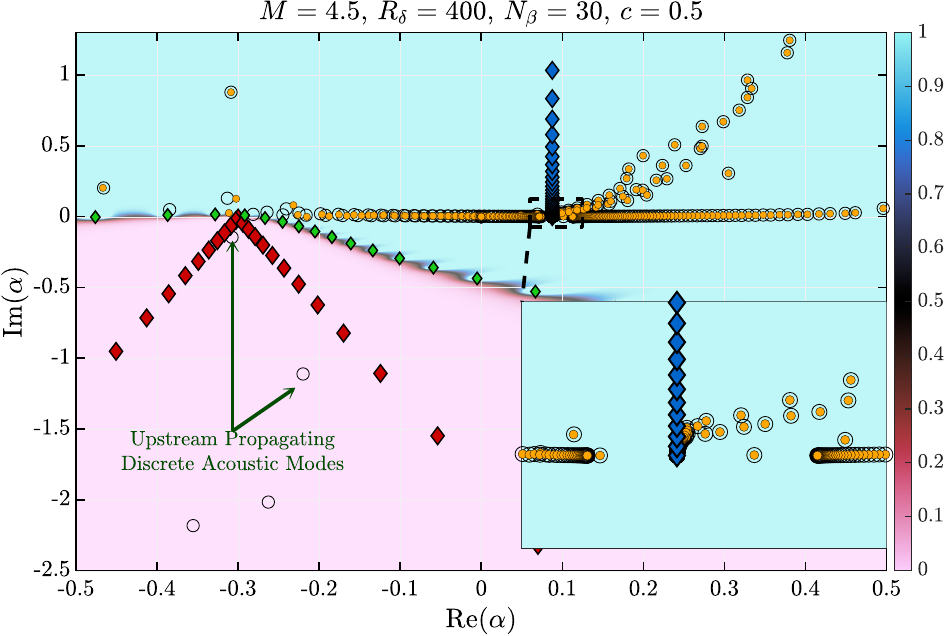}
    \caption{Numerical spectrum $\sigma({\mathbf{M}})$ and parabolised spectrum
    $\sigma(\mathbf{P}_{N_\beta}{\mathbf{M}})$ for a supersonic flat-plate
    boundary-layer ($M = 4.5$, $R_\delta = 400$, $N_\beta = 30$, $c = 0.5$).}
    \label{fig:spec_super}
\end{figure}

\end{appendices}

\bibliography{bib/tcfdowns}

\end{document}